\documentclass[a4paper,11pt]{article}
\pdfoutput=1 
\usepackage{jcappub} 
\usepackage{amssymb, amsmath}
\usepackage{amsmath}
\usepackage{pifont}

\usepackage[T1]{fontenc} 

\title{\boldmath g$\delta N$ formalism}

\author[a, b]{Takahiro Tanaka,}
\emailAdd{t.tanaka@tap.scphys.kyoto-u.ac.jp}
\author[c, d, e]{Yuko Urakawa}
\emailAdd{yukour@post.kek.jp}
\affiliation[a]{Department of Physics, Kyoto University, Kyoto 606-8502, Japan}
\affiliation[b]{Center for Gravitational Physics and Quantum Information, Yukawa Institute for Theoretical Physics, Kyoto University, Kyoto 606-8502, Japan}
\affiliation[c]{Institute of Particle and Nuclear Studies, High Energy Accelerator Research Organization (KEK), Oho 1-1, Tsukuba 305-0801, Japan}
\affiliation[d]{The Graduate University for Advanced Studies (SOKENDAI), Tsukuba 305-0801, Japan}
\affiliation[e]{International Center for Quantum-field Measurement Systems for Studies of the
Universe and Particles (QUP), High Energy Accelerator Research Organization
(KEK), Tsukuba, Ibaraki 305-0801, Japan}

\newcommand{\Mp}{M_{\rm pl}}
\newcommand{\sDif}{$[$sDiff$]~$}
\newcommand{\locality}{$[$locality$]~$}
\newcommand{\approxGe}{$[$approx${\cal G}_{\epsilon}$$]~$}
\newcommand{\bm}[1]{\hbox{\boldmath{$#1$}}}
\newcommand{\sbm}[1]{\hbox{\boldmath{\scriptsize$#1$}}}
\newcommand{\dNex}{g$\delta N$ formalism}
\newcommand{\Lee}{{{\cal L}\hspace{-6pt}\raise 1.5pt \hbox{-}\,}}
\newcommand{\If}{\hat{\cal I}_{(\alpha)f}}

\abstract{The $\delta N$ formalism has been the major computational tool to study the superhorizon evolution of the scalar type perturbation sourced by scalar fields. Recently, this formalism was generalized to compute an arbitrary scalar, vector, and tensor type perturbations, including the gravitational waves (GWs), sourced by an arbitrary bosonic fields. In this paper, we explain how to use the generalized $\delta N$ formalism (the g$\delta N$ formalism), considering a model with U(1) gauge fields as a concrete example. Several new findings on this model and prospects on future gravitational wave experiments are also discussed, including the condition for the two linear polarizations of GWs to have different amplitudes. This paper provides a detailed explanation of our previous paper published in Physical Review Letters. We also discuss the Weinberg's adiabatic mode for an anisotropic background, showing a qualitative difference from the one for the FLRW background.}

\begin{document}
\maketitle
\flushbottom

\section{Introduction}
\label{sec:intro}
Since the first detection of gravitational waves (GWs) by LIGO in 2015~\cite{LIGOScientific:2016fpe}, the observations of GWs have significantly deepened our understanding of the universe, marking a pivotal shift in the field of astrophysics. Subsequent detections by LIGO, along with its partners VIRGO and KAGRA, have cataloged numerous events, including black hole mergers and neutron star mergers. These observations have tested Einstein's theory of general relativity under extreme conditions and provided new insights on the origin of black holes and their impacts on the cosmological history. Last year, NANOGrav reported a potential signal of a nano-hertz gravitational wave background~\cite{NANOGrav:2023gor}, stimulating intensive discussions about possible new physics~\cite{NANOGrav:2023hvm}. Gravitational waves are also expected to uncover the detailed model of cosmic inflation (see, e.g., Ref.~\cite{LiteBIRD:2022cnt}).

To connect the theoretical prediction of inflation to various observations, we need to solve the evolution during the epoch when the scale of interest was much larger than the accessible scale by any causal propagation. The hierarchical difference between these two scales introduces a new expansion parameter $\epsilon$, providing a different expansion scheme from the standard cosmological perturbation theory. This expansion scheme is called the gradient expansion~\cite{Salopek:1990jq, Deruelle:1994iz, Shibata:1999zs}. Different local regions in the universe which are more distant than the scale of the causal communication evolve mutually independently. In the gradient expansion, the inhomogeneous universe is identified with glued numerous local regions which evolve independently. This is called the separate universe approach~\cite{Salopek:1990jq, Wands:2000dp, Lyth:2004gb}. This approach provides a swift and simple method to compute the long-wavelength evolution, namely at non-linear order of perturbations, which is necessary to evaluate the primordial non-Gaussianity.

Based on the gradient expansion, the $\delta N$ formalism \cite{Starobinsky:1982ee, Starobinsky:1986fxa, Sasaki:1995aw, Sasaki:1998ug, Lyth:2004gb} was developed to calculate the primordial spectrum of the adiabatic curvature perturbation $\zeta$. (See Refs.~\cite{Tanaka:2006zp, Weinberg:2008nf, Weinberg:2008si, Takamizu:2010xy, Naruko:2012fe} for a discussion about the next to leading order of the gradient expansion and Refs.~\cite{Tanaka:2010km, Abolhasani:2019cqw} for a review.) The $\delta N$ formalism not only enables a simple computation of the superhorizon dynamics, but also provides an intuitive understanding on the evolution of primordial fluctuations. However, their application was limited to a system which contains only scalar fields, and it could not be used to compute vector and tensor type perturbations. Therefore, the conventional $\delta N$ formalism was not applicable for a computation of primordial GWs.

Recently, we have shown that the separate universe approach can be generically applied, as long as the theory under consideration is local and preserves the spatial diffeomorphism invariance~\cite{Tanaka:2021dww}. Based on this, we can generalize the $\delta N$ formalism to compute arbitrary scalar, vector, and tensor type perturbations, including GWs sourced by arbitrary bosonic fields. We dubbed this generalized formalism as the generalized $\delta N$ formalism (\dNex). In Ref.~\cite{Tanaka:2023gul}, we have explicitly shown that the \dNex~is indeed applicable to compute primordial GWs, considering a concrete model of inflation. The purpose of this paper is to provide a more detailed explanation beyond Ref.~\cite{Tanaka:2023gul}.

Cosmological observations with an enhanced precision may uncover not only the property of the inflaton but also those of the spectator fields, which were energetically subdominant during inflation. During inflation, spectator fields with various spins might have been excited, leaving a possible imprint in the primordial perturbations. The generation of a vector field through the kinetic coupling with the inflaton $\phi$~\cite{Ratra:1991bn, Caldwell_2011, Martin:2007ue} or non-minimal coupling with gravity~\cite{Turner:1987bw}, was discussed as a mechanism of the primordial magnetogenesis. While a too large backreaction spoils inflation~\cite{Demozzi:2009fu}, a mild backreaction may lead to a sustainable anisotropic inflation with shear on large scales~\cite{Watanabe:2009ct, Watanabe:2010fh, Kanno:2010nr, Soda:2012zm}, providing a counterexample of cosmic no-hair conjecture~\cite{Wald:1983ky}. (See also Refs.~\cite{Gong:2019hwj,Graham:2015rva, Nakayama:2019rhg, Kitajima:2023fun}.) It was later extended to an inflaton with a more general kinetic term \cite{Ohashi:2013pca}, a coupling with a two-form field \cite{Ohashi:2013qba} and a SU(2) gauge field \cite{Murata:2011wv, Maeda:2013daa} (see also Ref.~\cite{Maleknejad:2011jw}). In Ref.~\cite{Kehagias:2017cym}, an enhancement of general integer spin fields through the kinetic coupling with the inflaton was studied as a framework of cosmological collider physics~\cite{Arkani-Hamed:2015bza, Lee:2016vti, Ghosh:2014kba}. The imprints of these non-zero spin fields encoded in the primordial curvature perturbation can be explored through the observations of the cosmic microwave background (CMB)~\cite{Akrami:2019izv, Bartolo:2017sbu, Bordin:2019tyb,Franciolini:2018eno, Sohn:2024xzd} and the large scale struction (LSS)~\cite{MoradinezhadDizgah:2018pfo, MoradinezhadDizgah:2018ssw, Schmidt:2015xka, Chisari:2016xki, Kogai:2018nse, Kogai:2020vzz}.

In this paper, as a simple example, we consider U(1) gauge fields which have the kinetic mixing with scalar fields such as the inflaton and the spectator fields. Such gauge fields ubiquitously appear, e.g., in the 4D low energy effective field theory of string theory. We calculate the primordial curvature perturbation and GWs sourced by both the scalar fields and the gauge fields. (Earlier studies of the gravitational waves in Bianchi background can be found, e.g., in Refs.~\cite{Gumrukcuoglu:2007bx, Gumrukcuoglu:2008gi, Pereira:2007yy}.) In this model, the two linear polarization modes of GWs generically evolve differently. Using the \dNex, we can easily understand the condition that the linear polarization modes have different power spectra. Several qualitative difference between models with a single gauge field and the ones with multiple gauge fields also becomes manifest.

This paper is organized as follows. In Sec.~\ref{Sec:deltaN}, we provide a brief introduction of the \dNex, explaining the basic formulae to compute the curvature perturbation and the GWs. In Sec.~\ref{Sec:PExpansion}, we perturbatively compute the mapping between the horizon crossing and the final time, e.g., at the reheating surface. In Sec.~\ref{Sec:PowerS}, using the formulae derived in Sec.~\ref{Sec:PExpansion}, we compute the power spectrums of the curvature perturbation, GWs, and also entropy perturbation. Finally, in Sec.~\ref{SSec:LiteBIRD}, we point out that there is an interesting parameter range in which a detection of the primordial GWs can play a crucial role. In Appendix \ref{Sec:WAM}, we discuss the Weinberg's adiabatic mode in an anisotropic background, which can depend on time unlike the one in the FLRW background.

\section{g\texorpdfstring{$\delta N$}{delta N} formalism} \label{Sec:deltaN}
In this subsection, after reviewing the basics of the g$\delta N$ formalism, developed in Ref.~\cite{Tanaka:2021dww}, we apply it to compute the long-wavelength evolution in an inflation model with U(1) gauge fields.

\subsection{Brief review of g\texorpdfstring{$\delta N$}{delta N} formalism}
In this subsection, we briefly summarize the basic conditions for the g$\delta N$ formalism to be applicable, following Ref.~\cite{Tanaka:2021dww}.

\subsubsection{Basic conditions}
The \dNex~computes the evolution of the large scale fluctuations at the leading order of the gradient expansion \cite{Salopek:1990jq, Shibata:1999zs}, which is an expansion scheme with respect to the spatial gradient at each given time. This provides a useful tool to address the long wavelength evolution of the fluctuations in a cosmological setup (see also Ref.~\cite{Deruelle:1994iz}). The gradient expansion starts with smoothing the small scale fluctuations. As a consequence of the smoothing, operating the spatial gradient gives rise to the suppression by a small parameter $\epsilon$, which is usually characterized by the spatial variation of $\varphi^a$ within each causally connected. At the leading order of the gradient expansion, $\epsilon$ is simply sent to 0.

The \dNex~generically applies to a theory which satisfies
\begin{itemize}
    \item \sDif : The Lagrangian density for the coarse-grained fields $\{\varphi^a (t,\, \bm{x})\}$, ${\cal L}(t, x^i)$, remains invariant under the $d$-dim spatial diffeomorphism (Diff),  
\begin{align}
     x^i \to \tilde{x}^i(t,\, x^i).
\end{align}
 \item \locality : After solving all the constraint equations except for gauge constraints, the effective dynamics of the coarse-grained fields is described by a local Lagrangian density ${\cal L}(t, x^i)$, which is a function of fields at $x^\mu=(t, x^i)$. Taking variation of the corresponding action gives local field equations for the coarse-grained fields.
\end{itemize}
We call a constraint which generates a gauge symmetry a gauge constraint, expressing it as ${\cal G}$. Gauge constraints can be derived by taking derivative with respect to Lagrange multipliers associated with the gauge symmetry. The variation of the action for ${\cal L}(t, x^i)$ gives field equations for the set of the fields $\{ \varphi^a \}$ which remain after all the non-gauge constraints are solved and the gauge conditions that remain even in the homogeneous universe are employed. 

Under these two conditions, the separate universe evolution~\cite{Salopek:1990jq, Lyth:2004gb}, {\it i.e.},
\begin{align*}
&[{\rm Separate~universe~evolution}~(\star)]:\cr
    &\quad \,\,(\mbox{Solving the large scale dynamics of an inhomogeneous universe}) \cr
& \quad \,\,\,= (\mbox{Solving the dynamics of glued nearly homogeneous patches independently})\,
\end{align*}
holds. When $(\star)$ holds, the time evolution of the inhomogeneous universe is determined solely by solving a set of the corresponding ordinary differential equations, where the spatial gradient terms are simply dropped, for a corresponding initial condition. The spatial inhomogeneity of $\{\varphi^a\}$ is incorporated by assigning different initial conditions, correspondingly, to these fields in different causal patches as
\begin{align}
   \lim_{\epsilon \to 0}  \varphi^a (t,\, \bm{x}) = \varphi^a (t; \{\varphi^{a'}(t_*)= \varphi^{a'}(t_*,\, \bm{x} ) \}')\,, \label{Cond:asym}
\end{align}
where $t_*$ is the initial time of the superhorizon evolution, which is typically set right after the horizon crossing. Here, $\varphi^a (t)$ in the right hand side denotes the time evolution of a set of fields which satisfy the field equations with a given initial conditions in the homogeneous universe.

The key to verify the separate universe evolution ($\star$), enabling an application of the \dNex, is to solve the enhanced set of the fields, maintaining additional degrees of freedom which can be removed by solving all the constraint equations and imposing gauge conditions. This enhanced set should be distinguished from the set of the physical degrees of freedom $\{ \varphi^{a_{\rm phys}} \}_{\rm phys}$, from which all the gauge degrees of freedom are removed. To clarify the definition of the enhanced set of the fields, let us classify the gauge constraints ${\cal G}$ into those whose terms are all suppressed by at least one spatial gradient, {\it i.e.}, 
$$
{\rm a~constraint~} \chi \in  {\cal G}_{\epsilon} \qquad \quad {\rm if~} \chi \in {\cal G} {\rm ~and~} \chi = {\cal O}(\epsilon)\,,
$$
and the others ${\cal G} \cap {\cal G}_{\epsilon}^c$. Here, we put the superscript $c$ to denote a complementary set. Similarly, let us also classify the whole set of gauge conditions, ${\cal X}$,  to those which vanish in the limit $\epsilon \to 0$, ${\cal X}_{\epsilon}$ and the others ${\cal X} \cap {\cal X}_{\epsilon}^c$. Now, let us define the three different sets of the metric and matter fields as 
\begin{align*}
     & \{ \varphi^a \}: \mbox{All metric and matter fields which remain after solving}~ {\cal C} \cap {\cal G}^c~ \&~ \mbox{imposing}~{\cal X} \cap {\cal X}_{\epsilon}^c \cr
     &\xrightarrow[ 
 \mbox{solving}~({\cal G} \cap {{\cal G}}_{\epsilon}^c)]{} ~~~ 
 \{\varphi^{a'}\}': \mbox{variables that specify initial conditions of separate universes}~~~ \cr
 & \xrightarrow[\mbox{solving}
 {\cal G}_{\epsilon}~\mbox{and removing residual gauge DOFs}]{}~~~  \{\varphi^{a_{\rm phys}}\}_{\rm phys}\,,
\end{align*}
where ${\cal C}$ denotes all the constraints in the system under consideration. Removing the residual gauge degrees of freedom (DOFs) at the last step, in general, requires imposing a gauge condition ${\cal X}_{\epsilon}$. In Sec.~\ref{SSec:constraints}, considering a concrete example, we will classify the constraints and the gauge conditions.

\subsubsection{Noether charge} \label{SSec:Noether}
In the \dNex, the Noether charge of a global transformation, 
\begin{align}
    x^i \to \tilde{x}^i = x^i + {M^i}_j \, x^j \,, \label{Exp:LGT}
\end{align}
which is a large spatial gauge transformation~\cite{Urakawa:2010it, Urakawa:2010kr} (see also Ref.~\cite{Hinterbichler:2013dpa}), plays an important role. Here, ${M^i}_j$ denotes an infinitesimal constant rank $(1,\, 1)$ tensor with $9$ independent components. The trace part of ${M^i}_j$ describes the scale transformation and the traceless part describes the shear transformation and the rotation. 

Here, we focus on the traceless part of $M^i{}_j$, i.e., we assume $M^i{}_i=0$. 
Then, the invariance of the action under Eq.~(\ref{Exp:LGT}) implies 
\begin{align}
    \frac{d}{dt} \int d^3\bm{x}\,  {Q^i}_{j\sbm{x}}  = 0\,,  \label{Exp:Noetherglobal}
\end{align}
where ${Q^i}_{j\sbm{x}}$ is the Noether charge density given by~\cite{Tanaka:2021dww} 
\begin{align}
  {Q^i}_{j\sbm{x}}\equiv \frac{1}{\Mp^2} \left( \left[- 2 {\pi^i}_j  + \sum_{\alpha} \pi^{\alpha}_{\rm mat}  \frac{\partial  \delta_{M} \varphi_{\rm mat}^{\alpha}}{\partial {M^j}_i} \right]^{\rm TL} \right)\,. 
   \label{Eq:Noether}
\end{align}
Here, $\Mp \equiv 1/(8 \pi G)$ and $\pi^{ij}$ and $\pi^{\alpha}_{\rm mat}$ are conjugate momenta defined as
\begin{align}
  \pi^{ij} \equiv \frac{\partial (N \sqrt{g} {\cal L})}{\partial \dot{\gamma}_{ij}}\,, \qquad 
  \pi^{\alpha}_{\rm mat} \equiv   \frac{\partial (N\!\sqrt{g} {\cal L})}{\partial \dot{\varphi}_{\rm mat}^{\alpha}}\,,  \label{Exp:piij} 
\end{align}
with ${\cal L}$ being the total Lagrangian density, and $\delta_{M} \varphi_{\rm mat}^{\alpha}$ denotes the change of the fields under Eq.~(\ref{Exp:LGT}). Here and hereafter, the superscript TL indicates the operation to pick up the traceless part. For instance, a scalar field $\phi$ and a vector field $A_\mu$ transform the transformation (\ref{Exp:LGT}) as
\begin{align}
    & \delta_M \phi = \delta_M A_0 = 0\,, \qquad \delta_M A_i = -  {M^k}_i A_k \,,
\end{align}
where higher order terms of ${M^i}_j$ are ignored, considering  infinitesimally small $|{M^i}_j|$. For our later convenience, we have inserted the constant factor $8 \pi G =1/\Mp^2$ in the definition of ${Q^i}_{j \sbm{x}}$. The scale transformation, which is described by the trace part of ${M^i}_j$ does not give a conservation because of the  contribution of the Jacobian factor. Instead, the invariance of the action under the scale transformation gives the equation which corresponds to the trace part of the $(i,\, j)$ component of the Einstein equation.

The conservation of the global quantity, (\ref{Exp:Noetherglobal}), itself is not very useful to solve the separate universe evolution. As shown in Eq.~(\ref{Eq:Noether}), the Noether charge density ${Q^i}_{j \sbm{x}}$ is expressed in terms of the set of the coarse-grained fields $\{ \varphi^a_{\sbm{x}} \}$ at each Hubble patch. Under the \locality condition, we can show that the Noether charge density ${Q^i}_{j \sbm{x}}$ is approximately conserved, satisfying 
\begin{align}
    \frac{d}{dt} {Q^i}_{j \sbm{x}} = {\cal O}(\epsilon)\,. \label{Exp:approxcsv}
\end{align}
The detailed derivation can be found in Sec. 3.4.1 of Ref.~\cite{Tanaka:2021dww}. In Sec.~\ref{SSec:formulagdN}, we will derive the basic equation in the \dNex, using Eq.~(\ref{Exp:approxcsv}). 

\subsection{Basic formulae in g\texorpdfstring{$\delta N$}{delta N} formalism} \label{SSec:formulagdN}
In this subsection, we derive the basic formulae in \dNex, which gives the mapping of the fluctuations between the horizon crossing, {\it i.e.}, $t = t_*$, and the final time, at which we want to evaluate the correlation functions. Generalizing the conventional $\delta N$ formalism, the \dNex~can provide the mapping for the curvature perturbation and the GWs which depend also on the vector and tensor fields at $t = t_*$. In this paper, we express the 4-dimensional line element as
\begin{align}
    ds^2 = - N^2 dt^2 + g_{ij} (dx^i + N^i dt) (dx^j + N^j dt)\,, 
\end{align}
with $i, j = 1, \cdots, 3$. We express the spatial metric as
\begin{align}
    g_{ij} = e^{2 \psi}\, \gamma_{ij}\,,
\end{align}
where $\gamma_{ij}$ satisfies $\det[\gamma]=1$. Using $\psi$, with which the determinant of $g_{ij}$ is given by $g= e^{6\psi}$. As a local gauge condition, which validates the separate universe evolution ($\star$), we impose  
\begin{align}
   N_i (t,\, \bm{x})=0 \,,\label{Cond:Ni}
\end{align}
as a spatial gauge condition.

In $N_i=0$ gauge, the expansion $K$ and the shear ${A^i}_j$, given by the trace part and traceless part of the extrinsic curvature, read
\begin{align}
    & K = \frac{3}{N} \dot{\psi}\,, \label{Exp:KS} \\
    & {A^i}_j = \frac{1}{2N} \left( \delta^i_k \delta^l_j - \frac{1}{3} \delta^i_j \delta^l_k \right) \gamma^{km} \dot{\gamma}_{ml} = \frac{1}{2N} \gamma^{im} \dot{\gamma}_{mj} = - \frac{1}{2N} \dot{\gamma}^{im} \gamma_{mj}\,. \label{Exp:AijS}
\end{align}
Integrating Eq.~(\ref{Exp:KS}) between $t_*$ and the final time $t_f$, we obtain the time evolution of $\psi$ as 
\begin{align}
    \psi(t_f,\, \bm{x}) - \psi(t_*,\, \bm{x}) = \frac{1}{3} \int^{t_f}_{t_*} dt N(t,\, \bm{x}) K(t,\, \bm{x}) \,. \label{Exp:deltaN2} 
\end{align}
In this gauge, $\psi$ directly corresponds to the non-perturbative $e$-folding number. Computing the $e$-folding number spent at each local homogeneous Hubble patch with different initial conditions by using Eq.~(\ref{Exp:deltaN2}), one can evaluate the long-wavelength evolution of $\psi$. At the linear perturbation, the curvature perturbation $\zeta$ is given by the linear combination of $\psi$ and the longitudinal part of $\gamma_{ij}$ (see Eq.~(\ref{Def:zetalinear})).

One may wonder if we can compute the gravitational waves similarly just by solving the evolution of the background homogeneous universe. In Ref.~\cite{Tanaka:2021dww}, it was shown that the separate universe evolution $(\star)$ can be verified as well as in the presence of large scale gravitational waves. The corresponding equation for GWs can be obtained by integrating Eq.~(\ref{Exp:AijS}) as
\begin{align}
    \gamma_{ij}(t_f,\,\bm{x}) - \gamma_{ij}(t_*,\,\bm{x}) = 2 \int^t_{t_*} dt N(t,\, \bm{x}) \gamma_{\{i|l} (t,\, \bm{x}) {A^l}_{|j\}}  (t,\, \bm{x})   \,,\label{Exp:deltagamma} 
\end{align}
where the indices $i$ and $j$ enclosed in braces are symmetrized. Using Eq.~(\ref{Exp:deltagamma}), one can compute GWs and the metric contribution in $\zeta$. These mapping formulae (\ref{Exp:deltaN2}) and (\ref{Exp:deltagamma}) are purely geometrical relations which do not depend on models under consideration.

We determine the time slicing at the horizon crossing, requiring 
\begin{align}
    \delta \psi (t_*,\, \bm{x})=0\,, \label{Cond:slicingini}
\end{align}
and the one at $t=t_f$, requiring 
\begin{align}
    \delta K (t_f,\, \bm{x})=0\,, \label{Cond:slicingfin}
\end{align}
both of which do not vanish in the limit $\epsilon \to 0$, {\it i.e.}, are not in ${\cal X}_\epsilon$. The residual gauge degrees of freedom in the coordinates can be eliminated by imposing additional gauge conditions, e.g., at $t=t_*$, if one wishes. We introduce $\delta {\gamma^i}_j$ as
\begin{align}
    \gamma_{ij}(t,\, \bm{x}) \equiv \bar{\gamma}_{ik}(t) \left[ e^{\delta \gamma(t,\, \sbm{x})} \right]{}^k\!_j \,, 
\end{align}
and we also define $\delta \gamma_{ij}(t,\, \bm{x}) \equiv  \bar{\gamma}_{ik}(t) \delta {\gamma^k}_j(t,\, \bm{x})$. Here and hereafter, we put a bar on background fields.

\subsection{Models}  \label{SSec:models}
To provide a more concrete idea about how to use the \dNex, in the following, let us consider a more specific model whose Lagrangian density is given by
\begin{align}
    {\cal L}_{\rm mat} = {\cal P}(X^{IJ},\, \phi^I) - \sum_{\alpha=1}^{D'} \frac{f_{(\alpha)}^2(X^{IJ},\,\phi^I)}{4} F_{\mu \nu(\alpha)} F_{(\alpha)}^{\mu \nu} -\sum_{\alpha=1}^{D'} \frac{g_{(\alpha)}(X^{IJ},\, \phi^I)}{4} F_{\mu \nu(\alpha)} \tilde{F}^{\mu \nu}_{(\alpha)} \,,  \label{Exp:Lmattergeneral}
\end{align}
which includes neutral scalar fields $\phi^I$ with $I=1,\, ....,\, D$  and U(1) gauge fields with $\alpha = 1,\, \cdots,\, D'$. Here, $X^{IJ} \equiv - \partial_\mu \phi^I \partial^\mu \phi^J/2$ and $F_{\mu \nu(\alpha)}$ and $\tilde{F}^{\mu \nu}_{(\alpha)}$ denote the field strengths of the gauge fields and their dual fields. A model with multiple U(1) gauge fields was studied, e.g.,, in Ref.~\cite{Yamamoto:2012tq}. Here, an interaction of the gauge fields with the kinetic terms of the scalar fields is also allowed. In this paper, we do not consider charged matters, because a non-zero condensate of charged fields screens the gauge fields at large scales, leaving only little impact on the homogeneous separate universe evolution (see, e.g., \cite{Tanaka:2021dww}). As a theory of gravitation which satisfies the \sDif and \locality conditions, we consider general relativity. For the gauge fields, let us impose gauge conditions, 
\begin{align}
    A_{0(\alpha)}(t,\, \bm{x})=0 \,, \label{Cond:A0}
\end{align}
which does not vanish in the long wavelength limit.

In order to relate purely geometrical relations, (\ref{Exp:deltaN2}) and (\ref{Exp:deltagamma}), with the matter fields, we need to specify the model. In our setup, the Einstein equations at the leading order of the gradient expansion are given by 
\begin{align}
    & \frac{2}{3} K^2 - A^i\!_j A^j\!_i  -  \frac{2}{\Mp^2} \rho  = {\cal O} (\epsilon^2)\,, \label{Eq:Hc_GR} \\
    &  \frac{1}{N} \partial_t K  + \frac{3}{2} {A^i}_j {A^j}_i = - \frac{3}{2 \Mp^2}  (\rho + P) +  {\cal O}(\epsilon^2)\,, \label{Eq:ijT_GR}
\end{align}
and
\begin{align}
    \frac{1}{N \sqrt{g}} \partial_t \left( \sqrt{g} {A^i}_j \right) = \frac{1}{\Mp^2}\,{\Pi^i}_j  + {\cal O}(\epsilon^2)\,. \label{Eq:ijS}
\end{align}
The energy density $\rho$, the isotropic pressure $P$, and the anisotropic pressure $\Pi^{ij}$ for the matter Lagrangian (\ref{Exp:Lmattergeneral}) are given by
\begin{align}
   &  \rho = 2 X^{IJ} \left( {\cal P}_{IJ} + \sum_{\alpha=1}^{D'} \frac{\partial \ln f_{(\alpha)}}{\partial X^{IJ}} \frac{f_{(\alpha)}^2}{N^2} g^{ij}  \dot{A}_{i(\alpha)} \dot{A}_{j(\alpha)} \right) \cr
   & \qquad \qquad \qquad \qquad \qquad - {\cal P}  + \sum_{\alpha=1}^{D'}  \frac{f_{(\alpha)}^2}{2N^2} g^{ij}  \dot{A}_{i(\alpha)} \dot{A}_{j(\alpha)}  + {\cal O}(\epsilon^2) \,, \label{Exp:rho_U1} \\
   &  P ={\cal P} + \sum_{\alpha=1}^{D'}   \frac{f_{(\alpha)}^2}{6N^2} g^{ij} \dot{A}_{i(\alpha)} \dot{A}_{j(\alpha)}   + {\cal O}(\epsilon^2)\,, \label{Exp:P_U1} \\
   & {\Pi}_{ij} = - \sum_{\alpha=1}^{D'} \frac{f^2_{(\alpha)}}{N^2} \left[ \dot{A}_{i(\alpha)} \dot{A}_{j(\alpha)} \right]^{\rm TL}  + {\cal O}(\epsilon^2) \,,  \label{Exp:Pij}
\end{align}
with ${\cal P}_{IJ} \equiv \partial {\cal P}/\partial X^{IJ}$. As will be discussed in Sec.~\ref{SSec:constraints}, when we solve the superhorizon evolution, the momentum constraints, which are in ${\cal G}_\epsilon$, need not to be solved as long as they are solved properly at $t=t_*$.

The Klein-Gordon (KG) equations are given by
\begin{align}
    &\frac{1}{N \sqrt{g}} \partial_t \left[ \sqrt{g} \left( {\cal P}_{IJ} + \sum_{\alpha=1}^{D'}   \frac{\partial \ln f_{(\alpha)}}{\partial X^{IJ}} \frac{f^2_{(\alpha)}}{N^2} g^{ij}  \dot{A}_{i(\alpha)} \dot{A}_{j(\alpha)} \right) \frac{\dot{\phi}^J}{N} \right] \cr
    & \qquad \qquad \qquad \qquad  - \frac{\partial {\cal P}}{\partial \phi^I}- \sum_{\alpha=1}^{D'}  \frac{\partial \ln f_{(\alpha)}}{\partial \phi^I}  \frac{f_{(\alpha)}^2}{N^2} g^{ij}  \dot{A}_{i(\alpha)} \dot{A}_{j(\alpha)}     = {\cal O}(\epsilon)\,. \label{Eq:KG}
\end{align}
The U(1) gauge constraints are given by  
\begin{align}
    \partial_i \pi^i_{(\alpha)} = 0  \label{Eq:U1gc}
\end{align}
with 
\begin{align}
    \pi^i_{(\alpha)} \equiv \frac{\partial (N \sqrt{g} {\cal L}_{\rm mat}) }{\partial \dot{A}_{i(\alpha) }} =   \sqrt{g} f_{(\alpha)}^2(X^{IJ},\, \phi^I) g^{ij} \frac{\dot{A}_{j(\alpha)}}{N}  + {\cal O}(\epsilon) \,. \label{Exp:piAex_U1}
\end{align}
Repeating the same argument as the one in Sec.~\ref{SSec:Noether} also for the U(1) gauge symmetries, we also obtain   
\begin{align}
    \partial_t \pi_{(\alpha)}^i = {\cal O}(\epsilon)\,, \label{Eq:Maxwell_neutral}
\end{align}
where $\pi_{(\alpha)}^i$ corresponds to the Noether charge density. The same equation can be derived also from the Maxwell equations.

Using $\pi_{(\alpha)}^i$, which remains constant at the leading order of the gradient expansion, we express (the electric part of) the energy density of the gauge fields as
\begin{align}
     \rho_A = \sum_{\alpha=1}^{D'} \rho_{A(\alpha)}\,,  \qquad \rho_{A(\alpha)} \equiv \sum_{\alpha=1}^{D'} \frac{e^{-4 \psi}}{2 f_{(\alpha)}^2} \gamma_{ij} \pi_{(\alpha)}^i \pi_{(\alpha)}^j + {\cal O}(\epsilon^2)\,.  \label{Exp:rhoA}
\end{align}
As discussed in Refs.~\cite{Watanabe:2009ct, Watanabe:2010fh} and also generalized in Ref.~\cite{Tanaka:2021dww}, ${A^i}_j/K$ is bounded by the slow-roll parameters. When the shear $|{A^i}_j|/K$ is smaller than ${\cal O}(1)$, the time variation of $\gamma_{ij}$ remains small in the time scale of the cosmic expansion, which is also an observational requirement. Then, when $f_{(\alpha)}$ decreases faster than $e^{-2\psi}$, $\rho_{A(\alpha)}$ definitely increases.

\subsection{Constraints to be and not to be solved for separate universe evolution} \label{SSec:constraints}
The Lagrangian density (\ref{Exp:Lmattergeneral}) and the one of general relativity obviously satisfy the [locality] condition and the \sDif condition. In the model under consideration, the constraints are all gauge constrains. There are three kinds of gauge constraints: the Hamiltonian constraint ${\cal H}$, the momentum constraints ${\cal H}_i$, and the U(1) gauge constraints ${\cal H}_{\rm U(1)}$. Among them, ${\cal H}_i$, and ${\cal H}_{\rm U(1)}$ are ${\cal G}_\epsilon$s, {\it i.e.}, the gauge constraints which vanish in the limit $\epsilon \to 0$.

The enhanced set of the fields, $\{ \varphi^a \}$, consists of the remaining (inhomogeneous) fields after imposing the gauge conditions which do not vanish in the large scale limit, {\it i.e.}, ${\cal X} \cap {\cal X}_\epsilon^c$. For instance, at $t =t_*$, on which $\delta \psi=0$, $\{ \delta \varphi_*^a \}$ is given by the fluctuations of 
\begin{align}
 & \quad \{\varphi^a \} =  K,\, \phi^I [D],\, A_{(\alpha)i} [3 D'],\,\gamma_{ij} [5],\, \pi_I [D],\, \pi_{(\alpha)}^i [3 D'],\, \pi^{ij} [5]\,,  \label{Exp:varphia}
\end{align}
with
\begin{align}
\label{eq:piij}
   \pi_I \equiv \frac{\partial (N\!\sqrt{g} {\cal L})}{\partial \dot{\phi}^I} \,, \qquad  \pi^{ij} \equiv \frac{\partial (N\!\sqrt{g} {\cal L})}{\partial \dot{g}_{ij}} \,,
\end{align}
and $\pi^i_{(\alpha)}$ is given by Eq.~(\ref{Exp:piAex_U1}). The numbers inside the subsequent square brackets in Eq.~(\ref{Exp:varphia}) denote the number of degrees of freedom for respective fields. Using Eq.~(\ref{Exp:KS}), the expansion $K$ and the lapse function $N$ are mutually related on the time slicing $\delta \psi = 0$. Since the determinant of $\gamma_{ij}$ is set to 1, the remaining degree of freedom of $\gamma_{ij}$ is 5.

By solving the Hamiltonian constraint ${\cal H}$, which is ${\cal G} \cap {\cal G}_\epsilon^c$ since it does not vanish in the large scale limit, we can eliminate the expansion $K$. Then, $\{ \delta \varphi^{a'}_* \}'$, which describes the initial condition of the separate universe evolution, is given by the fluctuations of 
\begin{align}
     \{\varphi^{a'} \}' &= \phi^I [D],\, A_{i(\alpha)} [3D'],\,\gamma_{ij} [5],\, \pi_I [D],\, \pi_{(\alpha)}^i [3 D'],\, \pi^{ij} [5]\,,  \label{List:dash}
\end{align}
which counts $2(D+3D'+ 5)$. If we solve $D'+3$ gauge constraints with $\epsilon$ suppression ${\cal G}_\epsilon$, {\it i.e.}, the momentum constraints and the U(1) gauge constraints and impose $D'+3$ (residual) gauge conditions with $\epsilon$ suppression, ${\cal X}_\epsilon$, the fields in Eq.~(\ref{List:dash}) can be reduced to $2 (D+ 2D' + 2)$ physical ones, e.g.,
\begin{align}
     \{\varphi^{a_{\rm phys}} \}_{\rm phys} &= \phi^I [D],\, A_{i(\alpha)} [2D'],\,\gamma_{ij} [2],\, \pi_I [D],\, \pi_{(\alpha)}^i [2D'],\, \pi^{ij} [2]\,.\label{List:phys}
\end{align}

As long as we provide the initial conditions at $t=t_*$, satisfying consistently ${\cal H}_i$, ${\cal H}_{\rm U(1)}$, the obtained solution remains to be a physical one, even though we solve the separate universe evolution, maintaining the redundant degrees of freedom~\cite{Tanaka:2021dww}. In other words, the power spectrums (as well as the higher order correlation functions) of $\{ \varphi^{a'} \}'$ at $t=t_*$ should be determined, so as to satisfy ${\cal G}_\epsilon$. The residual gauge degrees of freedom can be removed by imposing ${\cal X}_\epsilon$ at $t=t_*$. For example, when the background spatial metric is set to (or is approximately given by) $\bar{\gamma}_{ij}(t_*) = \delta_{ij}$, the residual gauge degrees of freedom, which were not removed by Eqs.~(\ref{Cond:Ni}) and (\ref{Cond:A0}), can be removed by imposing 
\begin{align}
    \delta^{ij} \partial_i \delta \gamma_{jk} (t_*,\, \bm{x}) = 0\,,     \label{Cond:trans_approx}
\end{align}
and 
\begin{align}
  \delta^{ij} \partial_i A_{j(\alpha)} (t_*,\, \bm{x}) = 0 \,, \label{Cond:Coulomb_approx}
\end{align}
which are ${\cal X}_\epsilon$. As is discussed in Appendix \ref{SSec:WAM}, Eq.~(\ref{Cond:trans_approx}) should be modified for $\bar{\gamma}_{ij}(t_*) \neq \delta_{ij}$.

In summary, the constraint equations and the gauge conditions are classified as
\begin{align} 
   & {\cal C} \cap {\cal G}^c : \emptyset \cr
   & {\cal G}_\epsilon:{\cal H}_i,\,  {\cal H}_{\rm U(1)}  \cr
   & {\cal G} \cap  {\cal G}_\epsilon^c : {\cal H} \label{Summary:symbols} \\
   & {\cal X} \cap {\cal X}_\epsilon^c : (\ref{Cond:Ni}),\, (\ref{Cond:A0}),\, (\ref{Cond:slicingini}),~{\rm and}~(\ref{Cond:slicingfin})\cr
   & {\cal X}_\epsilon: (\ref{Cond:trans_approx})~{\rm and}~ (\ref{Cond:Coulomb_approx}) \nonumber\,.
\end{align}
One can compute the long-wavelength evolution of $2(D+3D'+5)$ fields, $\{ \varphi^a\}$, e.g., by solving $D+3D'+5$ second order differential equations in time, ${\cal E}_{\rm ev}$
\begin{itemize}
    \item KG equations (\ref{Eq:KG}): $D$ second-order differential equations
    \item Maxwell equations (\ref{Eq:Maxwell_neutral}): $3D'$ second-order differential equations in time 
    \item Traceless $(i, j)$ component of Einstein equations (\ref{Eq:ijS}): 5 second-order differential equations
\end{itemize}
together with 
\begin{itemize}
    \item Hamiltonian constraint (\ref{Eq:Hc_GR}): 1 gauge constraint in ${\cal G} \cap {\cal G}_\epsilon^c$.  \label{Eqs:gdeltaN}
\end{itemize}
Solving these equations under various initial conditions specified by $\{ \varphi^{a'}_* \}'$, we obtain the mapping between the fields at the horizon crossing $t_*$ and the final time $t_f$.

One of the Klein-Gordon equations can be substituted by the trace $(i,\, j)$ component of the Einstein equations, (\ref{Eq:ijT_GR}). In the next subsection, we argue that Eq.~(\ref{Eq:ijS}) can be easily integrated, introducing the Noether charge density ${Q^j}_i$.

In Ref.~\cite{Talebian-Ashkezari:2016llx}, the authors attempted to generalize the $\delta N$ formalism to a Bianchi background. They argued that the shift vector and the $(0,\, i)$ component of the energy-momentum tensor need to be suppressed by ${\cal O}(\epsilon)$ to satisfy the momentum constraints. As shown in Ref.~\cite{Tanaka:2021dww} and also as has been summarized here, the momentum constraints can be satisfied without requiring this suppression.

\subsection{Long-wavelength evolution of the shear}  \label{SSec:shear}
The long-wavelength evolution of the shear, which is necessary to relate the right hand side of Eq.~(\ref{Exp:deltagamma}) to the matter fields, can be computed by using the traceless part of the $(i,\, j)$ Einstein equation (\ref{Eq:ijS}). In the following, we show that this becomes easier by using the Noether charge (density). In general relativity, rewriting the symmetric part of the Noether charge density, we can express the shear as
\begin{align}
    {A^j}_i =  - \frac{{Q^{j~{\rm sym}}}_{\hspace{-15pt}i}}{\sqrt{g}}\hspace{5pt}  - \sum_{\alpha = 1}^{D'} \frac{1}{\Mp^2\sqrt{g}} \left[  \pi^j_{(\alpha)} A_{i(\alpha)}   
    \right]^{\rm symTL}\hspace{-10pt} + {\cal O}(\epsilon)\,,  \label{Eq:Noetherex}
\end{align}
where the $(i,\, j)$ indices of $A_{ij}$ are symmetrized. In what follows, since we only consider the symmetric part of the Noether charge density, we omit the superscript sym. The first term in the right hand side of Eq.~(\ref{Eq:Noetherex}) describes the well-known contribution of the shear which falls off with the inverse of the physical spatial volume. As discussed in Ref.~\cite{Tanaka:2021dww}, this decaying contribution of the shear plays a crucial role in discussing the existence of the Weinberg's adiabatic mode. Integrating Eq.~(\ref{Eq:Noetherex}) formally, we obtain
\begin{align}
    \gamma_{ij} (t,\, \bm{x}) &=  \gamma_{ij} (t_*,\, \bm{x}) - 2  \int^t_{t_*} d t' N(t',\, \bm{x}) \frac{\gamma_{il} (t',\, \bm{x})}{\sqrt{g(t',\, \bm{x})}}  {Q^{l}}_j(\bm{x}) \cr
    & \quad - \sum_{\alpha=1}^{D'} \frac{2}{\Mp^2}\hspace{-2pt} \int^t_{t_*} \hspace{-2pt} d t'  \frac{N(t',\, \bm{x})}{e^{3\psi(t',\, \sbm{x})}} \!\left[ \gamma_{\{i|l|} \pi^l_{(\alpha)}  A_{j\}(\alpha)} - \frac{\gamma_{ij}}{3}  \pi^l_{(\alpha)}  A_{l(\alpha)}
    \right]_{\!(t',\, \sbm{x})}  + {\cal O}(\epsilon).\,\,  \label{Sol:gammaij_Noether}
\end{align}
The second term corresponds to the decaying solution and the term in the second line is the sourced contributions. Compared to the solution obtained by integrating Einstein equation, Eq.~(\ref{Sol:gammaij_Noether}) gives a formal expression of $\gamma_{ij}$ with one less time integral.

Integrating Eq.~\eqref{Exp:piAex_U1}, we obtain a formal solution of $A_{i(\alpha)}$ as
\begin{align}
    A_{i(\alpha)} (t,\, \bm{x}) =  A_{i(\alpha)*} (\bm{x}) + \pi_{(\alpha)}^j(\bm{x}) \int^t_{t_*} \frac{d t' N}{\sqrt{g} f_{(\alpha)}^2} g_{ij} (t',\, \bm{x}) + {\cal O}(\epsilon)\,.   \label{Sol:Ai}  
\end{align}
Inserting this solution (\ref{Sol:Ai}) into Eq.~(\ref{Sol:gammaij_Noether}) and dropping trivial decaying contributions in an expanding universe, we obtain
\begin{align}
    \gamma_{ij} (t,\, \bm{x}) &= \gamma_{ij\, *}(\bm{x}) - 2 \sum_{\alpha=1}^{D'} \frac{\pi_{(\alpha)}^l(\bm{x}) \pi_{(\alpha)}^m(\bm{x})}{\Mp^2} \int^t_{t_*} \frac{d t' N}{e^{3 \psi(t')}}  \int^{t'}_{t_*} \frac{dt'' N(t'')}{e^{\psi(t'')} f_{(\alpha)}^2(X^{IJ}(t''),\, \phi^I(t''))} \cr
    & \qquad \qquad \qquad \qquad  \times \left[ \gamma_{il} (t') \gamma_{jm} (t'') - \frac{1}{3} \gamma_{ij}(t') \gamma_{lm} (t'') \right]^{\rm sym}+ {\cal O}(\epsilon)\,, \label{Sol:gamma_formal}
\end{align}
where we have abbreviated $\bm{x}$-dependence in the integrand.

In this paper, we assume that the shear remains much smaller than the expansion, requiring 
\begin{align}
    \frac{{A^i}_j {A^j}_i}{K^2} = \frac{\dot{\gamma}^{ij} \dot{\gamma}_{ij}}{(2NK)^2} \ll 1\,. \label{Cond:smallshear}
\end{align}
As shown in Ref.~\cite{Maleknejad:2012as}, the amplitude of the shear is bounded by the slow-roll parameters under the slow-roll approximation. In Ref.~\cite{Tanaka:2021dww}, this bound was generalized to an inflation model in which not all slow-roll parameters remain small as
\begin{align}
    \sqrt{\frac{{A^i}_j {A^j}_i}{K^2}} \leq \sqrt{6} \varepsilon\,,  \label{Exp:boundshear}
\end{align}
with $\varepsilon$ being
\begin{align}
    \varepsilon \equiv - 3 \frac{\dot{K}}{N K^2}\,.
\end{align}
Using Eq.~(\ref{Exp:boundshear}), one can find that our assumption (\ref{Cond:smallshear}) generically holds during inflation, since $\varepsilon \ll 1$. For the fluctuations which correspond to the CMB scales at present, the stringent constraints on the statistical anisotropy of $\zeta$ \cite{Kim:2013gka, Akrami:2018odb} also bound the amplitude of the shear. 

Under the assumption (\ref{Cond:smallshear}), the time variation of $\gamma_{ij}$ remains small in the time scale of the cosmic expansion, enabling us to rewrite Eq.~(\ref{Sol:gamma_formal}) as
\begin{align}
   & \gamma_{ij} (t,\, \bm{x}) = \gamma_{ij*}(\bm{x}) 
    -  2 \left[ \gamma_{il*}(\bm{x}) \gamma_{jm*}(\bm{x}) \right]^{\rm TL}  \sum_{\alpha=1}^{D'} \hat{\pi}_{(\alpha)}^l(\bm{x}) \hat{\pi}_{(\alpha)}^m(\bm{x}) \, \hat{{\cal I}}_{(\alpha)}(t;\{\varphi^{a'}_*\}')+ {\cal O}(\epsilon^2) \,, \label{Sol:gamma}
\end{align}
with $\hat{{\cal I}}_{(\alpha)}$ and $\hat{\pi}^i_{(\alpha)}$ being
\begin{align}
    & \hat{{\cal I}}_{(\alpha)}(t;\{\varphi^{a'}_*\}') \equiv  \frac{\gamma_{ij*} \pi_{(\alpha)}^i \pi^j_{(\alpha)}}{\Mp^2} \!\int^t_{t_*}\! \frac{d t' N}{e^{3 \psi(t')}} \! \int^{t'}_{t_*} \! \frac{dt'' N}{e^{\psi(t'')} f_{(\alpha)}^2(X^{IJ}(t''),\, \phi^I(t''))}\,,  \label{Def:calI} 
\end{align}
and 
\begin{align}
    \hat{\pi}_{(\alpha)}^i \equiv \frac{\pi_{(\alpha)}^i}{\sqrt{\gamma_{ij*} \pi_{(\alpha)}^i \pi_{(\alpha)}^j}} = \frac{\gamma^{ij}_* \dot{A}_{j(\alpha)*}}{\sqrt{\gamma^{kl}_* \dot{A}_{k(\alpha)*} \dot{A}_{l(\alpha)*}}}\,,  \label{Def:hatpi} 
\end{align}
where $\{\varphi^{a'}_*\}'$ denotes the set of the fields which provide the initial conditions at the horizon crossing. Here, TL denotes the traceless part with respect to the $(i, j)$ indices defined by using the spatial metric $\gamma_{ij*}$. Equation (\ref{Sol:gamma}) describes the mapping between $\gamma_{ij}$ evaluated at $t$ and the fluctuations at $t=t_*$. This equation also gives the longitudinal part of $\gamma_{ij}$, which is necessary to compute $\zeta$.

A recursive computation determines the higher order terms with respect to $\hat{{\cal I}}$. Our assumption (\ref{Cond:smallshear}) can be rephrased as
\begin{align}
    \hat{{\cal I}} \ll 1\,. 
\end{align}
When we ignore the time variation of the expansion $K$, considering inflation, we can further rewrite $\If \equiv \hat{\cal I}_{(\alpha)}(t_f)$ with $t_f$ being the reheating surface as 
\begin{align}
    \If \simeq \frac{1}{\rho_{*}} \int^{\psi_f}_{\psi_*}  \frac{d \psi''  \gamma_{ij} \pi_{(\alpha)}^i \pi_{(\alpha)}^j }{e^{4\psi''}f_{(\alpha)}^2(X^{IJ}(\psi''),\,\phi^I(\psi''))} \simeq \frac{2}{\rho_{*}} \int^{\psi_f}_{\psi_*} d \psi'' \rho_{A(\alpha)} (\psi'')\,,  \label{Exp:If_inf}
\end{align}
with $\rho_*$ being the total energy density at $t=t_*$. On the first equality, $\gamma_{ij*} \pi_{(\alpha)}^i \pi_{(\alpha)}^j $ is moved to the inside of the integral, since the time variation of $\gamma_{ij}$ becomes higher order under the assumption (\ref{Cond:smallshear}). Then, we have integrated over $t'$ after exchanging the order of the integrals and ignored the exponentially suppressed term.  Using Eq.~(\ref{Exp:rhoA}), one can find that the integrand corresponds to the energy density of each gauge field. Therefore, denoting the $e$-folding number during which $\rho_{A(\alpha)}$ takes a maximum value $\rho_{A(\alpha)}^{\rm max}$ after the horizon crossing by $\Delta \psi_{(\alpha)}$, $\If$ is roughly given by $\If \sim 2 \Delta \psi_{(\alpha)} \rho_{A(\alpha)}^{\rm max}/\rho_*$.

\section{Perturbative expansion and application of g\texorpdfstring{$\delta N$}{delta N} formalism}  \label{Sec:PExpansion}
While the \dNex~is a non-perturbative formalism, we are often interested in the case where the inhomogeneity is perturbatively small, to be consistent with observations. In this section, we calculate the curvature perturbation and the gravitational waves perturbatively. 

\subsection{Perturbative expansion in g\texorpdfstring{$\delta N$}{delta N} formalism}
 Perturbing $\varphi^a (t_f,\, \bm{x})$ with respect to the initial inhomogeneities, evaluated at $t=t_*$, we obtain 
\begin{align}
    \delta \varphi^a (t_f,\, \bm{x}) &= \sum_{n=1}^{\infty} \frac{1}{n!} {\cal C}^a_{b_1' \cdots b_n'*} \delta \varphi^{b_1'}_* (\bm{x})  \cdots \delta \varphi^{b_n'}_*(\bm{x})   \,, \label{Exp:expansion}  
\end{align}
where the coefficients of the expansion are given by
\begin{align}
    {\cal C}^a_{b_1' \cdots b_n'*} \equiv \frac{\partial^n \varphi^a(t_f; \{ \varphi^{b'}_* \}' )}{\partial \varphi^{b'_1}_* \cdots  \partial \varphi^{b'_n}_*} \label{Exp:coeffieient}
\end{align}
with $\varphi^a(t; \{ \varphi^{b'}_* \}' )$ being the solution at the leading order of the gradient expansion under the initial condition $ \{\varphi^{b'}_*\}$, which is simply obtained by solving the background equations varying the initial conditions. The set of fields $\{\varphi^{a'}\}'$ is listed in Eq.~(\ref{List:dash}) and the correlators of their fluctuations at $t=t_*$ will be given by a QFT computation.

For example, the two-point functions at the linear perturbation is given by 
\begin{align}
   & \langle \Omega\, | \delta \varphi^{a_1} (t_f,\, \bm{x}_1) \delta \varphi^{a_2} (t_f,\, \bm{x}_2) |\, \Omega \rangle = \sum_{b_1',\, b_2'} {\cal C}^{a_1}_{b_1'*} {\cal C}^{a_2}_{b_2'*}  \langle \delta \varphi^{b_1'}_* (\bm{x}_1) \delta \varphi^{b_2'}_* (\bm{x}_2) \rangle \,, 
\end{align}
where $\langle \cdots \rangle$ in the right hand side denotes the correlators of classical stochastic fields at $t=t_*$. Here, assuming that the quantum fluctuations have already decohered at $t=t_*$, we identify the quantum correlators with the corresponding classical statistical correlators. While the coefficients ${\cal C}^{a}_{b'*}$ are determined by solving only the local equations, listed in Eq.~(\ref{Eqs:gdeltaN}), the correlation functions $ \langle \delta \varphi^{b_1'}_* (\bm{x}_1) \delta \varphi^{b_2'}_* (\bm{x}_2) \rangle$ are determined by solving all the equations, including the constraints ${\cal H}_i$ and ${\cal H}_{\rm U(1)}$. By using these correlation functions, the obtained solution is ensured to satisfy all the equations, being physical~\cite{Tanaka:2021dww}. The mixing among different fields results in non-vanishing off-diagonal components of ${\cal C}^{a}_{b'*}$ ($a \neq b'$). As a consequence of the global anisotropy, the cross-correlations between the scalar, vector, and tensor perturbations do not vanish already at the linear order of perturbation. Similarly, we can compute the three-point functions as 
\begin{align}
   & \langle \Omega\, | \delta \varphi^{a_1} (t_f,\, \bm{x}_1) \delta \varphi^{a_2} (t_f,\, \bm{x}_2) \delta \varphi^{a_2} (t_f,\, \bm{x}_2) |\, \Omega \rangle \cr
   & = \sum_{b_1',\, b_2',\, b_3'}  {\cal C}^{a_1}_{b_1'*} {\cal C}^{a_2}_{b_2'*}  {\cal C}^{a_3}_{b_3'*} \langle \delta \varphi^{b_1'}_* (\bm{x}_1) \delta \varphi^{b_2'}_* (\bm{x}_2) \delta \varphi^{b_3'}_* (\bm{x}_3) \rangle \cr
  & \quad + \frac{1}{2} \hspace{-2pt} \sum_{b_1',\,c_1',\,  b_2',\, b_3'}  \hspace{-5pt} {\cal C}^{a_1}_{b_1' c_1'*} {\cal C}^{a_2}_{b_2'*}  {\cal C}^{a_3}_{b_3'*}  \langle \delta \varphi^{b_1'}_* (\bm{x}_1) \delta \varphi^{c_1'}_* (\bm{x}_1) \delta \varphi^{b_2'}_* (\bm{x}_2) \delta \varphi^{b_3'}_* (\bm{x}_3) \rangle + (2{\rm perms})\,, \label{Exp:3pt}
\end{align}
where we have ignored the loop corrections. 

Equation (\ref{Exp:expansion}) implies that when the initial correlations at $t=t_*$ are parity invariant, satisfying\footnote{A rank $m$ tensor, $\varphi^m(\bm{x})$, transforms under the parity transformation as $\varphi_P^m(t,\, \bm{x}) = (-1)^m \varphi^m(\bm{-x})$ and the parity invariance requires $\varphi_P^m(t,\, \bm{x}) = \varphi^m(t,\, \bm{x})$.} 
\begin{align}
    \langle \delta \varphi_*^{b_1'}(\bm{x}_1) \cdots \delta \varphi_*^{b_n'}(\bm{x}_n)  \rangle = (-1)^{s_b'} \langle \delta \varphi_*^{b_1'}(-\bm{x}_1) \cdots \delta \varphi_*^{b_n'}(-\bm{x}_n)  \rangle\,, \label{Exp:parityinv_ti}
\end{align}
the final correlations remain parity invariant, {\it i.e.}, 
\begin{align}
    \langle \delta \varphi^{a_1}(t_f,\,\bm{x}_1) \cdots \delta \varphi^{a_n}(t_f,\,\bm{x}_n)  \rangle = (-1)^{s_a} \langle \delta \varphi^{a_1}(t_f,\,-\bm{x}_1) \cdots \delta \varphi^{a_n}(t_f,\,-\bm{x}_n)  \rangle\,,
\end{align}
or in the Fourier space, 
\begin{align}
    \langle \delta \varphi^{a_1}(t_f,\,\bm{k}_1) \cdots \delta \varphi^{a_n}(t_f,\,\bm{k}_n)  \rangle = (-1)^{s_a} \langle \delta \varphi^{a_1}(t_f,\,-\bm{k}_1) \cdots \delta \varphi^{a_n}(t_f,\,-\bm{k}_n)  \rangle\,,\label{Exp:parityinv_tf}
\end{align}
where $s_b' \equiv \sum_{i=1}^n s_{b'_i} $ with $s'_{b_i}$ being the corresponding rank of the tensor $\varphi^{b'_i}$ and the same for $s_a$ except for $b_i$ being replaced with $a_i$. When the Chern-Simons terms in the Lagrangian density (\ref{Exp:Lmattergeneral}) are not negligible at $t=t_*$, explicitly violating the parity, Eq.~(\ref{Exp:parityinv_ti}) does not hold (see, e.g., Refs.~\cite{Bartolo:2014hwa, Dimopoulos:2012av, Bartolo:2015dga, Tripathy:2023aha}).

\subsection{Polarization bases}
When the background shear is not negligible, $\bar{\gamma}_{ij}$ changes in time. In this paper, we set the background spatial metric at the reheating surface $t_f$ as 
\begin{align}
    \bar{\gamma}_{ij}(t_f) = \delta_{ij} \,.  \label{Eq:bargamma}
\end{align}
Then, we can introduce the polarization bases of the gauge fields and the gravitational waves as usual at $t=t_f$. When the background shear had already become negligible at $t=t_f$, the background spatial metric remains  $\bar{\gamma}_{ij}(t) = \delta_{ij}$ all the time after $t_f$, ensuring the linear decomposition among the scalar, vector, and tensor type perturbations.

\subsubsection{Linear polarization bases}
We can introduce the orthonormal bases in the Gaussian coordinates $(\hat{\bm{k}},\, \bm{e}^{(1)}(\hat{\bm{k}}),\, \bm{e}^{(2)}(\hat{\bm{k}}))$, which satisfy 
\begin{align}
     \delta^{ij} \hat{k}_j e^{(\lambda)}_i(\hat{\bm{k}})=0\,, \qquad \delta^{ij} e^{(\lambda)}_i(\hat{\bm{k}}) e^{(\lambda')}_j(\hat{\bm{k}}) = \delta^{\lambda \lambda'}
\end{align}
at $t=t_f$. Here, $\hat{k}_i$ denotes the normalized vector $\hat{k}_i \equiv k_i/k$, which satisfies $\delta^{ij} \hat{k}_i \hat{k}_j =1$. Let us also introduce the linear polarization bases of rank-2 tensors as
\begin{align}
   & e_{ij}^{(+)} (\hat{\bm{k}}) \equiv \frac{1}{\sqrt{2}} \left[ e_i^{(1)} (\hat{\bm{k}}) e_j^{(1)}(\hat{\bm{k}})   - e_i^{(2)}(\hat{\bm{k}})  e_j^{(2)}(\hat{\bm{k}})  \right]  \label{Def:eplus} \,, \\
   &  e_{ij}^{(\times)} (\hat{\bm{k}}) \equiv \frac{1}{\sqrt{2}} \left[ e_i^{(1)} (\hat{\bm{k}}) e_j^{(2)}(\hat{\bm{k}})   + e_i^{(2)}(\hat{\bm{k}})  e_j^{(1)}(\hat{\bm{k}})  \right]\,,  \label{Def:ecross}
\end{align}
which satisfies
\begin{align}
    \delta^{ij} \hat{k}_i e_{jl}^{(\lambda_{\rm gw})}  (\hat{\bm{k}})= 0\,, \quad \delta^{ij}e_{ij}^{(\lambda_{\rm gw})}  (\hat{\bm{k}})= 0 \,, \quad
    \delta^{il} \delta^{jm} e_{ij}^{(\lambda_{\rm gw})}  (\hat{\bm{k}}) e_{lm}^{(\lambda'_{\rm gw})}  (\hat{\bm{k}}) = \delta^{\lambda_{\rm gw} \lambda'_{\rm gw}}
\end{align}
for $\lambda_{\rm gw}= +,\, \times$. In what follows, we lower and raise the indices $e_i^{(\lambda)}$ with $\lambda=1,\,2$ and $e_{ij}^{(\lambda_{\rm gw})}$ by using $\bar{\gamma}_{ij}(t_f) = \delta_{ij}$ and $\bar{\gamma}^{ij}(t_f) = \delta^{ij}$.

In general, we need two parameters to characterize the direction of $\bar{\hat{\bm{\pi}}}_{(\alpha)}$. Let us introduce $\Theta_\alpha$ and $\Psi_\alpha$ as
\begin{align}
    & \hat{\bm{k}} \cdot \bar{\hat{\bm{\pi}}}_{(\alpha)} = \cos \Theta_\alpha(\hat{\bm{k}}) \,, \cr
    & \bm{e}^{(1)}(\hat{\bm{k}})  \cdot \bar{\hat{\bm{\pi}}}_{(\alpha)} = \sin \Theta_\alpha(\hat{\bm{k}})  \cos \Psi_\alpha(\hat{\bm{k}}) \,,\label{Exp:PhiPsi} \\
    & 
    \bm{e}^{(2)}(\hat{\bm{k}})  \cdot \bar{\hat{\bm{\pi}}}_{(\alpha)} = \sin \Theta_\alpha(\hat{\bm{k}})  \sin \Psi_\alpha(\hat{\bm{k}}) \,. \nonumber 
\end{align}
For an arbitrary $\hat{k}^i$, we can choose $e_i^{(1)}$ in the 2D plane spanned by $\hat{k}_i$ and $\bar{\pi}_{(\alpha)}^i$ for one of the gauge fields, e.g., $\alpha=1$, setting $\Psi_1 = 0$. Then, we obtain 
 \begin{align}
     \bm{e}^{(1)}(\hat{\bm{k}})  \cdot \bar{\hat{\bm{\pi}}}_{(1)} = \sin \Theta_1(\hat{\bm{k}}) \,, \qquad \bm{e}^{(2)}(\hat{\bm{k}})  \cdot \bar{\hat{\bm{\pi}}}_{(1)} = 0\,  \label{Exp:Psi1}
\end{align}
for the corresponding gauge field.  When we flip the signature of $\bm{k}$, the linear polarization bases change as
\begin{align}
    e^{(1)}_i (- \hat{\bm{k}}) = e^{(1)}_i (\hat{\bm{k}})\,, \qquad 
    e^{(2)}_i (- \hat{\bm{k}}) = - e^{(2)}_i (\hat{\bm{k}})\,. \label{Exp:parityLPB}
\end{align}
Then, correspondingly, $\Theta_\alpha$ and $\Psi_\alpha$ changes as
\begin{align}
    \Theta_\alpha(- \hat{\bm{k}}) = \pi - \Theta_\alpha(\hat{\bm{k}}), \qquad \Psi_\alpha(- \hat{\bm{k}}) = - \Psi_\alpha(\hat{\bm{k}})\,. \label{Ep:ThetaPsiparity}
\end{align}
In what follows, for a notational brevity, we sometimes drop the argument $(\hat{\bm{k}})$ of the polarization bases, $\Theta_\alpha$, and $\Psi_\alpha$.

By using $e_{ij}^{(\lambda_{\rm gw})}$, the two polarization modes of the gravitational waves are given by 
\begin{align}
    \gamma^{(\lambda_{\rm gw})}(t_f,\, \bm{k}) \equiv e_{ij}^{(\lambda_{\rm gw})}(\hat{\bm{k}})\, \delta \gamma_{ij} (t_f,\, \bm{k}) \,.
\end{align}
At $t=t_f$, the adiabatic curvature perturbation is also given by the usual definition, e.g., at the linear perturbation as 
\begin{align}
    \zeta(t_f,\, \bm{k}) = \delta \psi(t_f,\, \bm{k}) - \frac{1}{4} \hat{k}_i \hat{k}_j \delta \gamma_{ij}(t_f,\, \bm{k})\,. \label{Def:zetalinear}
\end{align}

\subsubsection{Circular polarization bases}
One may want to use circular polarization bases. Using $e_i^{(\lambda)}$, the circular polarization bases are given by 
\begin{align}
    e^{(\pm_c)}_i(\hat{\bm{k}}) \equiv \frac{1}{\sqrt{2}} (e_i^{(1)}(\hat{\bm{k}}) \mp i e_i^{(2)}(\hat{\bm{k}}))\,.
\end{align}
Using $e^{(\pm_c)}_i$, one can construct a symmetric and traceless rank $n\, (=2, 3,\, \cdots)$ tensor. In particular, $n=2$ gives the bases of the gravitational waves in the circular polarization bases as
\begin{align}
    e^{(\pm_c)}_{ij}(\hat{\bm{k}}) \equiv e^{(\pm_c)}_i(\hat{\bm{k}}) e^{(\pm_c)}_j(\hat{\bm{k}})\,,
\end{align}
which satisfy
\begin{align}
    e^{(\pm_c)}_{ij}(\hat{\bm{k}}) e^{(\pm_c)}_{ij}(\hat{\bm{k}}) = 0\,, \qquad  e^{(\pm_c)}_{ij}(\hat{\bm{k}}) e^{(\mp_c)}_{ij}(\hat{\bm{k}}) = e^{(\pm_c)}_{ij}(\hat{\bm{k}}) e^{(\pm_c)*}_{ij}(\hat{\bm{k}}) = 1\,,
\end{align}
where the upper asterisk denotes the complex conjugate. Using Eq.~(\ref{Exp:parityLPB}), we find
\begin{align}
    e^{(\pm_c)}_i (\hat{\bm{k}}) = e^{(\pm_c)*}_i (- \hat{\bm{k}}) = e^{(\mp_c)}_i (-\hat{\bm{k}})\,. \label{Eq:epm} 
\end{align}
As usual, we can define the circular polarization modes at $t=t_f$ as
\begin{align}
    \gamma^{(\pm_c)}(t_f,\, \bm{k}) \equiv e_{ij}^{(\mp_c)} (\hat{\bm{k}}) \delta \gamma_{ij} (t_f,\, \bm{k}) \,. 
\end{align}
The GWs in the two porlarization bases are related as 
\begin{align}
    \gamma^{(\pm_c)}(t_f,\, \bm{k}) = \frac{1}{\sqrt{2}} (\gamma^{(+)}(t_f,\, \bm{k}) \pm i \gamma^{(\times)}(t_f,\, \bm{k}) ),  \label{Eq:2polarizations}
\end{align}
where we have used $e_{ij}^{(\pm_c)} = (e_{ij}^{(+)} \mp i e_{ij}^{(\times)})/\sqrt{2}$. Using Eq.~(\ref{Eq:2polarizations}), we find that their power spectrums are related as
\begin{align}
    \langle \gamma^{(\pm_c)} (t_f,\, \bm{k}) \gamma^{(\pm_c)} (t_f,\, \bm{p}) \rangle &= \frac{1}{2} \! \left( \langle \gamma^{(+)} (t_f,\, \bm{k}) \gamma^{(+)} (t_f,\, \bm{p}) \rangle - \langle \gamma^{(\times)} (t_f,\, \bm{k}) \gamma^{(\times)} (t_f,\, \bm{p}) \rangle \right) \cr
    & \quad \pm  \frac{i}{2} \! \left( \langle \gamma^{(+)} (t_f,\, \bm{k}) \gamma^{(\times)} (t_f,\, \bm{p}) \rangle + \langle \gamma^{(\times)} (t_f,\, \bm{k}) \gamma^{(+)} (t_f,\, \bm{p}) \rangle\right)\!\!. \label{Eq:power2polarizations}
\end{align}
When the cross-correlation between the $+$ and $\times$ modes does not vanish, the two circular polarization modes have in general different amplitudes.

\subsubsection{Parity and circular polarization}  \label{SSSec:parity}
When all the fluctuations at $t=t_*$ are parity invariant, satisfying Eq.~(\ref{Exp:parityinv_ti}), the circular polarization modes of the GWs at $t=t_f$ satisfy 
\begin{align}
    \langle \gamma^{(\lambda_1)}(t_f,\, \bm{k}_1) \cdots \gamma^{(\lambda_n)}(t_f,\, \bm{k}_n) \rangle  =  \langle \gamma^{(- \lambda_1)}(t_f,\, - \bm{k}_1) \cdots \gamma^{(- \lambda_n)}(t_f,\, -\bm{k}_n) \rangle \label{Exp:gammapm_tf}
\end{align}
for $\lambda_1,\, \cdots,\, \lambda_n= \pm_c$, where we have used Eqs.~(\ref{Exp:parityinv_tf}) and (\ref{Eq:epm}). specially for the power spectrum with $n=2$, we obtain
\begin{align}
    \langle \gamma^{(\lambda_1)}(t_f,\, \bm{k}) \gamma^{(\lambda_2)}(t_f,\, -\bm{k}) \rangle  =  \langle  \gamma^{(- \lambda_2)}(t_f,\, \bm{k}) \gamma^{(- \lambda_1)}(t_f,\, - \bm{k}) \rangle \,,   \label{Eq:Ppmc}
\end{align}
where we have used Eq.~(\ref{Exp:gammapm_tf}) and the commutation relation. We have also taken into account the momentum conservation. Therefore, as a direct consequence of the parity invariance, the auto power spectrums for the two polarization modes agree with each other. 

Meanwhile, even if the correlators at $t=t_*$ are all parity invariant, satisfying Eq.~(\ref{Exp:parityinv_ti}), the background contributions of the gauge fields may result in  
\begin{align}
    \langle \gamma^{(\lambda_1)}(t_f,\, \bm{k}_1) \cdots \gamma^{(\lambda_n)}(t_f,\, \bm{k}_n) \rangle  \neq  \langle \gamma^{(- \lambda_1)}(t_f,\, \bm{k}_1) \cdots \gamma^{(- \lambda_n)}(t_f,\, \bm{k}_n) \rangle 
\end{align}
for $n \geq 3$, while the parity is still conserved at $t=t_f$. In this sense, checking the validity of Eq.~(\ref{Exp:gammapm_tf}) becomes a more direct test of the parity violation. The non-Gaussian spectrums for the $+$ mode and the $-$ mode can have the different amplitudes without violating the parity symmetry. Similarly, the parity invariance at $t=t_*$ leads to 
\begin{align}
    \langle \zeta(t_f,\, \bm{k}) \gamma^{(\lambda)} (t_f,\, - \bm{k})  \rangle= \langle \zeta(t_f,\, -\bm{k}) \gamma^{(-\lambda)} (t_f,\, \bm{k})  \rangle  \,,  \label{Eq:crosszetaGWcp}
\end{align}
while this is not necessarily equal to $\langle \zeta(t_f,\, \bm{k}) \gamma^{(-\lambda)} (t_f,\, - \bm{k})  \rangle$. The discussion in Sec.~\ref{SSSec:parity} applies generically, when the separate universe evolution $(\star)$ holds and the fluctuations are parity invariant at $t=t_*$, satisfying Eq.~(\ref{Exp:parityinv_ti}).

\subsection{GWs at linear perturbation}
Perturbing Eq.~(\ref{Sol:gamma}), the dominant contribution in the linear perturbation under the assumption (\ref{Cond:smallshear}) is given by 
\begin{align}
   & \delta \gamma_{ij} (t_f,\, \bm{x})  = \delta \gamma_{ij*}(\bm{x}) 
    -  2 \left[ \bar{\gamma}_{il*} \bar{\gamma}_{jm*}\right]^{\rm TL} \sum_{\alpha=1}^{D'} \left\{  2\bar{\hat{\pi}}^{\{l}_{(\alpha)} \delta \hat{\pi}^{m\}}_{(\alpha)}(\bm{x})  \bar{\hat{{\cal I}}}_{(\alpha)f} + \bar{\hat{\pi}}^l_{(\alpha)} \bar{\hat{\pi}}^m_{(\alpha)} \delta \hat{{\cal I}}_{(\alpha)f} \right\}  + \cdots\,,  \label{Sol:gamma2}
\end{align}
where the indices $l$ and $m$ are symmetrized. The first term in the right hand side is the usual vacuum contribution. The second term corresponds to the fluctuation of the shear sourced by the fluctuations in the directions of $\pi^i_{(\alpha)}$, given by 
\begin{align}
   \delta \hat{\pi}_{(\alpha)}^i  = \frac{\delta \pi^i_{(\alpha)}}{\Bar{\pi}_{(\alpha)}} -  \bar{\hat{\pi}}^i_{(\alpha)} \Bar{\gamma}_{jl*} \bar{\hat{\pi}}^j_{(\alpha)}\frac{\delta \pi^l_{(\alpha)}}{\Bar{\pi}_{(\alpha)}} + \cdots\,. \label{Exp:deltapi}
\end{align}
Here, we have introduced $\Bar{\pi}_{(\alpha)} \equiv \sqrt{\bar{\gamma}_{kl*} \Bar{\pi}^k_{(\alpha)} \Bar{\pi}^l_{(\alpha)}}$. 
The second term corresponds to the one sourced by the fluctuations in the amplitudes of $\pi^i_{(\alpha)}$. In Eqs.~(\ref{Sol:gamma2}) and (\ref{Exp:deltapi}), the contributions of $\delta \gamma_{ij*}$ are abbreviated except for the first term in Eq.~(\ref{Sol:gamma2}), since they are subdominant compared to the first term in the square brackets in Eq.~(\ref{Sol:gamma2}) as long as Eq.~(\ref{Cond:smallshear}) is satisfied. This is because the fluctuations of $\gamma_{ij*}$ and $\hat{\pi}^i_{(\alpha)*}$ amount to 
$$
 \left| \frac{\delta \gamma_{ij*}}{\delta \hat{\pi}^i}_{\hspace{-5pt}(\alpha)} \right| \sim \sqrt{\frac{\rho_{A(\alpha)*}}{\rho_*}}\quad  (\ll 1) \,,
$$
due to the difference in their canonical normalization factors. In what follows, we ignore these abbreviated terms. Since $\delta \hat{\pi}_{(\alpha)}^i$ satisfies $\bar{\gamma}_{ij*} \delta \hat{\pi}_{(\alpha)}^i \bar{\hat{\pi}}^j_{(\alpha)}=0$, the second term in Eq.~(\ref{Sol:gamma2}) is sourced by the fluctuations of the gauge fields which are orthogonal to their background directions. The fluctuations $\delta \pi^i_{(\alpha)}$ along their background directions contribute through $\delta \hat{{\cal I}}_{(\alpha)f}$ in Eq.~(\ref{Sol:gamma2}).

Even if the gauge field had been negligible until the horizon crossing $t_*$, strictly speaking, we cannot choose $\bar{\gamma}_{ij}(t_*)=\delta_{ij}$ simultaneously with Eq.~(\ref{Eq:bargamma}). However, as discussed in Sec.~\ref{SSec:shear}, since the time variation of $\gamma_{ij}$ amounts to 
\begin{align}
    |\gamma_{ij}(t_f) - \gamma_{ij}(t_*) | \sim |\hat{{\cal I}}(t_f)| \ll 1\,, \label{Assum:background}
\end{align}
we use $\bar{\gamma}_{ij}(t_*) \sim \delta_{ij}$ in the following, ignoring the time variation of the background spatial metric.

\subsubsection{Linear polarization of GWs} \label{SSSec:linearP}
Operating $e_{ij}^{(\lambda_{\rm gw})}$ on Eq.~(\ref{Sol:gamma2}), we obtain
\begin{align}
    \gamma^{(+)}(t_f,\, \bm{k}) &= \gamma_*^{(+)}(\bm{k}) - 2 \sqrt{2} \sum_{\alpha=1}^{D'} \bar{\hat{{\cal I}}}_{(\alpha)f} \{ e_i^{(1)} \bar{\hat{\pi}}^i_{(\alpha)} e_j^{(1)} 
 \delta \hat{\pi}^j_{(\alpha)}(\bm{k}) - e_i^{(2)} \bar{\hat{\pi}}^i_{(\alpha)} e_j^{(2)} 
 \delta \hat{\pi}^j_{(\alpha)} (\bm{k})  \} \cr
  & \qquad \quad - \sqrt{2} \sum_{\alpha=1}^{D'}  \{ ( e_i^{(1)} \bar{\hat{\pi}}^i_{(\alpha)})^2 - ( e_i^{(2)} \bar{\hat{\pi}}^i_{(\alpha)})^2 \} \delta \hat{{\cal I}}_{(\alpha)f}(\bm{k}) \,, \label{Exp:gamma+multi} \\
  \gamma^{(\times)}(t_f,\, \bm{k}) &= \gamma_*^{(\times)}(\bm{k}) - 2 \sqrt{2} \sum_{\alpha=1}^{D'} \bar{\hat{{\cal I}}}_{(\alpha)f} \{ e_i^{(1)} \bar{\hat{\pi}}^i_{(\alpha)} e_j^{(2)} 
 \delta \hat{\pi}^j_{(\alpha)}(\bm{k}) + e_i^{(2)} \bar{\hat{\pi}}^i_{(\alpha)} e_j^{(1)} 
 \delta \hat{\pi}^j_{(\alpha)} (\bm{k})  \} \cr
  & \qquad \quad - 2 \sqrt{2} \sum_{\alpha=1}^{D'}   e_i^{(1)} \bar{\hat{\pi}}^i_{(\alpha)} e_j^{(2)} \bar{\hat{\pi}}^i_{(\alpha)} \delta \hat{{\cal I}}_{(\alpha)f}(\bm{k}) \,. \label{Exp:gammacrmulti}
\end{align}
For $\bm{k}$ which is parallel to $\bar{\bm{\pi}}_{(\alpha)}$, the GWs sourced by the corresponding gauge field in Eqs.~(\ref{Exp:gamma+multi}) and (\ref{Exp:gammacrmulti}) vanish, since they are proportional to $e_i^{(\lambda)}\bar{\hat{\pi}}^i_{(\alpha)}$ with $\lambda=1,\, 2$.  At higher orders of $\hat{{\cal I}}$, each gauge field can contribute to the GWs for all directions of $\bm{k}$.

By using Eq.~(\ref{Exp:deltapi}), the fluctuation of $e_i^{(\lambda)}(\hat{\bm{k}}) \delta \hat{\pi}^i_{(\alpha)}$ for $\lambda=1,\, 2$ is given by 
\begin{align}
    e_i^{(\lambda)}(\hat{\bm{k}})  \delta \hat{\pi}^i_{(\alpha)} (\bm{k}) &= \left\{ 1- (e_i^{(\lambda)}(\hat{\bm{k}}) \bar{\hat{\pi}}^i_{(\alpha)})^2 \right\} \frac{e_j^{(\lambda)}(\hat{\bm{k}}) \delta \pi^j_{(\alpha)}(\bm{k})}{\bar{\pi}_{(\alpha)}} \cr
    & \qquad \qquad - e_i^{(1)}(\hat{\bm{k}}) \bar{\hat{\pi}}^i_{(\alpha)} e_i^{(2)} (\hat{\bm{k}})\bar{\hat{\pi}}^i_{(\alpha)} \frac{e_j^{(\lambda')}(\hat{\bm{k}}) \delta \pi^j_{(\alpha)}(\bm{k})}{\bar{\pi}_{(\alpha)}}\,, \label{Exp:deltapi_pr}
\end{align}
where $\lambda'=2$ for $\lambda=1$ and $\lambda'=1$ for $\lambda=2$. With Eq.~(\ref{Eq:U1gc}), we have dropped the longitudinal modes of $\delta \pi^i_{(\alpha)}(\bm{k})$. Inserting Eq.~(\ref{Exp:deltapi_pr}) into Eqs.~(\ref{Exp:gamma+multi}) and (\ref{Exp:gammacrmulti}) and using Eq.~(\ref{Exp:PhiPsi}), we obtain
\begin{align}
     \gamma^{(+)}(t_f,\, \bm{k}) &= \gamma_*^{(+)}(\bm{k}) -  \sqrt{2} \sum_{\alpha=1}^{D'} \sin^2 \Theta_\alpha \cos 2 \Psi_\alpha \delta \hat{{\cal I}}_{(\alpha)f}(\bm{k}) \cr
     & \quad - 2 \sqrt{2} \sum_{\alpha=1}^{D'}  \sin \Theta_\alpha \frac{\delta \pi^j_{(\alpha)}(\bm{k})}{\Bar{\pi}_{(\alpha)}}  \bar{\hat{{\cal I}}}_{(\alpha)f} \bigl[ \cos \Psi_\alpha (1 - \sin^2 \Theta_\alpha \cos 2 \Psi_\alpha) e_j^{(1)}(\hat{\bm{k}})  \cr
     & \qquad \qquad \qquad \qquad \qquad \qquad\qquad - \sin \Psi_\alpha (1 + \sin^2\Theta_\alpha \cos 2 \Psi_\alpha)  e_j^{(2)}(\hat{\bm{k}})  \bigr]\!,\label{Exp:gammapl_linear}
\end{align}
and
\begin{align}
     \gamma^{(\times)}(t_f,\, \bm{k}) &= \gamma_*^{(\times)}(\bm{k}) -  \sqrt{2} \sum_{\alpha=1}^{D'} \sin^2 \Theta_\alpha \sin 2 \Psi_\alpha \delta \hat{{\cal I}}_{(\alpha)f}(\bm{k}) \cr
     & \quad - 2 \sqrt{2} \sum_{\alpha=1}^{D'}  \sin \Theta_\alpha \, \bar{\hat{{\cal I}}}_{(\alpha)f}\,  \frac{\delta \pi^j_{(\alpha)}(\bm{k})}{\Bar{\pi}_{(\alpha)}}  \bigl[ \sin \Psi_\alpha (1 - 2 \sin^2 \Theta_\alpha \cos^2 \Psi_\alpha) e_j^{(1)}(\hat{\bm{k}})  \cr
     & \qquad \qquad \qquad \qquad \qquad \qquad + \cos \Psi_\alpha (1 - 2 \sin^2\Theta_\alpha \sin^2 \Psi_\alpha)  e_j^{(2)}(\hat{\bm{k}})  \bigr]\!.\label{Exp:gammacr_linear}
\end{align}
For $D'=1$, setting $\Psi_1=0$ as discussed around Eq.~(\ref{Exp:Psi1}) and ignoring $\delta \If$, one can reproduce the result obtained in Ref.~\cite{Tanaka:2023gul}.

\subsubsection{Alternative expression and condition for linear polarization} 
As shown in Eqs.~(\ref{Exp:gammapl_linear}) and (\ref{Exp:gammacr_linear}), in general, the two linear polarization modes of the GWs evolve differently. In fact, this is generically the case where the backreaction of the generated gauge fields on the scalar fields is not negligible. To show this, let us rewrite $\gamma_{ij}$, given in Eq.~(\ref{Sol:gamma}), by using $\pi^i_{(\alpha)}$, as 
\begin{align}
   & \gamma_{ij} (t,\, \bm{x}) = \gamma_{ij*}(\bm{x}) 
    -  2 \left[ \gamma_{il*}(\bm{x}) \gamma_{jm*}(\bm{x}) \right]^{\rm TL}  \sum_{\alpha=1}^{D'} \pi_{(\alpha)}^l(\bm{x}) \pi_{(\alpha)}^m(\bm{x}) \, {\cal I}_{(\alpha)}(t;\{\varphi^{a'}_*\}')+ {\cal O}(\epsilon) \,, \label{Sol:gammaLP} 
\end{align}
where ${\cal I}_{(\alpha)}(t;\{\varphi^{a'}_*\}') \equiv \hat{{\cal I}}_{(\alpha)}(t;\{\varphi^{a'}_*\}')/(\gamma_{ij*} \pi^i_{(\alpha)} \pi^j_{(\alpha)})$, which is given by 
\begin{align}
    {\cal I}_{(\alpha)}(t;\{\varphi^{a'}_*\}') &= \frac{1}{\Mp^2}\! \int^{t}_{t_*}  \!\frac{dt' N}{e^{3 \psi(t')}} \! \int^{t'}_{t_*}\! \frac{dt'' N(t'')}{e^{\psi(t'')} f_{(\alpha)}^2(X^{IJ}(t''),\, \phi^I(t''))}\,.
\end{align}
${\cal I}_{(\alpha)}(t;\{\varphi^{a'}_*\}')$ does not explicitly depend on the gauge field(s), while it does implicitly when the backreaction of the gauge fields on the evolution of the scalar fields is not negligible.

Considering the linear perturbation of Eq.~(\ref{Sol:gammaLP}) and ignoring $\delta \gamma_{ij}$ in the sourced contribution, one can show that when the fluctuation of ${\cal I}_{(\alpha)}$ is negligible, satisfying 
\begin{align}
     \left| \frac{\delta {\cal I}_{(\alpha)}}{\bar{{\cal I}}_{(\alpha)}} \right| \ll  \left| \frac{\delta \pi^i_{(\alpha)}}{\Bar{\pi}_{(\alpha)}} \right|\,, \label{Cond:AbsenseLP}
\end{align}
and the two linear polarization modes of the gauge fields have the same amplitude at $t=t_*$, satisfying 
\begin{align}
    \langle e_i^{(1)}(\hat{\bm{k}}) \delta \pi^i_{(\alpha)}(\bm{k}) e_j^{(1)}(\hat{\bm{p}}) \delta \pi^j_{(\alpha)}(\bm{p})\rangle=\langle e_i^{(2)}(\hat{\bm{k}}) \delta \pi^i_{(\alpha)}(\bm{k}) e_j^{(2)}(\hat{\bm{p}}) \delta \pi^j_{(\alpha)}(\bm{p}) \rangle
\end{align}
for all $\alpha$s, the power spectrums of $\gamma^{(+)}(t_f,\, \bm{k})$ and $\gamma^{(\times)}(t_f,\, \bm{k})$ agree. Otherwise the linear polarization can be generated through the superhorizon evolution. The fluctuation of ${\cal I}_{(\alpha)}$ consists of the fluctuations of the metric and the fluctuations of the scalar fields. The condition (\ref{Cond:AbsenseLP}) requires that both intrinsic fluctuations of $\bm{\Phi} \equiv \{\phi^I,\, \pi_I\}$ and the fluctuations of $\bm{\Phi}$ sourced by the gauge fields need to be negligible. When the backreaction of the gauge fields on the scalar fields becomes important, generating a non-negligible sourced fluctuations of $\bm{\Phi}$, the condition (\ref{Cond:AbsenseLP}) fails to be satisfied (related to this point, a non-trivial example will be discussed in Sec.~\ref{SSSec:single}). When the contribution of $\delta {\cal I}_{(\alpha)}$ is not negligible, the two linear polarization modes evolve differently.

Here, we expand the fluctuation of ${\cal I}_{(\alpha)}(t;\{\varphi^{a'}_*\}')$ in terms of $\delta \pi^i_{(\alpha)}$ and $\delta \bm{\Phi}$ as
\begin{align}
    \delta \If (\bm{k}) = 2 \left( \frac{\partial \bar{\hat{{\cal I}}}_{(\alpha)f}}{\partial \bar{\rho}_{A(\alpha)*}}  \right)_{\!\!\sbm{\Phi}_*}  \bar{\rho}_{A(\alpha)*} \Bar{\gamma}_{ij*} \Bar{\hat{\pi}}^i_{(\alpha)} \frac{\delta \pi^j_{(\alpha)}(\bm{k})}{\Bar{\pi}_{(\alpha)*}} + \left( \frac{\partial \bar{\hat{{\cal I}}}_{(\alpha)f}}{\partial \bar{\bm{\Phi}}_*}  \right)_{\!\!\pi_{(\alpha)}}\hspace{-5pt} \delta \bm{\Phi}_*(\bm{k})\,,  \label{Exp:deltaI}
\end{align}
where the metric perturbations are ignored at $t=t_*$. Then, Eqs.~(\ref{Exp:gammapl_linear}) and (\ref{Exp:gammacr_linear}) can be recast into
\begin{align}
     &\gamma^{(+)}(t_f,\, \bm{k}) - \gamma_*^{(+)}(\bm{k}) \cr
     &= -  \sqrt{2} \sum_{\alpha=1}^{D'} \sin^2 \Theta_\alpha \cos 2 \Psi_\alpha \left( \frac{\partial \bar{\hat{{\cal I}}}_{(\alpha)f}}{\partial \bm{\bar{\Phi}}_*}  \right)_{\!\!\pi_{(\alpha)}} \hspace{-5pt} \delta \bm{\Phi}_*(\bm{k}) \cr
     &  \quad - 2 \sqrt{2} \sum_{\alpha=1}^{D'}  \sin \Theta_\alpha   \cos \Psi_\alpha   \left( \bar{\hat{{\cal I}}}_{(\alpha)f}  + {\cal D} \bar{\hat{{\cal I}}}_{(\alpha)f} \sin^2 \Theta_\alpha \cos2 \Psi_\alpha \right) \frac{\delta \pi^j_{(\alpha)}(\bm{k})}{\Bar{\pi}_{(\alpha)}}   e_j^{(1)}(\hat{\bm{k}}) \cr
     &  \quad + 2 \sqrt{2} \sum_{\alpha=1}^{D'}  \sin \Theta_\alpha   \sin \Psi_\alpha   \left( \bar{\hat{{\cal I}}}_{(\alpha)f}  - {\cal D} \bar{\hat{{\cal I}}}_{(\alpha)f} \sin^2 \Theta_\alpha \cos2 \Psi_\alpha \right) \frac{\delta \pi^j_{(\alpha)}(\bm{k})}{\Bar{\pi}_{(\alpha)}}  e_j^{(2)}(\hat{\bm{k}}),  \label{Exp:gammapl_linear2}
\end{align}
and
\begin{align}
     &\gamma^{(\times)}(t_f,\, \bm{k}) - \gamma_*^{(\times)}(\bm{k}) \cr
     &= -  \sqrt{2} \sum_{\alpha=1}^{D'} \sin^2 \Theta_\alpha \sin 2 \Psi_\alpha \left( \frac{\partial \bar{\hat{{\cal I}}}_{(\alpha)f}}{\partial \bm{\bar{\Phi}}_*}  \right)_{\!\!\pi_{(\alpha)}} \hspace{-5pt} \delta \bm{\Phi}_*(\bm{k}) \cr
     &  \quad - 2 \sqrt{2} \sum_{\alpha=1}^{D'}  \sin \Theta_\alpha   \sin \Psi_\alpha   \left( \bar{\hat{{\cal I}}}_{(\alpha)f}  + 2 {\cal D} \bar{\hat{{\cal I}}}_{(\alpha)f} \sin^2 \Theta_\alpha \cos^2 \Psi_\alpha \right) \frac{\delta \pi^j_{(\alpha)}(\bm{k})}{\Bar{\pi}_{(\alpha)}}  e_j^{(1)}(\hat{\bm{k}}) \cr
     &  \quad - 2 \sqrt{2} \sum_{\alpha=1}^{D'}  \sin \Theta_\alpha   \cos \Psi_\alpha   \left( \bar{\hat{{\cal I}}}_{(\alpha)f}  + 2  {\cal D} \bar{\hat{{\cal I}}}_{(\alpha)f} \sin^2 \Theta_\alpha \sin^2 \Psi_\alpha \right) \frac{\delta \pi^j_{(\alpha)}(\bm{k})}{\Bar{\pi}_{(\alpha)}}   e_j^{(2)}(\hat{\bm{k}})\,, \label{Exp:gammacr_linear2}
\end{align}
with
\begin{align}
    {\cal D} \bar{\hat{{\cal I}}}_{(\alpha)f} \equiv  \bar{\rho}_{A(\alpha)*}^2 \!\! \left( \frac{\partial }{\partial \bar{\rho}_{A(\alpha)*}} \frac{ \bar{\hat{{\cal I}}}_{(\alpha)f}}{\bar{\rho}_{A(\alpha)*}} \right)_{\!\!\sbm{\Phi}_*}\,.
\end{align}
Here, the condition (\ref{Cond:AbsenseLP}) corresponds to requiring 
\begin{align}
    \left| \frac{1}{\bar{\hat{{\cal I}}}_{(\alpha)f}} \left( \frac{\partial \bar{\hat{{\cal I}}}_{(\alpha)f}}{\partial \bar{\bm{\Phi}}_*}  \right)_{\!\!\pi_{(\alpha)}} \hspace{-15pt} \delta \bm{\Phi} \right| \ll \left| \frac{\delta \pi^i_{(\alpha)}}{\Bar{\pi}_{(\alpha)}} \right| \,, \qquad   \left| \frac{{\cal D} \bar{\hat{{\cal I}}}_{(\alpha)f}}{\bar{\hat{{\cal I}}}_{(\alpha)f}} \right| \ll 1 \,. \label{Cond:AbsenseLP2}
\end{align}
The former corresponds to ignoring the intrinsic fluctuations of the scalar fields and the latter corresponds to ignoring their fluctuations sourced by the fluctuations of the gauge fields\footnote{When the sourced fluctuation is negligible but the intrinsic one is not, the power spectrums of the two linear polarization modes can still coincide when $\Psi_\alpha$ takes specific values with $\sin 2 \Psi_\alpha = \cos 2 \Psi_\alpha$. For $D'=1$, one can choose $\Psi_\alpha$ so that this is satisfied, while for $D' \geq 2$, this is not possible unless the background gauge fields are in fine-tuned spatial configurations. \label{footnote:Psialpha}}. In fact, using Eq.~(\ref{Def:calI}), the latter condition of (\ref{Cond:AbsenseLP2}) can be rephrased as requiring that $f_{(\alpha)}^2(X^{IJ}(\psi),\,\phi^I(\psi))$  as a function of $\psi$ for $\psi_* \leq \psi \leq \psi_f$ in the integrand should be independent of the value of $\rho_{A(\alpha)*}$, which holds only when the backreaction is negligible.

One can also expand $\gamma^{(\lambda_{\rm gw})}(t_f,\, \bm{k})$ in terms of $\delta \dot{A}_{i(\alpha)*}$ (and $\delta \bm{\Phi}_*$) instead  of $\delta \pi^i_{(\alpha)}$ by using Eq.~(\ref{Def:hatpi}). A straightforward computation shows that in the expressions (\ref{Exp:gammapl_linear2}) and (\ref{Exp:gammacr_linear2}), one can replace $\delta \pi^i_{(\alpha)}/\Bar{\pi}$ with $\Bar{\gamma}^{ij}_* \delta \dot{A}_{j(\alpha)*}/\dot{\Bar{A}}_{(\alpha)*}$, where $\dot{\Bar{A}}_{(\alpha)*} \equiv \sqrt{\Bar{\gamma}^{ij}_* \dot{\Bar{A}}_{i(\alpha)*}\dot{\Bar{A}}_{j(\alpha)*}}$~.\footnote{To be more specific, Eqs.~(\ref{Exp:gammapl_linear2}) and (\ref{Exp:gammacr_linear2}) still hold as well as after the following replacement,
\begin{align}
  \left( \frac{\partial \bar{\hat{{\cal I}}}_{(\alpha)f}}{\partial \bar{\bm{\Phi}}_*}  \right)_{\!\!\pi_{(\alpha)}}  \to \left( \frac{\partial \bar{\hat{{\cal I}}}_{(\alpha)f}}{\partial \bar{\bm{\Phi}}_*}  \right)_{\!\!\dot{A}_{(\alpha)*}} , \qquad \qquad   \frac{\delta \pi^j_{(\alpha)}(\bm{k})}{\Bar{\pi}_{(\alpha)}}   \to \frac{\Bar{\gamma}^{jl}_* \delta \dot{A}_{l(\alpha)*}(\bm{k})}{\dot{\Bar{A}}_{(\alpha)*}}  \,.\label{Exp:replacement}
\end{align}
\label{footnote:replacement}.} Because of that, as long as the metric perturbation can be neglected, the condition for the absence of the linear polarization does not depend on which variable we use for the expansion.

\subsection{Adiabatic curvature perturbation}
As shown in Eq.~(\ref{Def:zetalinear}), the (linear) adiabatic perturbation is given by the summation of the fluctuations of the $e$-folding number and the longitudinal part of $\delta \gamma_{ij}$ at $t=t_f$. Repeating a similar computation to GWs, we obtain the latter as
\begin{align}
    \frac{1}{4} \hat{k}_i \hat{k}_j \delta \gamma_{ij}(t_f,\, \bm{k}) & = - \sum_{\alpha=1}^{D'} \hat{k}_i \bar{\hat{\pi}}^i_{(\alpha)} \hat{k}_j \delta \hat{\pi}^j_{(\alpha)} (\bm{k}) \bar{\hat{{\cal I}}}_{(\alpha)f} - \frac{1}{2}\sum_{\alpha=1}^{D'} \left( \cos^2 \Theta_\alpha - \frac{1}{3} \right) \delta \hat{{\cal I}}_{(\alpha)f}(\bm{k}) \,,  \label{Eq:gammaL}
\end{align}
where the fluctuation of the spatial metric, which only gives the subdominant contributions, is again ignored. Since $\hat{\pi}^i_{(\alpha)}$ is given by $\dot{A}_{i(\alpha)*}$ as in Eq.~(\ref{Def:hatpi}), $\hat{k}_j \delta \hat{\pi}^j_{(\alpha)} (\bm{k})$ in the first term can be written as 
\begin{align}
    \hat{k}_j \delta \hat{\pi}^j_{(\alpha)} (\bm{k}) = - \hat{k}_i \bar{\hat{\pi}}^i_{(\alpha)} \Bar{\hat{\pi}}^j_{(\alpha)} \frac{\delta \dot{A}_{j(\alpha)*}}{\dot{\Bar{A}}_{(\alpha)*}}\,, \label{Eq:gammaL2}
\end{align}
where we have used $k_i \delta \pi^i_{(\alpha)} =0$. Meanwhile, the fluctuation of the $e$-folding number depends significantly on the detail of the models. Once the power spectrums at $t=t_*$ are given, one can compute the power spectrums of the adiabatic curvature perturbation and GWs, using the formulae derived in this section.

\section{Power spectrums of GWs}  \label{Sec:PowerS}
In the previous section, the mapping between $t=t_*$ and $t=t_f$ was derived only by imposing the assumption (\ref{Cond:smallshear}). In what follows, we consider the case where the contributions of the gauge fields to the background evolution had been negligibly small until $t=t_*$, which allows to ignore the cross-correlation among the two polarization modes of GWs, the gauge fields, and the scalar fields at $t=t_*$. To be precise, since we have chosen the background spatial metric as $\bar{\gamma}_{ij}(t_f) = \delta_{ij}$, we cannot simultaneously set $\bar{\gamma}_{ij}(t_*) = \delta_{ij}$ in the presence of the background shear, while the corrections due to this become higher orders with respect to $\If$. This difference can be taken into account by performing the corresponding coordinate transformation after computing the quantum correlations in another set of coordinates with $\bar{\gamma}_{ij}(t_*) = \delta_{ij}$. These higher order corrections regarding $\If$ are ignored in this paper. We also ignore $\delta \pi_{I*}(\bm{k})$, imposing the slow-roll approximation at $t=t_*$.  

We express the power spectrums of the scalar fields and (the conjugate momenta of) the gauge fields at $t=t_*$ as
\begin{align}
   & \langle \delta \phi^I_* (\bm{k}) \delta \phi^J_* (\bm{p}) \rangle=  P^{IJ}_*(k) \delta(\bm{k} + \bm{p} )\,, \label{Exp:PPhi} \\
   & \langle e_i^{(\lambda)}(\hat{\bm{k}}) \delta \dot{A}_{i(\alpha)}(\bm{k}) e_j^{(\lambda')}(\hat{\bm{p}}) \delta \dot{A}_{j(\alpha)}(\bm{p}) \rangle = \delta^{\lambda \lambda'} \delta_{\alpha \alpha'} P_{(\alpha)*}^{\lambda}(k) \, \delta(\bm{k} + \bm{p} )\,, \label{Exp:PA}
\end{align}
where $P_{(\alpha)*}^{\lambda}(k) $ denotes the power spectrums of $\delta \dot{A}_{i(\alpha)}$ in the linear polarization bases with $\lambda=1,\,2$.

\subsection{General formulae of GWs}  \label{SSec:generalformulaPS}
\subsubsection{Power spectrums}
Using the formulae derived in this paper, we compute the power spectrums of GWs. In the main text, we consider the case with $g_{(\alpha)} =0$, choosing the linear polarization bases. (A discussion for $g_{(\alpha)} \neq 0$ can be found in Appendix \ref{Sec:circualrP}.) In this case, the two polarization modes of the gauge fields have the same amplitude at $t=t_*$. Using Eqs.~(\ref{Exp:gammapl_linear2}) and (\ref{Exp:gammacr_linear2}) with the replacement (\ref{Exp:replacement}), we obtain 
\begin{align}
    &\langle \gamma^{(\lambda_{\rm gw})} (t_f,\, \bm{k}) \gamma^{(\lambda_{\rm gw})} (t_f,\, \bm{p}) \rangle \cr
    &= \delta (\bm{k} + \bm{p}) \frac{2}{k^3} \left( \frac{H_{*}}{\Mp} \right)^2 \left( 1 +  \sum_{\alpha=1}^{D'} g_{t (\alpha)}^{\lambda_{\rm gw}} (\bm{k}) \sin^2 \Theta_\alpha (\hat{\bm{k}})+ \Delta^{\lambda_{\rm gw}}_\phi(\bm{k})   \right)  \,, \label{Def:gt}
\end{align}
with the contribution from the gauge field(s) being
\begin{align}
    & g_{t(\alpha)}^{+} (\bm{k})= g_{t(\alpha)} \cos 2 \Psi_\alpha(\hat{\bm{k}}) \! \left(1 + 2 \frac{{\cal D}\bar{\hat{{\cal I}}}_{(\alpha)f}}{\bar{\hat{{\cal I}}}_{(\alpha)f}} \sin^2 \Theta_\alpha(\hat{\bm{k}}) + \left( \frac{{\cal D}\bar{\hat{{\cal I}}}_{(\alpha)f}}{\bar{\hat{{\cal I}}}_{(\alpha)f}} \right)^{\!\!2} \hspace{-2pt}  \sin^4 \Theta_\alpha(\hat{\bm{k}}) \cos^2 2 \Psi_\alpha(\hat{\bm{k}}) \! \right)\!\!, \label{Exp:gtpl2} \\
    & g_{t(\alpha)}^{\times}(\bm{k}) =g_{t(\alpha)} \cos 2 \Psi_\alpha(\hat{\bm{k}}) \! \left(1 - \left( \frac{{\cal D}\bar{\hat{{\cal I}}}_{(\alpha)f}}{\bar{\hat{{\cal I}}}_{(\alpha)f}} \right)^{\!\!2} \hspace{-2pt} \sin^4 \Theta_\alpha(\hat{\bm{k}}) \sin^2 2\Psi_\alpha(\hat{\bm{k}}) \! \right)\!\!, \label{Exp:gcr2}
\end{align}
and the one from the scalar field(s) being
\begin{align}
    &\Delta_{\phi}^+ (\bm{k})  = \sum_{\alpha=1}^{D'} \sum_{\beta=1}^{D'}  g_{t(\phi;\alpha\beta)} \sin^2 \Theta_\alpha(\hat{\bm{k}}) \cos 2 \Psi_\alpha(\hat{\bm{k}}) \sin^2 \Theta_\beta(\hat{\bm{k}}) \cos 2 \Psi_\beta(\hat{\bm{k}}) \,, \\
    & \Delta_{\phi}^\times (\bm{k})  = \sum_{\alpha=1}^{D'} \sum_{\beta=1}^{D'}  g_{t(\phi;\alpha\beta)} \sin^2 \Theta_\alpha(\hat{\bm{k}}) \sin 2 \Psi_\alpha(\hat{\bm{k}}) \sin^2 \Theta_\beta(\hat{\bm{k}}) \sin 2 \Psi_\beta(\hat{\bm{k}}) \,. 
\end{align}
Here, we have introduced $g_{t(\alpha)}$ and $g_{t(\phi;\alpha\beta)}$ as
\begin{align}
    &  g_{t(\alpha)} \equiv 4 \bar{\hat{{\cal I}}}_{(\alpha)f}^2  \frac{P_{(\alpha)*}^{\lambda}(k) k^3}{\dot{\bar{A}}_{(\alpha)*}^2} \left( \frac{\Mp}{H_*} \right)^2,  \label{Exp:gt} \\
    & g_{t(\phi;\alpha\beta)} \equiv  \left( \frac{\partial \bar{\hat{{\cal I}}}_{(\alpha)f}}{\partial \bar{\phi}^I_*}  \right)_{\!\!\dot{A}_{(\alpha)*}}   \left( \frac{\partial \bar{\hat{{\cal I}}}_{(\beta)f}}{\partial \bar{\phi}^J_*}  \right)_{\!\!\dot{A}_{(\alpha)*}} P^{IJ}_*(k) \,k^3 \left( \frac{\Mp}{H_*} \right)^2\,. \label{Exp:gtphi}
\end{align}

Repeating a similar computation, we obtain the cross-correlation between the two polarization modes as
\begin{align}
     \langle \gamma^{(+)} (t_f,\, \bm{k}) \gamma^{(\times)} (t_f,\, \bm{p}) \rangle = 
     \delta (\bm{k} + \bm{p}) \frac{2}{k^3} \left( \frac{H_{*}}{\Mp} \right)^2 \left(  \,\sum_{\alpha=1}^{D'} g_{t (\alpha)}^{\rm cr} (\bm{k}) \sin^2 \Theta_\alpha (\hat{\bm{k}})+ \Delta^{\rm cr}_\phi(\bm{k})   \right)
\end{align}
with
\begin{align}
   & g_{t (\alpha)}^{\rm cr} (\bm{k}) = - g_{t(\alpha)} \sin 2 \Psi_\alpha(\hat{\bm{k}})  \left( 1 + \frac{{\cal D}\bar{\hat{{\cal I}}}_{(\alpha)f}}{\bar{\hat{{\cal I}}}_{(\alpha)f}} \sin^2 \Theta_\alpha(\hat{\bm{k}})  +\left( \frac{{\cal D}\bar{\hat{{\cal I}}}_{(\alpha)f}}{\bar{\hat{{\cal I}}}_{(\alpha)f}} \right)^{\!\!2} \hspace{-2pt}   \sin^4 \Theta_\alpha(\hat{\bm{k}})  \cos^2 2 \Psi_\alpha(\hat{\bm{k}})  \right),\\
   & \Delta_{\phi}^{\rm cr} (\bm{k})  = - \sum_{\alpha=1}^{D'} \sum_{\beta=1}^{D'}  g_{t(\phi;\alpha\beta)} \sin^2 \Theta_\alpha(\hat{\bm{k}}) \cos 2 \Psi_\alpha(\hat{\bm{k}}) \sin^2 \Theta_\beta(\hat{\bm{k}}) \sin 2 \Psi_\beta(\hat{\bm{k}}). 
\end{align}

Here, we can find a couple of qualitative differences between single gauge field models with $D'=1$ and multi gauge field models with $D' \geq 2$. For the former, we can choose the orthonomal bases so that $\Psi=0$, while for the latter, we cannot set all $\Psi_\alpha$ to 0 simultaneously. Therefore, the cross correlation between the two linear polarization modes of GWs takes a non-vanishing value only for $D' \geq 2$. Similarly, the cross-correlation between $\zeta$ and $\gamma^{\times}$ does not vanish either for $D' \geq 2$. In addition, for $D' \geq 2$, the spectrums depend also on the azimuthal direction of $\bm{k}$ in any choice of coordinates.

Inserting the power spectrums obtained here into Eq.~(\ref{Eq:power2polarizations}), one can also compute the power spectrums of GWs in the circular polarization basis. While the cross-correlation between $\gamma^{(+)}$ and $\gamma^{(\times)}$ does not vanish for $D' \geq 2$, the amplitudes of the two circular polarization modes become the same. This is because the two terms in the second line of Eq.~(\ref{Eq:power2polarizations}) are canceled with each other since $g_{t (\alpha)}^{\rm cr} (-\bm{k}) = - g_{t (\alpha)}^{\rm cr} (\bm{k})$ and $\Delta_{\phi}^{\rm cr} (-\bm{k}) = - \Delta_{\phi}^{\rm cr} (\bm{k})$ as a consequence of $\Psi_\beta(-\hat{\bm{k}}) = - \Psi_\beta(\hat{\bm{k}})$. (See also Appendix \ref{Sec:circualrP}.) Meanwhile, a non-zero cross-correlation between $\zeta$ and $\gamma^{\times}$ results in $\langle \zeta(t_f,\, \bm{k}) \gamma^{(\pm_c)}(t_f,\, \bm{p}) \rangle \neq \langle \zeta(t_f,\, \bm{k}) \gamma^{(\mp_c)}(t_f,\, \bm{p}) \rangle$, while the theory is parity invariant. (See also Eq.~(\ref{Eq:crosszetaGWcp}).)

\subsubsection{Order estimation}
For a concrete evaluation, in what follows, let us compute the mode function of the gauge fields, ignoring he metric perturbations and the mixing with the scalar fields until $t=t_*$. Then, the mode functions of $\delta A_{i(\alpha)}$ in the subhorizon limit is given by 
\begin{align}
    v^{(\alpha)} (t,\, k) = \frac{1}{f_{(\alpha)}(t)} \frac{1}{\sqrt{2 k}} e^{- i k \eta(t)}  \,,
\end{align}
for the two polarization modes in the linear polarization bases, where $\eta$ denotes the corresponding conformal time. Choosing $t_*$ as the horizon crossing time, {\it i.e.}, $k = H_* e^{\Bar{\psi}_*}$ with $H \equiv \dot{\bar{\psi}}$ being the Hubble parameter, we obtain
\begin{align}
    \frac{P_{(\alpha)*}(k)}{\dot{\bar{A}}_{(\alpha)*}^2} & = \frac{1}{2 k} \left( 1 + \left( \frac{\dot{\bar{f}}_{(\alpha)*}}{H_* \Bar{f}_{(\alpha)*}} \right)^2 \right) \left( \frac{H_{*}}{\bar{f}_{(\alpha)*} \dot{\bar{A}}_{(\alpha)*} } \right)^2 \cr
    & = \frac{1}{12} \left( 1 + \left( \frac{\dot{\bar{f}}_{(\alpha)*}}{H_*\Bar{f}_{(\alpha)*}} \right)^2 \right) \frac{1}{k^3} \frac{\Bar{\rho}_{*}}{\Bar{\rho}_{A (\alpha)*}} \left( \frac{H_{*}}{\Mp} \right)^2 \,.   \label{Eq:ampA}
\end{align}
Using this expression, we obtain $g_t$, defined in Eq.~(\ref{Exp:gt}), as
\begin{align}   
    g_{t(\alpha)} =  \frac{\Bar{\rho}_*}{3 \Bar{\rho}_{A(\alpha)*}} \bar{\hat{{\cal I}}}_{(\alpha)f}^2 
    \left( 1 + \left( \frac{\dot{\bar{f}}_{(\alpha)*}}{H_{*}\Bar{f}_{(\alpha)*}} \right)^{\!\!2} \right)\,. 
\end{align}
To validate the perturbation, 
\begin{align}
   1 \gg  \frac{\Bar{\rho}_*}{\Bar{\rho}_{A(\alpha)*}} \left( \frac{H_*}{\Mp}\right)^2 \simeq \frac{H_*^4}{\Bar{\rho}_{A(\alpha)*}}  \label{Cond:perturbation}
\end{align}
needs to be fulfilled.

When ${\cal D}\bar{\hat{{\cal I}}}_{(\alpha)f}$ does not vanish, the amplitudes of the two linear polarization modes become different (see also footnote \ref{footnote:Psialpha}). The information of each model is encoded through $\If$ and the time variation of $f_{(\alpha)}$, which can be determined just by solving the background evolution. The smaller the background energy fraction of the gauge field $(\alpha)$ becomes, the more the fluctuation of the corresponding gauge field generates GWs. This is because the amplitude of the fluctuation of the gauge field, $\delta \dot{A}_{(\alpha)}$, is determined by the Hubble parameter. Therefore, the normalized fluctuation, $|\delta \dot{A}_{(\alpha)}/\dot{\Bar{A}}_{(\alpha)}|$ becomes larger as $|\dot{\Bar{A}}_{(\alpha)}|$ or $\Bar{\rho}_{A(\alpha)}/\Bar{\rho}$ decreases. When we roughly estimate $\partial \hat{{\cal I}}_{(\alpha)f}/\partial \bar{\phi}_*$ as $(f_{(\alpha), \phi}/f_{(\alpha)}) \hat{{\cal I}}_{(\alpha)f}$ and $P(k) k^3 \sim (H_*/\Mp)^2/c_s$ with $P(k)$ and $c_s$ being the power spectrum and the sound speed of the scalar field, the ratio between the contributions of the gauge field and the scalar field is given by 
\begin{align}
    \frac{g_{t(\alpha)}}{g_{t(\phi;\alpha\beta)}} \sim c_s \frac{f_{(\alpha)}}{f_{(\alpha), \phi}}  \frac{f_{(\beta)}}{f_{(\beta), \phi}} \frac{\Bar{\rho}_*}{ \Bar{\rho}_{A(\alpha)*}}\,. 
\end{align}
Here, the different scalar fields are not distinguished just to provide an order estimation.

\subsection{Case studies}  \label{SSec:Models}
So far, we have derived the power spectrums of GWs without assuming a concrete model. In this subsection, as a classification of models, we focus on whether 
\begin{align}
  \left| \frac{\delta \If}{\If} \right| \ll |\delta \hat{\pi}_{(\alpha)}^i|\, \label{Cond:deltaI}
\end{align}
holds or not. When it does, the power spectrums of GWs can be obtained by setting  
\begin{align}
    g_{t(\phi;\alpha\beta)} = 0, \qquad {\cal D}\bar{\hat{{\cal I}}}_{(\alpha)f} = - \bar{\hat{{\cal I}}}_{(\alpha)f}\, \label{Eqs:deltaI0}
\end{align}
in the formulae derived in the previous subsection. When $\delta \If$ is negligible, it is more convenient to use Eqs.~(\ref{Exp:gammapl_linear2}) and (\ref{Exp:gammacr_linear2}) after the replacement (\ref{Exp:replacement}).

As discussed after Eq.~(\ref{Exp:If_inf}), when $\rho_{A(\alpha)}$ reaches and sustains the maximum value $\rho_{A(\alpha)}^{\rm max}$ during the period of $\Delta \psi_{(\alpha)}$ after the horizon crossing, $\If$ is roughly given by $\If \sim 2 \Delta \psi_{(\alpha)} \rho_{A(\alpha)}^{\rm max}/\rho_*$. In this case, Eq.~(\ref{Cond:deltaI}) requires
\begin{align}
   \left| \frac{\delta \rho_{A(\alpha)}^{\rm max}}{\rho_{A(\alpha)}^{\rm max}} \right| = \left| - 2 \frac{\delta f^{\rm max}_{(\alpha)}}{f^{\rm max}_{(\alpha)}} + \frac{\delta(\gamma_{ij} \pi^i_{(\alpha)}\pi^j_{(\alpha)})}{\gamma_{ij} \pi^i_{(\alpha)}\pi^j_{(\alpha)}} \right| \ll |\delta \hat{\pi}_{(\alpha)}^i|\,. \label{Cond:deltaI2}
\end{align}
Since the amplitude of the second term in the middle expression is roughly comparable to $|\delta \hat{\pi}_{(\alpha)}^i|$, the condition (\ref{Cond:deltaI}) does not hold, unless the scalar fields and the gauge fields are tightly correlated with each other. Namely, the condition (\ref{Cond:deltaI2}) requires a reduction of degrees of freedom in the phase space.

\begin{figure}
    \centering
     \includegraphics[
      width=8cm]{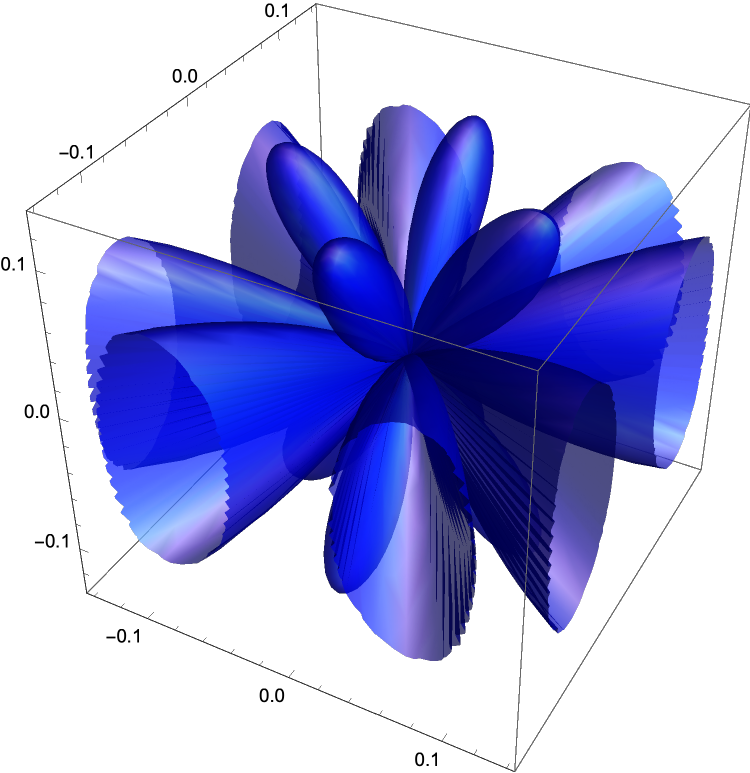}
     \includegraphics[
     width=7.2cm]{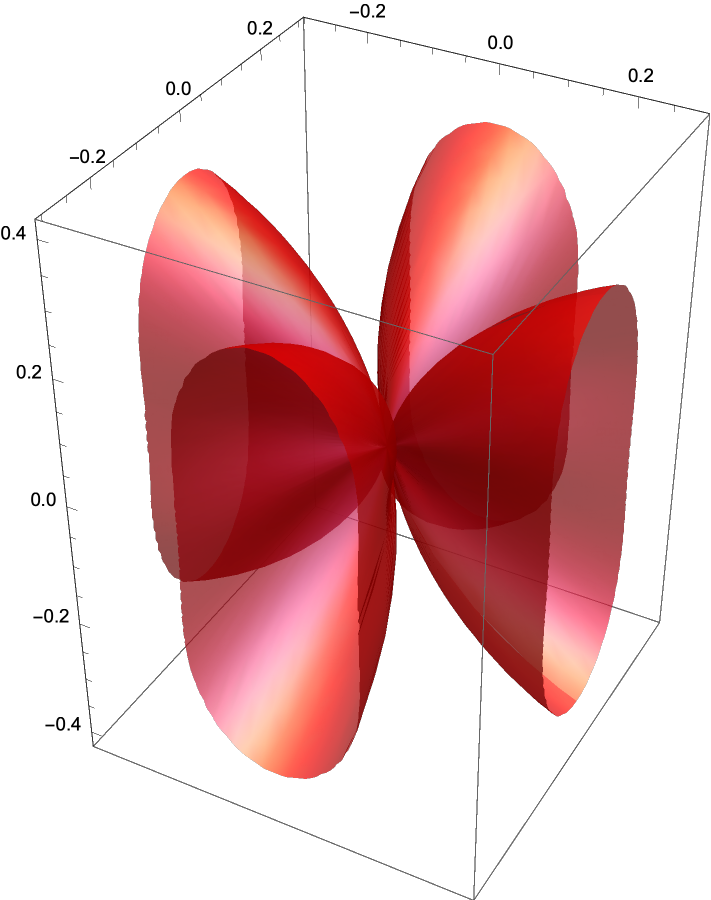}
    \hfill
    \caption{\label{fig:gt}This plot is to demonstrate the difference between $g_t^+(\bm{k})$ and $g_t^{\times}(\bm{k})$ given by Eqs.~(\ref{Exp:gtpldeltaI0multi}) and (\ref{Exp:gtcrdeltaI0multi}). The left panel shows $\sin^2 \Theta(\hat{\bm{k}})\, g_t^+(\bm{k})$ and the right one shows $\sin^2 \Theta(\hat{\bm{k}})\, g_t^{\times}(\bm{k})$ for each direction of $\bm{k}$. Here, the index of the gauge field, $\alpha$, is ignored. The wavenumber $\hat{\bm{k}}$ which is parallel to the background gauge field corresponds to the $z$ axis $(\Theta=0)$. Meanwhile, the wavenumbers which are orthogonal to the background gauge field correspond to the $x\,$-$\,y$ plane $(\Theta= \pi/2)$.} 
\end{figure}
When the condition (\ref{Cond:deltaI}) holds, inserting Eq.~(\ref{Eq:ampA}) into Eqs.~(\ref{Exp:gammapl_linear2}) and (\ref{Exp:gammacr_linear2}) after the replacement (\ref{Exp:replacement}), we obtain the power spectrums of GWs in the linear polarization basis as Eq.~(\ref{Def:gt}) with\cite{Tanaka:2023gul} 
\begin{align}
    & g_{t(\alpha)}^{+} (\bm{k})= g_{t(\alpha)} \cos 2 \Psi_\alpha(\hat{\bm{k}}) \! \left(1 - 2  \sin^2 \Theta_\alpha(\hat{\bm{k}}) +   \sin^4 \Theta_\alpha(\hat{\bm{k}}) \cos^2 2 \Psi_\alpha(\hat{\bm{k}}) \! \right)\!\!, \label{Exp:gtpldeltaI0multi} \\
    & g_{t(\alpha)}^{\times}(\bm{k}) =g_{t(\alpha)} \cos 2 \Psi_\alpha(\hat{\bm{k}}) \! \left(1 - \sin^4 \Theta_\alpha(\hat{\bm{k}}) \sin^2 2\Psi_\alpha(\hat{\bm{k}}) \! \right)\!\!, \label{Exp:gtcrdeltaI0multi}\\
    & \Delta^{+}_{\phi}=\Delta^{\times}_{\phi}=0\,. 
\end{align}
For $D'=1$, setting $\Psi=0$, we can further simplify $g_{t(\alpha)}^{\lambda_{\rm gw}}$ as $g_{t}^{+} = g_{t} \cos^4 \Theta$ and $g_{t}^{\times} =g_{t}$~\cite{Fujita:2018zbr}. The angular dependence of the power spectrums for these two linear polarization modes is shown in Fig.~\ref{fig:gt}.

\subsubsection{Condition (\ref{Cond:deltaI})}
In the rest of this subsection, let us consider a model whose Lagrangian density is given by 
\begin{align}
    {\cal L}_{\rm mat} = {\cal P}(X^I) - V(\phi^I) - \sum_{\alpha=1}^{D'} \frac{f_{(\alpha)}^2(\phi^I)}{4} F_{\mu \nu(\alpha)} F_{(\alpha)}^{\mu \nu} \,,  \label{Exp:Lmatter_single}
\end{align}
with $X^I \equiv - \partial^\mu \phi^I \partial_\mu \phi^I/2\, (I=1,\, \cdots D)$. 
The condition (\ref{Cond:deltaI}) can be satisfied, for instance, when $\rho_{A(\alpha)}$ approaches and sustains the maximum value, keeping the time variation of $\rho_{A(\alpha)}$ negligible during $\Delta \psi$. In what follows, let us find out such an example. At the leading order of the gradient expansion, the KG equation for each scalar field under the slow-roll approximation is given by 
\begin{align}
    3 \dot{\psi} {\cal P}_{X^I} \dot{\phi}^I + V_I -2 \sum_{\alpha=1}^{D'} (\ln f_{(\alpha)})_{,\phi^I} \, \rho_{A(\alpha)} = 0\,, \label{Eq:KGanimulti}
\end{align}
with $V_I \equiv \partial V/\partial \phi^I$. Here, we solve the leading order of the gradient expansion to determine the evolution of separate universes whose initial conditions at $t=t_*$ differs from the background evolution.

Dividing the KG equation (\ref{Eq:KGanimulti}) by the Friedman equation $3 \dot{\psi}^2 = V/\Mp^2$, where the kinetic terms of the scalar fields and the energy density of the gauge field are ignored, we obtain
\begin{align}
    {\cal P}_{X^I} \frac{d \phi^I}{d \psi} + \Mp^2 \frac{V_I}{V} - 2 \sum_{\alpha=1}^{D'} (\ln f_{(\alpha)})_{, \phi^I} \Mp^2 \frac{\rho_{A(\alpha)}}{V} = 0\,. \label{Eq:KGani2multi}
\end{align}
Multiplying $(\ln f_{(\alpha)})_{, \phi^I}/{\cal P}_{X^I}$, summing over $I=1,\,\cdots,\, D$, and using 
\begin{align}
    \sum_{I=1}^D (\ln f_{(\alpha)})_{, \phi^I} \frac{d \phi^I}{d \psi} = - \frac{1}{2} \frac{d \ln \rho_{A(\alpha)}}{d \psi} - 2 \,,
\end{align}
we obtain
\begin{align}
    \frac{d}{d \psi} \frac{\rho_{A(\alpha)}}{V} +  \sum_{\beta=1}^{D'} S_{\alpha \beta}(\phi^I, X^I)  \frac{\rho_{A(\beta)}}{V}\frac{\rho_{A(\alpha)}}{V} - \Lambda_\alpha(\phi^I, X^I) \frac{\rho_{A(\alpha)}}{V}  = 0\,, \label{Eq:eomrhoAmulti}
\end{align}
with
\begin{align}
    &S_{\alpha \beta}(\phi^I, X^I) \equiv 4 \Mp^2 \sum_{I=1}^D \frac{(\ln f_{(\alpha)})_{, \phi^I}(\ln f_{(\beta)})_{, \phi^I}}{{\cal P}_{X^I}} \,,\label{Def:Lambdamulti} \\
    & \Lambda_\alpha (\phi^I, X^I) \equiv 2 \Mp^2 \sum_{I=1}^D \frac{(\ln f_{(\alpha)})_{,\phi^I} (\ln V)_{, \phi^I} }{{\cal P}_{X^I}} - 4 \,. 
\end{align}
Here, we have neglected the time variation of $V$. For clarification, we have explicitly written here the summation over $I$. As is clear from the definition, $S_{\alpha \beta}$ is a symmetric matrix, satisfying $S_{\alpha \beta} = S_{\beta \alpha}$. The positivity of ${\cal P}_{X^I}$ implies that the diagonal component of $S_{\alpha \beta}$ is positive definite. In the limit $\rho_{A(\alpha)}/V \ll 1$ for all $\alpha$s, we can ignore the second term of Eq.~(\ref{Eq:eomrhoAmulti}). Then, the energy density of each gauge field evolves independently, following $\rho_{A(\alpha)}/V \propto \exp[\int^{\psi} d\psi' \Lambda_{\alpha} (\psi')]$. Therefore, for $\Lambda_\alpha(\phi^I, X^I) > 0$, $\rho_A$ exponentially increases in time. This condition on $\Lambda_\alpha$ can be obtained also from 
\begin{align}
   - 2 > \frac{d \ln  f_{(\alpha)}}{d \psi} =  \sum_{I=1}^D (\ln f_{(\alpha)})_{,\phi^I} \frac{d\phi^I}{d \psi} = - \Mp^2 \sum_{I=1}^D \frac{(\ln f_{(\alpha)})_{,\phi^I} (\ln V)_{, \phi^I} }{{\cal P}_{X^I}}\,, 
\end{align}
which is necessary for $\rho_{A(\alpha)}$ to increase in time, where we have used the KG equation~\cite{Soda:2012zm}. If each scalar field interacts with only one gauge field, $\rho_{A(\alpha)}$ evolves independently without being mixed with other gauge fields, since $S_{\alpha \beta}$ becomes diagonal. Otherwise, when the second term of Eq.~(\ref{Eq:eomrhoAmulti}) starts to be important after the exponential growth of $\rho_{A(\beta)}$, different gauge fields start to interact through the backreaction on the scalar fields $\phi^I$.

To analyze the equation (\ref{Eq:eomrhoAmulti}), first let us consider a special case where $\Lambda_\alpha$ and $S_{\alpha \beta}$ are constant in time. If the symmetric matrix $S_{\alpha \beta}$ is invertible, there exists an asymptotic solution 
\begin{align}
   \frac{\rho_{A(\alpha)}}{V} \to S^{-1}_{\alpha \beta} \Lambda_\beta \left( \equiv \frac{\rho_{A(\alpha), {\rm asym}}}{V} \right) \,.  \label{Exp:rhoA_asympmulti}
\end{align}
The invertibility of $S_{\alpha \beta}$ requires $D \geq D'$, since the rank of $S_{\alpha \beta}$, which is $D$, needs to be equal to or larger than the dimension of the matrix $S_{\alpha \beta}$, $D'$. When we perturb Eq.~(\ref{Eq:eomrhoAmulti}) around this asymptotic solution as $\rho_{A(\alpha)}/V = (1 + \delta_\alpha)\rho_{A(\alpha), {\rm asym}}/V $, we obtain
\begin{align}
    \frac{d \delta_\alpha}{d \psi} + \sum_{\beta=1}^{D'} \tilde{S}_{\alpha \beta} \delta_\beta = 0\, 
\end{align}
with $\tilde{S}_{\alpha \beta} \equiv S_{\alpha \beta}\rho_{A(\beta), {\rm asym}}/V = \sum_{\gamma=1}^{D'}S_{\alpha \beta} S^{-1}_{\beta \gamma} \Lambda_\gamma$, which can be solved as
\begin{align}
    \delta_{\alpha} (\psi) = \left[ e^{- \Tilde{S}(\psi - \psi_i)} \right]_{\alpha \beta} \delta_{\beta} (\psi_i) \,.
\end{align}
Since all eigenvalues of $\tilde S_{\alpha\beta}$ are positive\footnote{Using a $D' \times D'$ orthogonal matrix $R={\rm diag}(\rho_1/V,\, \cdots,\, \rho_{D'}/V)$, one can express $\tilde{S}$ as $\tilde{S}= S R$, where the matrix indices are abbreviated. Introducing the eigenvalues and eigenvectors for $S$ and $\tilde{S}$, given by 
\begin{align}
    S \bm{v}_i = \lambda_i \bm{v}_i\, \qquad  \tilde{S} \tilde{\bm{v}}_i = \tilde{\lambda}_i \tilde{\bm{v}}_i \qquad (i=1,\, \cdots\,, D), \label{Eq:eigeneq}
\end{align}
we obtain
\begin{align}
    {\bm v}_j \cdot \tilde{S} \tilde{\bm v}_i = \lambda_j {\bm v}_j \cdot R  \tilde{\bm v}_i = \tilde{\lambda}_i {\bm v}_j \cdot \tilde{\bm v}_i\,.  \label{Eq:positivityoflambda}
\end{align}
Here, at the first equality we have used the first equation of Eq.~(\ref{Eq:eigeneq}) and at the second equality, we have used the second one. Multiplying $ \lambda_j \tilde{\bm{v}}_i \cdot \bm{v}_j$ on Eq.~(\ref{Eq:positivityoflambda}) and taking summation over $j$, we obtain
\begin{align}
    \sum_{j=1}^{D'} (\tilde{\bm{v}}_i \cdot \bm{v}_j) ({\bm v}_j \cdot R  \tilde{\bm v}_i ) = \tilde{\bm{v}}_i \cdot R  \tilde{\bm v}_i  =  \tilde{\lambda}_i \sum_{j=1} \lambda_j^{-1} (\bm{v}_j \cdot \bm{v}_i)^2\,,
\end{align}
where on the first equality, we have used $\sum_j \bm{v}_j \otimes \bm{v}_j = {\bm 1}$. Since $\lambda_i$ and the diagonal components of $R$ are all positive, $\tilde{\lambda}_i$ is shown to be all positive. 
}, $\rho_{A(\alpha)}$ all approach the asymptotic values (\ref{Exp:rhoA_asympmulti}). We define $\psi_{(\alpha){\rm asym}}$ as the $e$-folding number at which $\rho_{A(\alpha)}$ approaches the asymptotic value (\ref{Exp:rhoA_asympmulti})\footnote{For $\psi \geq \psi_{{\rm asym}(\alpha)}$, the KG equation (\ref{Eq:KGani2multi}) can be rewritten as 
\begin{align}
    {\cal P}_{X^I} \frac{d \phi^I}{d \psi} + \Mp^2 \frac{V_I}{V} - 2 \Mp^2 \sum_{\alpha, \beta=1}^{D'} (\ln f_{(\alpha)})_{, \phi^I} S_{\alpha \beta}^{-1} \Lambda_\beta= 0\,. \label{Eq:KGani3}
\end{align}
For example, for $D=1$ and $D'=1$, the ratio between the potential driven force (the second term) and the friction term due to the gauge field (the third term) is given by 
\begin{align}
    \frac{\Lambda}{2\Mp^2 (\ln V)_{,\phi}(\ln f)_{,\phi}/{\cal P}_X } = \frac{\Lambda}{\Lambda+4}\,. 
\end{align}
This indicates that when $\rho_A$ undergoes a sizable exponential growth in the time scale of the cosmic expansion, having $\Lambda \geq {\cal O}(1)$, the backreaction of the gauge field on the scalar field always become non-negligible for $\psi \geq \psi_{\rm asym}$.} . For $D'=1$, since $\tilde{S}= \Lambda$, for $\Lambda>0$, $\rho_A/V$ approaches the asymptotic value $\Lambda/S$ within the $e$-folding number $1/\Lambda$. (Here, we have dropped the index of the gauge field.) In fact, for $D'=1$, Eq.~(\ref{Eq:eomrhoAmulti}) can be solved as
\begin{align}
    \frac{\rho_A}{V} =  \frac{\rho_{A, {\rm asym}}}{V}  \left[ 1 + \left(  \frac{\rho_{A, {\rm asym}}}{\rho_{A*}}  - 1 \right) e^{- \Lambda(\psi - \psi_*)} \right]^{-1}\,,  \label{Sol:rhoA_asympgen}
\end{align}
which becomes constant in time once the exponential term in the denominator falls off.

Next, let us consider the time variation of $\Lambda_\alpha$ and $S_{\alpha \beta}$, expanding as
\begin{align}
    Q(\psi) = Q(\psi_*) + \frac{d Q}{d \psi}\Big|_{\psi= \psi_*} (\psi - \psi_*) + \cdots  \qquad (Q=\Lambda_\alpha, S_{\alpha \beta}) \,,\label{Eq:expansionQ}
\end{align}
where the expansion coefficients are suppressed by the slow-roll parameters, by definition, during the slow-roll evolution. Repeating a similar computation, we find that $\rho_{A(\alpha)}/V$ still approach $S^{-1}_{\alpha \beta *} \Lambda_{\beta*}$, evaluated at $t=t*$ with small corrections which roughly amount to the second terms in Eq.~(\ref{Eq:expansionQ})\footnote{When we expand $\rho_{A(\alpha)}(\psi)/V$, $Q(\psi)$ with $Q=\Lambda_\alpha, S_{\alpha \beta}$ as
\begin{align}
    \frac{\rho_{A(\alpha)}(\psi)}{V} = S^{-1}_{\alpha \beta*}\Lambda_{\beta*}(1 + \Delta_\alpha(\psi))\,, \qquad Q(\psi) = Q_* + \delta Q(\psi)\,,
\end{align}
one can solve Eq.~(\ref{Eq:eomrhoAmulti}) perturbatively as
\begin{align}
    \Delta_{\alpha} (\psi) = \left[ e^{- \tilde{S}_*(\psi - \psi_*)}\right]_{\alpha \beta} \Delta_{\beta*} + \int^{\psi}_{\psi_*} d \psi' \left[ e^{\tilde{S}_*(\psi' - \psi)} \right]_{\alpha \beta} \left( \delta \Lambda_\beta(\psi') - \delta S_{\beta \gamma}(\psi') S^{-1}_{\gamma \kappa*} \Lambda_{\kappa*} \right)\,,
\end{align}
where we have abbreviated the summations over $\beta$, $\gamma$, and $\kappa$. The second term is suppressed by the slow-roll parameters. }.

In the separate universe approach, the spatial inhomogeneity is described by different initial conditions, which are assigned to each separate universe based on the quantum correlators at $t=t_*$. When $\Lambda_\alpha \gtrsim 1$ and $\tilde{S}_{\alpha \beta} \gtrsim 1$ both hold, the averaged energy densities of the gauge fields in each separate universe typically approach the value (\ref{Exp:rhoA_asympmulti}) right after the horizon crossing. Then, we obtain 
\begin{align}
    \If \sim 2  \frac{\rho_{A(\alpha), {\rm asym}}}{\rho_*} \Delta \psi_{\alpha} \,, \label{Exp:Ifatt}
\end{align}
for which the fluctuation, $|\delta \If/\If|$, is suppressed by the slow-roll parameters  compared to $|\delta \hat{\pi}^i_{(\alpha)}|$ as long as $|\delta \Delta \psi/\Delta \psi|$ remains smaller, satisfying Eq.~(\ref{Cond:deltaI})\footnote{Let us consider an example where $\rho_{A(\alpha)}$ approaches to the constant value given in Eq.~(\ref{Exp:rhoA_asympmulti}) after the exponential growth $\rho_{A(\alpha)} \propto e^{\Lambda_\alpha \psi}$ and remains to take this constant value at least until the trajectory converges. In this case, let us evaluate the fluctuation of $\Delta \psi_{\alpha}$, assuming $\delta \Delta \psi_\alpha \sim - \delta \psi_{\alpha, {\rm asym}}$, where $\psi_{\alpha, {\rm asym}}$ denotes the $e$-folding number at which $\rho_{A(\alpha)}$ reached the asymptotic value. Since $\psi_{\alpha, {\rm asym}}$ can be roughly estimated by equating $\rho_{A(\alpha)*} e^{\Lambda_\alpha (\psi_{\alpha, {\rm asym}} - \psi_*)}$, as a crude estimation, one can obtain 
\begin{align}
    \delta \Delta \psi_\alpha \sim - \delta \psi_{\alpha, {\rm asym}} \sim \delta \ln \rho_{A(\alpha)*}/\Lambda_\alpha \sim |\delta \hat{\pi}^i_{(\alpha)}|/\Lambda_\alpha,  \label{Eq:DeltaPsi}
\end{align}
finding $|\delta \Delta \psi_\alpha/\Delta \psi_\alpha| \ll |\delta \hat{\pi}_{(\alpha)}|$, when the condition $\Lambda_\alpha \Delta \psi_\alpha \gg 1$, which should be satisfied for the statistical anisotropy to be observable, holds.}. As one can imagine from the fact that the backreaction of the growing gauge field on the motion of the scalar field becomes non-negligible, two linear polarization modes of GWs evolve differently in the large scale limit. In this case, the angular dependent contribution in the power spectrum is given by Eqs.~(\ref{Exp:gtpldeltaI0multi}) and (\ref{Exp:gtcrdeltaI0multi}).

As has been shown, during slow-roll inflation, one can widely find a solution of $\rho_{A(\alpha)}$ which approaches Eq.~(\ref{Exp:rhoA_asympmulti}) after the exponential growth. The non-trivial conditions is only the positivity of $\Lambda_\alpha$, which requires a large coupling between the gauge fields and the scalar fields. Using slow-roll parameters,
\begin{align}
   \varepsilon_I \equiv   \frac{1}{2{\cal P}_{X^I}} \left( \frac{\Mp V_I}{V} \right)^2  \,, \label{Def:verep}
\end{align}
we find that the condition $\Lambda_\alpha > 0$ typically requires that the coupling needs to be enhanced by $1/\sqrt{\varepsilon_I}$ or a larger. As an example, let us consider a dilatonic coupling with $\ln f_{(\alpha)} = \sum_{I=1}^D \lambda^{(\alpha)}_I  \phi^I/\Mp$, which ubiquitously appears in a 4D low energy effective field theory of string theory. In this model, the positivity of $\Lambda_{\alpha}$ requires 
\begin{align}
    \sum_{I=1}^D\lambda^{(\alpha)}_I \sqrt{\frac{\varepsilon_I}{{\cal P}_{X^I}}} > \sqrt{2} \,.
\end{align}

\subsubsection{With direct interaction between gauge field and inflaton}  \label{SSSec:single}

In the rest of this subsection, let us consider a couple of specific examples. In Refs.~\cite{Watanabe:2009ct, Watanabe:2010fh}, Watanabe, Kanno, and Soda considered a model with $D=D'=1$ where the inflaton $\phi$ directly interacts with the gauge field, assuming that $f(\phi)$ and $V(\phi)$ are related as 
\begin{align}
    \frac{d}{d \phi} \ln f = \frac{2c}{\Mp^2} \frac{V}{V_\phi}\,.  \label{Exp:fandV}
\end{align}
In this model, $\Lambda$ is given by $\Lambda = 4 (c/{\cal P}_X - 1)$, and  $\Lambda >0$ requires $ c/{\cal P}_X > 1$. When this condition holds, for $\psi- \psi_* \gg 1/(c/{\cal P}_{X} -1)$, the energy fraction of the gauge field approaches 
\begin{align}
    \frac{\rho_{A,{\rm asym}}}{V}  = \frac{c- {\cal P}_{X}}{4 c^2} \left( \frac{V_{\phi}}{\Mp V} \right)^2\,,  \label{Exp:rhoA_asymp}
\end{align}
after the exponential growth of $\rho_A$. (This exponentially growing period was focused in Ref.~\cite{Naruko:2014bxa}.)

In Refs.~\cite{Watanabe:2009ct, Watanabe:2010fh, Soda:2012zm}, Watanabe et al. addressed the case with $0< c-{\cal P}_X \ll 1$, assuming that the background trajectory already had reached this asymptotic value at $t=t_*$. In this case, as we will show shortly, the condition (\ref{Cond:deltaI}) does not hold\footnote{In Ref.~\cite{Fujita:2017lfu}, it was shown that approaching this asymptotic solution is more unlikely than previously thought due to the stochastic diffusion, which can dominate over the slow classical evolution. (See also Ref.~\cite{Talebian:2019opf}.) As a simple check of the \dNex, here we just consider the case where the background is given by this asymptotic solution. }. In the $\delta N$ formalism, the spatial inhomogeneity is encoded in the different initial conditions of the separate universes at $t=t_*$. For $c-{\cal P}_X \ll 1/60$, there is not enough time for the energy density of the gauge field in each universe to approach the asymptotic value given in Eq.~(\ref{Exp:rhoA_asymp}). Since the fluctuation of $\rho_A$ does not become negligible because of the second term in the denominator of Eq.~(\ref{Sol:rhoA_asympgen}), the condition (\ref{Cond:deltaI}) does not hold. In fact, using Eq.~(\ref{Sol:rhoA_asympgen}) with $\Lambda = 4 (c/{\cal P}_X - 1)$, as an order estimation, we obtain
\begin{align}
    \left| \frac{\delta \rho_A}{\rho_A} \right| \sim   \left| \frac{V}{\rho_A} \frac{\partial (\rho_A/V)}{\partial \pi} \delta \pi \right| \sim   \frac{\frac{\rho_{A,{\rm asym}}}{\rho_{A*}}  e^{-4 (c/{\cal P}_X - 1)(\psi - \psi_*)}}{ 1 + \left(  \frac{\rho_{A, {\rm asym}}}{\rho_{A*}}  - 1 \right) e^{- 4 (c/{\cal P}_X - 1)(\psi - \psi_*)}} \left| \delta \hat{\pi} \right|\,,
\end{align}
which implies that the condition (\ref{Cond:deltaI2}) does not hold for $(c/{\cal P}_X - 1)(\psi - \psi_*) \ll 1$. Therefore, the fluctuation of $\rho_A$ is not negligible, although the background value $\Bar{\rho}_A$ is almost constant. Since the background energy density of the gauge field is given by Eq.~(\ref{Exp:rhoA_asymp}), we can find that the two conditions (\ref{Cond:AbsenseLP2}) hold as long as $c_s$ and ${\cal P}_X$ are around ${\cal O}(1)$. Therefore, in this case, the amplitudes of the two linear polarization modes of GWs approximately coincide~\cite{Watanabe:2010fh, Soda:2012zm}. 

Next, let us compute the curvature perturbation for $0< c-{\cal P}_X \ll 1$. Inserting Eq.~(\ref{Sol:rhoA_asympgen}) into Eq.~(\ref{Eq:KGani2multi}) and integrating the obtained equation, we find
\begin{align}
 - \int^{\phi_f}_{\phi_*} \frac{d\phi}{\Mp^2} \frac{V}{V_{\phi}}    = \psi_f - \psi_* - \frac{{\cal P}_X}{4c} \ln \frac{e^{4 (c/{\cal P}_X-1) (\psi_f - \psi_*)} + \frac{\rho_{A, {\rm asym}}}{\rho_{A*}}  -1}{\frac{\rho_{A, {\rm asym}}}{\rho_{A*}}} \,.
\end{align}
Perturbing around the background value, we obtain the fluctuation of the $e$-foldings sourced by the gauge field as 
\begin{align}
    \left( \frac{\partial \bar{\psi}_f}{\partial \dot{\bar{A}}_*^i} \right)_{\!\!\bar{\phi}_*}  \delta \dot{A}_{i*} = \frac{\bar{{\cal P}}_X}{2c} \,\frac{e^{4 (c/\bar{{\cal P}}_X-1) (\Bar{\psi}_f - \Bar{\psi}_*)}-1}{  e^{4 (c/\bar{{\cal P}}_X-1)(\Bar{\psi}_f - \Bar{\psi}_*)}\frac{\bar{{\cal P}}_X}{c}  +  \frac{\rho_{A, {\rm asym}}}{\rho_{A*}}-1 } \hat{\Bar{\pi}}^j  \frac{\delta \dot{A}_{j*}}{\dot{\bar{A}}_{*}}  \label{Exp:dpsidA} \,.
\end{align}
Taking the limit $c/\bar{{\cal P}}_X-1 \ll 1$ in this expression and using Eq.~(\ref{Eq:ampA}), we obtain the contribution of the gauge field to the power spectrum as\footnote{The slow-roll parameter $\varepsilon$ introduced in Eq.~(\ref{Def:verep}) is different from the usual one, $- \ddot{\psi}/\dot{\psi}$ due to the contributions of the shear and the energy density of the gauge field. Here, we have ignored these higher order contributions.}
\begin{align}
    \langle \zeta(t_f,\, \bm{k}) \zeta(t_f,\, \bm{p}) \rangle = \delta(\bm{k} + \bm{p}) \frac{1}{4 c_{s*} \varepsilon_*} \frac{1}{k^3} \left( \frac{H_*}{\Mp} \right)^2 \left[ 1 + g_s \sin^2 \Theta(\hat{\bm{k}}) \right]  \label{Eq:powerzeta}
\end{align}
with 
\begin{align}
    g_s \sim   \frac{\rho_{A*}}{\rho_*} \frac{c_{s*}}{\varepsilon_*} (\bar{\psi}_f - \bar{\psi}_*)^2\,,  \label{Exp:gstarzeta_WKS}
\end{align}
reproducing the result obtained in Ref.~\cite{Watanabe:2010fh}. The longitudinal part of $\delta \gamma_{ij}$ also contributes to $\zeta$ but it turns out to be more suppressed by the slow-roll parameter. In the limit $\Lambda \ll 1$ (or $c/{\cal P}_X \sim 1$)
, the equation of motion of the inflaton is not modified by the gauge field (see Eq.~(\ref{Eq:KGani3})). However, the fluctuation of the $e$-foldings acquires the large angular dependent contribution sourced by the gauge field. As has been emphasized, this is because each separate universe has not yet reached the asymptotic solution, at which the averaged background has already arrived. An earlier attempt to reproduce the result in Ref.~\cite{Watanabe:2010fh} from the $\delta N$ formalism can be found, e.g., in Ref.~\cite{Abolhasani:2013zya}, where the origin of the angular dependent contribution was erroneously attributed to the fluctuation of $\rho_{A, {\rm asym}}$.

Meanwhile when $c/{\cal P}_X (>1)$ is not very close to 1, the exponentially decaying term in Eq.~(\ref{Sol:rhoA_asympgen}) can die off within several $e$-foldings. Once this term becomes negligible, the condition (\ref{Cond:deltaI}) starts to hold. Then, two linear polarization modes of GWs evolve differently in the large scale limit. In this case, the mixing between the scalar fields and the gauge fields, which is proportional to $\sqrt{\bar{\rho}_A/(\bar{V} \varepsilon )}$, is not suppressed, resulting in a large angular dependent contribution in the power spectrum of $\zeta$, which contradicts with the observations. One may think that this can be evaded if the fluctuations at the CMB scale had crossed the horizon scale before $\rho_A$ reached the asymptotic value, {\it i.e.}, when $\rho_A$ was much smaller than the one given in Eq.~(\ref{Exp:rhoA_asymp}). However, in this case, using Eq.~(\ref{Eq:DeltaPsi}), one can find that $g_s$ which appears from the fluctuation of $\psi_{\rm asym} - \psi_*$ amounts to $\varepsilon_* (\rho_*/\rho_{A*})$, which is now larger than 1. Here $\psi_{\alpha, {\rm asym}}$ denotes the $e$-folding number at which $\rho_{A(\alpha)}$ reached the asymptotic value.

\subsubsection{Without direct interaction between gauge field and inflaton}  \label{SSSec:multi}
Next, let us address the case where the inflaton $\phi$ does not directly interact with the gauge field, assuming that the coupling $f$ only depends on the other scalar fields $\chi^I$ with $I=1,\, \cdots,\, D-1$, whose energy density $\rho_\chi$ only occupies a small fraction, satisfying $\rho_{\chi} \ll \rho$. Here, let us consider the case where $\Lambda_\alpha$ are all positive and $\geq {\cal O}(1)$. Then, after the exponential growth, $\rho_{A(\alpha)}$ reaches the maximum value, given by Eq.~(\ref{Exp:rhoA_asympmulti}), and sustains during the period $\Delta \psi_\alpha$. This period ends, e.g., when the slow-roll condition is violated, and $\rho_{A(\alpha)}$ starts to decrease. In this case, $\If$ is given by $\If \simeq 2 (\bar{\rho}_{A(\alpha),{\rm asym}}/\bar{\rho}_*) \Delta \psi_\alpha$. We also assume that the background contribution of the gauge field was still negligible at $t=t_*$, ensuring that the power spectrum of the gauge field can be approximately given by Eqs.~(\ref{Exp:PA}) and (\ref{Eq:ampA}).

Inserting the expression of $\bar{\hat{{\cal I}}}_{(\alpha)f}$, given in Eq.~(\ref{Exp:Ifatt}), into Eq.~(\ref{Exp:gt}), we obtain the amplitude of the angular dependent contribution in the power spectrum of GWs as
\begin{align}
    g_{t(\alpha)} = \frac{4}{3}  \left( 1 + \left(\frac{\Lambda_{\alpha}}{2} + 2
    \right)^{\!\!2} \right)  \left(\frac{\Bar{\rho}_{A(\alpha), {\rm asym}}}{\Bar{\rho}_*} \Delta \psi_\alpha \right)^2 \frac{\Bar{\rho}_{*}}{\bar{\rho}_{A(\alpha)*}}
   \,,  \label{Exp:gt_rhoAconst}
\end{align}
where we have computed $\dot{\bar{f}}_{(\alpha)*}/(H_{*}\Bar{f}_{(\alpha)*})$ by using $f_{(\alpha)} \propto e^{- (\Lambda_\alpha + 4) \psi/2}$. 
Since the gauge field is sourced by the subdominant fields $\chi^I$, $\rho_{A}$ should be much smaller than $\rho$, satisfying $\rho_{A} \leq \rho_\chi \ll \rho$ and resulting in the suppression by $\Bar{\rho}_{A(\alpha), {\rm asym}}/\Bar{\rho}_* \ll 1$. However, $g_{t(\alpha)}$ acquires the exponential enhancement by $\Bar{\rho}_{A (\alpha), {\rm asym}}/\bar{\rho}_{A(\alpha)*} \sim e^{\Lambda_\alpha(\psi_{{\rm asym}} - \psi_*)}$. It is also enhanced by (the square of) $\Delta \psi_\alpha$, which is bounded above by $\sim 60$ for the CMB scales, while the validity of our analysis requires $ \Delta \psi_\alpha \ll \Bar{\rho}_* /\Bar{\rho}_{A(\alpha), {\rm asym}}$.  The exponential enhancement originates because the conversion from the fluctuation of the gauge field to the gravitational waves takes place mainly when $\bar{\rho}_{A(\alpha)}$ reaches the maximum value $\bar{\rho}_{A(\alpha), {\rm asym}}$, while the (normalized) power spectrum of the gauge field at $t=t_*$, given by Eq.~(\ref{Eq:ampA}), is inversely proportional to $\Bar{\rho}_{A(\alpha)*}/\Bar{\rho}_*$. Since $\bar{\rho}_{A(\alpha)}$ is almost constant and $\delta \Delta \psi_\alpha/\Delta \psi_\alpha$ is suppressed by the inverse of $\Delta \psi_\alpha$ for $\Delta \psi_\alpha > {\cal O}(1)$, the condition (\ref{Cond:deltaI}) holds in this case. Then, the power spectrums of GWs in the linear polarization basis are given by Eq.~(\ref{Def:gt}) with Eqs.~(\ref{Exp:gtpldeltaI0multi}), (\ref{Exp:gtcrdeltaI0multi}), and (\ref{Exp:gt_rhoAconst}).

Next, let us compute the adiabatic curvature perturbation, which is defined as Eq.~(\ref{Def:zetalinear}). Since the fluctuation of the $e$-foldings depends on the detail of the model, here for simplicity, let us consider the case with $D=2$ and $D'=1$, assuming that the two scalar fields have the canonical kinetic term and the subdominant scalar field $\chi$ also remains almost constant during $\psi_{\rm asym} \leq \psi \leq \psi_{\rm asym} + \Delta \psi$. If $\rho_A$ remains constant until the end of inflation, $\psi_{\rm asym} + \Delta \psi$ should be identified with $\psi_f$. Since $\phi$ does not directly interact with the gauge field, the KG equation for $\phi$ is given by 
\begin{align}
    \frac{d \phi}{d \psi} \simeq - \Mp^2  \frac{V_\phi}{V + \rho_{\chi} + \rho_A}.
\end{align}
Integrating this equation, we obtain
\begin{align}
   \psi_f - \psi_* \simeq  (\psi_f - \psi_*)_{\phi} + \frac{\rho_{\chi,{\rm asym}}}{V} \Delta \psi + \frac{\rho_{A, {\rm asym}}}{V} \Delta \psi\,,  \label{Exp:efoldinginex}
\end{align}
where $\rho_{\chi,{\rm asym}}$ denotes $\rho_\chi$ during $\Delta \psi$, and the first term denotes the $e$-folding number which is determined only by the dynamics of $\phi$. The subdominant scalar field $\chi$ can also provide an angular-independent sub-dominant contributions to $\psi_{\rm asym} - \psi_*$ and $\psi_f - (\psi_{\rm asym}+ \Delta \psi)$, which can be addressed by using the conventional $\delta N$ formalism. Here and hereafter, we ignore them, focusing on the leading angular dependent contribution.

Perturbing the above expression, one can compute the fluctuation of the $e$-foldings. The perturbation of the last term in Eq.~(\ref{Exp:efoldinginex}) roughly corresponds to $\delta \hat{\cal I}_f$. Therefore, this contribution becomes subdominant compared to the term with $\delta \hat{\pi}^i$ in $\hat{k}_i \hat{k}_j \delta \gamma_{ij}$  when Eq.~(\ref{Cond:deltaI}) holds. Consequently, the dominant angular dependent contribution appears from the perturbation of the second term in Eq.~(\ref{Exp:efoldinginex}). The perturbation of $ \rho_{\chi, {\rm asym}}$ is given by 
\begin{align}
   \delta \rho_{\chi, {\rm asym}} \simeq \bar{V}_{\chi, {\rm asym}} \delta \chi_{\rm asym} \simeq \frac{\bar{V}_{\chi, {\rm asym}}}{(\ln \bar{f})_{, \chi, {\rm asym}}} \left( \Bar{\hat{\pi}}^i \frac{\delta \dot{A}_{i*}}{\dot{\bar{A}}_*}  + 2 (\ln \Bar{f})_{, \chi *} \delta \chi_* \right)\,,  \label{Exp:rhochi}
\end{align}
where on the first equality we have ignored the kinetic term under the slow-roll approximation and on the second equality we have eliminated $\delta \chi_{\rm asym}$, using $\delta \rho_{A, {\rm asym}} \simeq 0$ with $\delta \pi^i$ being expressed by the gauge fields with the time derivative operator at $t=t_*$. Using the formulae derived here and Eqs.~(\ref{Def:zetalinear}), (\ref{Eq:gammaL}), (\ref{Eq:gammaL2}), and (\ref{Exp:Ifatt}), we obtain
\begin{align}
    \zeta(t_f,\, \bm{k}) &= - \frac{\bar{V}_*}{\Mp^2 \bar{V}_{\phi*}} \delta \phi_*(\bm{k}) + 2\frac{\bar{V}_{\chi, {\rm asym}}}{\bar{V}} \Delta \psi \delta \chi_* \nonumber \\
    & \qquad - 2 \frac{\bar{\rho}_{A, {\rm asym}}}{\bar{\rho}_*} \Delta \psi \frac{\bar{\hat{\pi}}^i \delta \dot{A}_{i*}(\bm{k})}{\dot{\bar{A}}_*}  \left(  \cos^2 \Theta(\hat{\bm{k}}) - \frac{1}{2} \frac{\bar{V}_{\chi, {\rm asym}}}{ (\ln \bar{f})_{, \chi, {\rm asym}} \Bar{\rho}_{A, {\rm asym}}} \right), \label{Eq:zeta}
\end{align}
where we have approximated $(\ln \bar{f})_{, \chi, {\rm asym}} \simeq (\ln \bar{f})_{, \chi*}$ since the time variation is assumed to be small. The first term in the parentheses is the model independent contribution which comes from the longitudinal mode of $\delta \gamma_{ij}$ and the second one comes from the fluctuation of the $e$-folding number.

Using Eq.~(\ref{Eq:zeta}), we obtain the power spectrum of $\zeta$ as 
\begin{align}
    &\langle \zeta(t_f,\, \bm{k}) \zeta(t_f,\, \bm{p}) \rangle \cr
    &= \delta(\bm{k} + \bm{p}) \frac{1}{4\varepsilon_*} \frac{1}{k^3} \left( \frac{H_*}{\Mp} \right)^2 \Biggl[ 1 + 8 \varepsilon_* \left( \frac{\Bar{V}_{\chi, {\rm asym}} \Mp}{\bar{V}} \Delta \psi \right)^2 \cr
    & \qquad \qquad \qquad  \qquad \qquad \qquad\qquad+ \varepsilon_* g_t \left(  \cos^2 \Theta(\hat{\bm{k}}) - \frac{1}{2} \frac{\bar{V}_{\chi, {\rm asym}}}{ (\ln \bar{f})_{, \chi, {\rm asym}} \Bar{\rho}_{A, {\rm asym}}} \right)^2 \Biggr],  \label{Eq:powerzeta_deltaI0}
\end{align}
where we have used $g_t$, given in Eq.~(\ref{Exp:gt_rhoAconst}). As is shown here, the ratio between the angular dependent contribution and the angular independent one in the power spectrum of $\zeta$ is suppressed by $\varepsilon_*$, compared to the one in the power spectrum of GWs. This simply originates from the ratio between the overall amplitudes of the (isotropic) power spectrums of $\zeta$ and GWs. Repeating a similar computation, we obtain the cross correlations between $\zeta$ and GWs as 
\begin{align}
   & \langle \zeta(t_f,\, \bm{k}) \gamma^{(+)}(t_f,\, \bm{p}) \rangle = \delta(\bm{k} + \bm{p})  \frac{\sqrt{2}}{8} \frac{g_t}{k^3} \left( H_* \over \Mp \right)^2  \sin^2 2 \Theta(\hat{\bm{k}}) \cr
   & \qquad \qquad \qquad \qquad  \qquad \qquad  \times \left(  \cos^2 \Theta(\hat{\bm{k}}) - \frac{1}{2} \frac{\bar{V}_{\chi, {\rm asym}}}{ (\ln \bar{f})_{, \chi, {\rm asym}} \Bar{\rho}_{A, {\rm asym}}} \right)\,, \\
    & \langle \zeta(t_f,\, \bm{k}) \gamma^{(\times)}(t_f,\, \bm{p}) \rangle = 0\,. 
\end{align}
When there exists only one gauge field, {\it i.e.}, $D'=1$, $\delta \If$ is negligible, and the cross correlation between the scalar fields and the two linear polarization modes of the gauge field all vanish at $t=t_*$, the cross correlation between $\zeta$ and $\gamma^{(\times)}$ generically vanishes. 

Here, we have left the functional form of the coupling $f$ and the Lagrangian density of $\chi$ are left general, while a slow-roll evolution is assumed. In Ref.~\cite{Fujita:2018zbr}, one of the authors studied an example where the subdominant scalar field $\chi$ goes through the linear scalar potential and has the dilatonic coupling with the gauge field as $\ln f(\chi) = \lambda (\chi/\Mp)$. The formulae derived here reproduces the result in Ref.~\cite{Fujita:2018zbr}\footnote{In this model, the contribution of the scalar field $\chi$, e.g., the second term in Eq.~(\ref{Eq:zeta}), becomes smaller than the contribution of the gauge field. With the notation in Ref.~\cite{Fujita:2018zbr}, the second term in the parenthesis in the second line of Eq.~(\ref{Eq:powerzeta_deltaI0}) becomes $- n/(n-2)$.}.

In the present example, the subdominant scalar field $\chi$ contributes to the entropy perturbation. Let us also show that this contribution can be computed easily by using the \dNex. We have introduced $\zeta$ as the curvature perturbation in the uniform expansion slicing, (\ref{Cond:slicingfin}), which corresponds to the uniform total energy density slicing after the large scale shear becomes negligible. Using the curvature perturbation defined in the $\delta \rho_\chi =0$ slicing, $\zeta_\chi$, let us introduce the (relative) entropy perturbation $S$ as \cite{Malik:2002jb}
\begin{align}
   S \equiv  3 \left( \zeta -  \zeta_\chi \right) \simeq 3 \frac{d \bar{\psi}}{d \bar{\rho}_\chi} \delta \rho_\chi \,. \label{Def:S}
\end{align}
For simplicity, let us assume that both $\zeta$ and $\zeta_\chi$ (and consequently $S$) do not change in time in the large scale limit during $\psi_{\rm asym} + \Delta \psi  \leq \psi \leq \psi_f$ or the asymptotic solution persists until the end of inflation, {\it i.e.}, $\psi_f = \psi_{\rm asym} + \Delta \psi$. Then, using Eq.~(\ref{Exp:rhochi}), we obtain the entropy perturbation as
\begin{align}
   S(t_f,\, \bm{k}) = - \frac{3}{2} \left( \Bar{\hat{\pi}}^i \frac{\delta \dot{A}_{i*}(\bm{k})}{\dot{\bar{A}}_*}  + 2 (\ln \Bar{f})_{, \chi *} \delta \chi_*(\bm{k}) \right)\,, 
\end{align}
where we have used $d \ln \Bar{f}_{\rm asym}/d \bar{\psi} = -2$. Using this expression, we obtain the power spectrum of the entropy perturbation $S$ as
\begin{align}
    \langle S(t_f,\, \bm{k}) S(t_f,\, \bm{p}) \rangle &= \delta(\bm{k} + \bm{p}) \frac{3}{16 k^3} \left( \frac{H_*}{\Mp} \right)^2  \cr
    & \quad\qquad \times \left[ \left( 1 + \left( \frac{\Lambda}{2} + 2 \right)^{\!\!2} \right) \!\frac{\bar{\rho}_{*}}{\bar{\rho}_{A*}} \sin^2 \Theta(\hat{\bm{k}}) + 24 \left( \frac{\Mp \Bar{f}_{\chi*}}{\Bar{f}_*}\right)^{\!\!2}  \right]\!\!.
\end{align}
Namely, the ratio between the power spectrum of $S$ sourced by the gauge field, {\it i.e.}, the first term here, and the power spectrum of $\zeta$, which we express as $\beta_A(\bm{k})$, is given by 
\begin{align}
    \frac{\beta_A(\bm{k})}{0.1} \simeq \left( \frac{0.02}{(\Bar{\rho}_{A {\rm asym}}/\Bar{\rho}_*) \Delta \psi} \right)^{\!2} \,  \frac{r_{\rm iso} g_t}{0.001} \sin^2 \Theta(\hat{\bm{k}}) \,,
\end{align}
where we have used Eq.~(\ref{Exp:gt_rhoAconst}) and introduced the tensor to scalar ratio for the isotropic contribution as $r_{\rm iso} = 16 \varepsilon_*$. Since our analysis is based on the perturbative expansion with respect to $\bar{\hat{{\cal I}}}_f$, the validity requires $\bar{\hat{{\cal I}}}_f \simeq 2 (\Bar{\rho}_{A {\rm asym}}/\Bar{\rho}_*) \Delta \psi \ll 1$. When the scalar field $\chi$ remains also after the reheating or the curvature perturbation $\zeta_\chi$ is conserved through the reheating (the $\chi$ constant surface should be rephrased as the corresponding relic constant surface), the entropy perturbation should be bounded by the observations.

\subsection{Primordial non-Gaussianity}
In the \dNex, the local type non-Gaussianity can be computed easily. Using Eq.~(\ref{Sol:gamma2}), we obtain
\begin{align}
    \delta \gamma_{ij}(t_f,\, \bm{k}) \simeq \delta \gamma_{ij*}(\bm{k}) + \frac{\partial \Bar{\gamma}_{ij}(t_f)}{\partial \Bar{\pi}_{(\alpha)}^i} \delta \pi^i_{(\alpha)}(\bm{
k}) + \frac{1}{2} \frac{\partial^2 \Bar{\gamma}_{ij}(t_f)}{\partial \Bar{\pi}_{(\alpha)}^i \partial \Bar{\pi}_{(\beta)}^j} (\delta \pi^i_{(\alpha)} \star \delta \pi^j_{(\beta)}) (\bm{k}) + \cdots \,,
\end{align}
with 
\begin{align}
   & \frac{\partial \Bar{\gamma}_{ij}(t_f)}{\partial \Bar{\pi}_{(\alpha)}^l} = - 4 (\Bar{\gamma}_{\{i|l|*} \Bar{\gamma}_{j\}m*} - \frac{1}{3} \Bar{\gamma}_{ij*} \Bar{\gamma}_{lm*}) \Bar{\hat{\pi}}^m_{(\alpha)} {\Bar{\hat{\cal I}}}_{(\alpha)f}\,, 
 \label{Exp:Coeff1}\\ 
   & \frac{\partial \Bar{\gamma}_{ij}(t_f)}{\partial \Bar{\pi}_{(\alpha)}^l \partial \Bar{\pi}_{(\beta)}^m} = - 4 (\Bar{\gamma}_{il*} \Bar{\gamma}_{jm*} - \frac{1}{3} \Bar{\gamma}_{ij*} \Bar{\gamma}_{lm*})  {\Bar{\hat{\cal I}}}_{(\alpha)f} \, \delta_{\alpha \beta}\,.\label{Exp:Coeff2}
\end{align}
Here, we have neglected the subleading contributions and $\delta \If$, imposing the condition (\ref{Cond:deltaI}). Using Eq.~(\ref{Exp:3pt}), we obtain the angular dependent contribution in the squeezed bispectrum of GWs in the limit $k_1/k_2,\, k_1/k_3 \ll 1$ as 
\begin{align}
    &\langle \gamma^{(\lambda_{\rm gw})} (t_f,\, \bm{k}_1) \gamma^{(\lambda_{\rm gw})} (t_f,\, \bm{k}_2) \gamma^{(\lambda_{\rm gw})} (t_f,\, \bm{k}_3) \rangle \cr
    & \supset \prod_{n=1}^3 e_{i_n j_n}^{(\lambda_{\rm gw})}(\bm{k}_n)  
 \biggl[ \frac{\partial \bar{\gamma}_{i_1j_1}}{\partial \bar{\hat{\pi}}^{l_1}_{(\alpha)*}} \frac{\partial^2 \bar{\gamma}_{i_2j_2}}{\partial \bar{\hat{\pi}}^{l_2}_{(\alpha)*} \partial \bar{\hat{\pi}}^{m_2}_{(\alpha)*}} \frac{\partial \bar{\gamma}_{i_3j_3}}{\partial \bar{\hat{\pi}}^{l_3}_{(\alpha)*}} \langle \delta \hat{\pi}^{l_1}_{(\alpha)*}(\bm{k}_1) (\delta \hat{\pi}^{l_2}_{(\alpha)*} \star \delta \hat{\pi}^{m_2}_{(\alpha)*})(\bm{k}_2)   \delta \hat{\pi}^{l_3}_{(\alpha)*}(\bm{k}_3) \rangle \cr
 & \qquad \qquad \qquad \qquad \qquad \qquad \qquad + (2 \leftrightarrow 3) \biggr]  \,. 
\end{align}
Inserting Eqs.~(\ref{Exp:Coeff1}) and (\ref{Exp:Coeff2}) to this formula, we obtain the contribution of the gauge fields in the squeezed bispectrum of GWs as
\begin{align}
    f_{\rm NL}^{\gamma, A} \sim \sum_{\alpha=1}^{D'} \left( \frac{\bar{\rho}_*}{\bar{\rho}_{A(\alpha) *}} \right)^2 \, \bar{\hat{{\cal I}}}_{(\alpha)f}^3 \sim \sum_{\alpha=1}^{D'} \left( \frac{\bar{\rho}_{A(\alpha) {\rm asym}}}{\bar{\rho}_{A(\alpha) *}} \right)^2 \frac{\bar{\rho}_{A(\alpha) {\rm asym}}}{\bar{\rho}_*} \Delta \psi_{(\alpha)}^3 \,,\label{fNLGW}
\end{align}
where we have dropped the angular dependence. In the definition of $f_{\rm NL}^{\gamma, A}$, the squeezed bispectrum of GWs is divided by the vacuum power spectrum. Similarly to the power spectrum, the bispectrum is also enhanced by $\bar{\rho}_{A(\alpha) {\rm asym}}/\bar{\rho}_{A(\alpha) *} \gg 1$. Repeating a similar computation, we find that the contribution of the gauge field to the local type non-Gaussianity of $\zeta$ roughly amounts to 
\begin{align}
f_{\rm NL}^{\zeta, A} \sim  \varepsilon^2 f_{\rm NL}^{\gamma, A} \,, 
\end{align}
which stems from the bispectrum of $\hat{k}^i \hat{k}^j \delta \gamma_{ij} (t_f,\, \bm{k})$. Apart from this contribution, there is also the contribution from the fluctuation of the $e$-folding number, which depends on the model.

\subsection{Prospects on future GW experiments} \label{SSec:LiteBIRD}
\begin{figure}
    \centering
     \includegraphics[
      width=7.5cm]{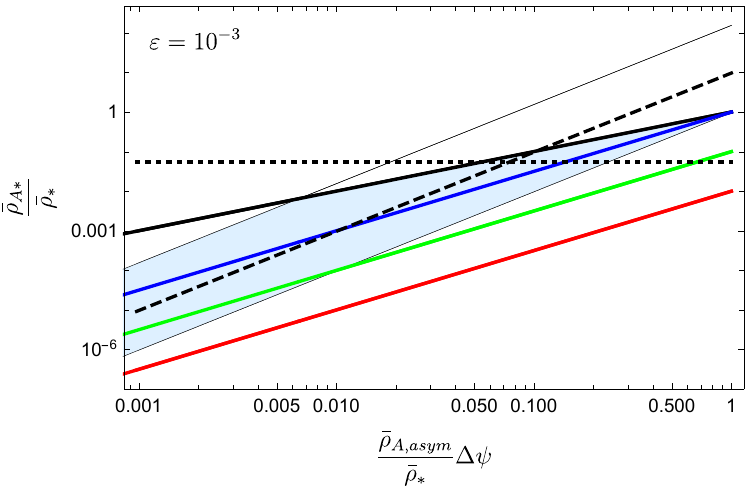}
     \includegraphics[
     width=7.5cm]{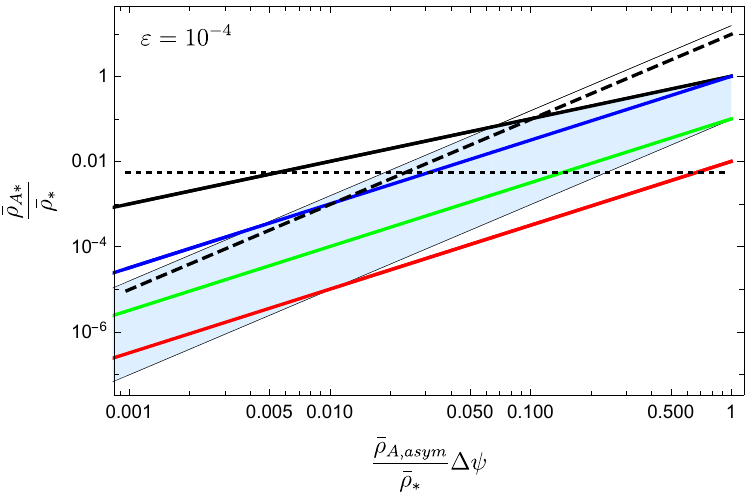}
    \hfill
    \caption{\label{fig:fNL}The horizontal axis denotes the product of $\bar{\rho}_{A, {\rm asym}}/\bar{\rho}_*$ and $\Delta \psi$, {\it i.e.}, the energy fraction of the gauge field during the $\rho_A=$const epoch and its duration $\Delta \psi$, and the vertical axis denotes the background energy fraction at the horizon crossing, $\bar{\rho}_{A*}/\bar{\rho}_*$. In the left panel, the slow-roll parameter is set to $\varepsilon= 10^{-3}$ and in the right panel it is set to $\varepsilon=10^{-4}$. Here, $D'=1$ and $\Lambda$ is set to 1. In the blue shaded region, $r_{\rm iso} g_t$ becomes larger than $10^{-3}$, which is expected to be accesible by LiteBIRD, while being consistent with the present constraints by CMB and LSS. Since $\rho_{A*} < \rho_{A, {\rm asym}}$ and $\Delta \psi \geq {\cal O}(1)$, we have excluded the parameter region with $(\bar{\rho}_{A, {\rm asym}}/\bar{\rho}_*) \Delta \psi < \bar{\rho}_{A*}/\bar{\rho}_*$ from the shaded region. Below the dashed line, $g_t$ becomes $g_t \geq 1$, {\it i.e.}, the power spectrum of GWs becomes highly anisotropic. The dotted line corresponds to the case where the amplitude of $\beta_A$ becomes 0.1. When the isocurvature perturbation $S$ remains in the Universe also after the reheating, the region below the dotted line should be explored by the constraints on the isocurvature perturbation. The blue, green, red lines correspond to $f_{\rm NL}^{\gamma, A} = 1,\, 10^2,\, 10^4$, respectively, where Eq.~(\ref{fNLGW}) was used for an order estimation of the tensor non-Gaussianity.} 
\end{figure}
So far, imposing the condition (\ref{Cond:deltaI}), we have computed the contributions of the gauge fields in the power spectrum of $\zeta$, GWs, and the isocurvature perturbation and conducted the order estimation of the bispectrum of $\zeta$ and GWs, which are determined by the two parameters 
$$
\frac{\Bar{\rho}_{A(\alpha),{\rm asym}}}{\Bar{\rho}_*} \Delta \psi_\alpha\,, \qquad \frac{\bar{\rho}_{A(\alpha)*}}{\Bar{\rho}_*} 
$$
for each gauge field and $\varepsilon$, which should be determined by the property of the inflaton. (To be precise, $\Lambda_\alpha = {\cal O}(1)$ can introduce an ${\cal O}(1)$ modulation.) For simplicity, here let us consider the case with one gauge field, {\it i.e.}, $D'=1$, for which these anisotropic contributions are determined by the three parameters. For example, for
$$
\frac{\Bar{\rho}_{A,{\rm asym}}}{\Bar{\rho}_*} \Delta \psi = 2 \times 10^{-2}, \qquad \frac{\bar{\rho}_{A*}}{\Bar{\rho}_*} = 10^{-4} , \qquad  \varepsilon = 5 \times 10^{-5}, 
$$
with $\Lambda=1$, we obtain the statistical anisotropy in the power spectrum as
$$
g_t \sim 3.9, \qquad  g_\zeta \sim 1.9 \times 10^{-4}, \qquad \beta_A \sim 0.31\,,
$$
where $g_\zeta \sim \varepsilon g_t$ denotes the coefficient of the cosine term in the power spectrum of $\zeta$, given in Eq.~(\ref{Eq:powerzeta_deltaI0}). Since the statistical anisotropy in the power spectrum of $\zeta$ is suppressed by the slow-roll parameter compared to that of GWs, the statistical anisotropy in the power spectrum of GWs can be as large as ${\cal O}(1)$, keeping the scalar spectrum consistent with the present observations~\cite{Fujita:2018zbr, Tanaka:2023gul}. In fact, the CMB observation provides the upper bound on the statistical anisotropy as $g_* \leq {\cal O}(10^{-2})$~\cite{Kim:2013gka, Akrami:2018odb}, while here $g_\zeta$ is well below this upper bound.  If the subdominant scalar field $\chi$ remains, this model can be examined also through the imprint of the isocurvature perturbation. In this model, the amplitude of GWs is given by 
$$
r_{\rm iso} \sim 8 \times 10^{-4}\, \qquad r_{\rm iso} g_t \sim 3.1 \times 10^{-3}\,,
$$
being within the reach of LiteBIRD. The contribution of the gauge field to the squeezed bispectrums of $\zeta$ and GWs are given by 
$$
f_{\rm NL}^{\gamma, A} \sim 8\,, \qquad f_{\rm NL}^{\zeta, A} \sim  2 \times 10^{-8}\,,
$$
predicting a fairly large tensor non-Gaussianity. Since $g_\zeta \sim r_{\rm iso} g_t/16$ is constrained by the bound on $g_s$ (the ansatz of the angular dependence is given in Eq.~(\ref{Eq:powerzeta})), the sweat spot at which LiteBIRD can play an important role for an exploration of the statistical anisotropy is roughly
\begin{align}
    10^{-3} \leq r_{\rm iso} g_t \leq 16 g_*  \sim 0.1\,. 
\end{align}

In Fig.~\ref{fig:fNL}, the blue shaded region denotes the parameter space in which
$r_{\rm iso} g_t$ for the GWs generated by the gauge field becomes larger than $10^{-3}$, being detectable by LiteBIRD, while satisfying the present constraints. As is shown, a smaller $\varepsilon$ leaves a broader parameter space, since the anisotropic contribution in the scalar power spectrum is more suppressed, evading the present constraints. Since $g_t$ becomes larger than 1 below the dashed line, one can see that GWs becomes highly anisotropic mostly in the shaded region. In the viable parameter range, $f_{\rm NL}^{\gamma, A}$ can be much larger than 1, especially for a smaller $\varepsilon$. Here, we have used Eq.~(\ref{fNLGW}) to roughly estimate the contribution of the gauge field to the squeezed bispectrum of GWs. However, we need a more careful computation for a more precise estimation.

\section{Conclusion}
In this paper, we have shown that the superhorizon evolution of the primordial curvature perturbation and the primordial GWs can be easily computed by using the \dNex, applying this formalism to an inflation model with U(1) gauge field. The \dNex~is useful also to develop an intuitive understanding of the non-linear evolution of long wavelength perturbations. The list of the new findings in this paper (and partly in our previous papers~\cite{Tanaka:2021dww, Tanaka:2023gul}) is
\begin{itemize}
    \item When the backreaction of the gauge fields on the evolution of the scalar fields become non-negligible, the two linear polarization modes of GWs, $\gamma^+$ and $\gamma^{\times}$, evolve differently in the superhorizon limit. 

    \item When there exist more than one gauge field, {\it i.e.} $D' \geq 2$,  there is no choice of the polarization bases for which the cross-correlations $\langle \gamma^+ \gamma^{\times} \rangle$ and $\langle \zeta \gamma^{\times} \rangle$ vanish. 

    \item The spectrums of $\zeta$, $\gamma^+$, and $\gamma^{\times}$ depend also on the azimuthal direction $(\Psi_\alpha)$ for $D' \geq 2$. 

    \item The number of the Weinberg's adiabatic mode in $\delta \gamma_{ij}$ is 4 for $D'=1$ and 5 for $D' \geq 2$ (See Appendix ~\ref{SSec:WAM}.). 

    \item The tensor (local-type) non-Gaussianities sourced by the gauge field(s) $f_{\rm NL}^{\gamma, A}$ can be much larger than ${\cal O}(1)$, while keeping the adiabatic perturbation consistent with observations.  
\end{itemize}
As was also pointed out in Ref.~\cite{Fujita:2018zbr}, when only subdominant scalar fields directly interact with the gauge fields, the power spectrum of GWs can acquire O(1) statistical anisotropy, while keeping the adiabatic perturbation consistent with the present observations. As was commented around Eq.~(\ref{Eq:crosszetaGWcp}), even with the parity symmetry unbroken, $\langle \zeta(t_f,\, \bm{k}) \gamma^{(+_c)} (t_f,\,  \bm{p})  \rangle$ does not necessarily coincide with $\langle \zeta(t_f,\, \bm{k}) \gamma^{(-_c)} (t_f,\,  \bm{p})  \rangle$.

Since the \dNex~generically applies to a model with the locality and the spatial diffeomorphism invariance, various applications are possible. An interesting application may be considering a non-abelian gauge field or a $n$-form field with $n \geq 2$.

\appendix
\section{Superhorizon evolution of GWs}  \label{Sec:WAM}
In this Appendix, we study the superhorizon evoluion of GWs, focusing on the Weinberg's adiabatic mode (WAM). 

\subsection{Weinberg's adiabatic mode} \label{SSec:WAM}
In Appendix \ref{SSec:WAM}, we will show that the WAM in the metric perturbation $\delta \gamma_{ij}$ still exists for an anisotropic background at the linear perturbation but it varies in time. Because of that, GWs evolve in time, even if the WAM is the dominant solution. A more detailed discussion about the WAM can be found in Ref.~\cite{Tanaka:2021dww}.

\subsubsection{Brief overview}
In Ref.~\cite{Tanaka:2021dww}, it was shown that when a system under consideration satisfies \locality and \sDif conditions and also a condition \approxGe$\!$,
\begin{itemize}
    \item \approxGe: When the equations to be solved to determine the separate universe evolution of $\{ \varphi^{a} \}$, described as ${\cal E}$, are satisfied, the gauge constraints that trivially vanish in the limit $\epsilon \to 0$, described as ${\cal G}_\epsilon$, are approximately satisfied at the leading order of $\epsilon$.  
\end{itemize}
the system has the WAM as an approximate solution beyond the horizon scale. By "approximately," we mean that ${\cal G}_\epsilon$ are satisfied, when we ignore the contributions that decay in time in an expanding universe. As summarized in Eq.~(\ref{Summary:symbols}), the system under consideration, whose matter Lagrangian density is given by Eq.~(\ref{Exp:Lmattergeneral}), has two ${\cal G}_\epsilon$s, which are the momentum constraints ${\cal H}_i$ and the U(1) gauge constraint, ${\cal H}_{\rm U(1)}$, (\ref{Eq:U1gc}). The existence of the Noether charge density ${Q^i}_j$ ensures that when ${\cal E}$ all hold, ${\cal H}_i$ is automatically satisfied, leaving aside decaying contributions which become negligible a few $e$-foldings after the horizon crossing. Meanwhile, satisfying ${\cal E}$ does not always ensure the approximate validity of ${\cal H}_{\rm U(1)}$. In this case, the WAM does not exist even as an approximate solution in the large scale limit. However, as we will see below, $\delta \gamma_{ij}$ has the WAM as an approximate solution at large scales exceptionally at the linear order of perturbation.

By contrast, in a system which only contains scalar fields or which also contains gauge fields but with non-vanishing background charged scalar fields, the WAM becomes approximate solutions at all orders of perturbation, satisfying the \approxGe condition~\cite{Sugiyama:2012tj, Garriga:2015tea}. Even in these cases, because of the contributions that decay inversely proportional to the physical volume, $1/\sqrt{g}$, the WAM is not the exact solution particularly for non-linear perturbations~\cite{Tanaka:2021dww}.  

For the isotropic background with $\bar{\gamma}_{ij}$ being time independent, the WAM corresponds to the time-independent (approximate) solution for both $\delta \psi$ and $\delta \gamma_{ij}$. The conservation of GWs in a scalar field system was first pointed out in the historic paper by Starobinsky~\cite{Starobinsky:1986fx}. On the other hand, for an anisotropic time dependent background $\bar{\gamma}_{ij}$, $\delta \gamma_{ij}$ corresponding to the WAM is not a time-independent (approximate) solution unlike $\delta \psi$, which remains to be time independent~\cite{Tanaka:2021dww}.

\subsubsection{Weinberg's adiabatic mode in anisotropic background}  \label{SSSec:formal}
Under the shear transformation, 
\begin{align}
    x^i \to x^i_C \equiv \bigl[ e^{- \frac{C}{2}} \bigr]{}^i_{\, j}\,  x^j\,,  \label{Exp:sheartrans}
\end{align}
with $C^i_{\,\,j}$ being a constant traceless tensor, the anisotropic spatial metric transforms as
\begin{align}
    \gamma_{ij\,C}(t,\, \bm{x}_C) = \bigl[e^{ \frac{C}{2}} \bigr]{}^k_{~ i}\, \bigl[e^{ \frac{C}{2}}\bigr]{}^l_{~ j}\,
    \gamma_{kl}(t,\, \bm{x})\,. \label{Exp:sheartrans}
\end{align}
Using the triad basis $e_i^{(\alpha)}$ with $\alpha=1,\, 2 ,\, 3$ with which the spatial metric is expressed as 
\begin{align}
    \gamma_{ij} = e_i^{(\alpha)} e_{j(\alpha)}\,, 
\end{align}
let us also introduce a rank-2 tensor 
\begin{align}
    {H^j}_i  \equiv 2  \ln \left( e_i^{(\alpha)} \bar{e}^{\,j (\alpha)} \right)\,, 
\end{align}
where $\bar{e}^{\,i (\alpha)}$ denotes the background triad basis. Since $\bar{e}_i^{(\alpha)} \bar{e}^{\,j (\alpha)} = {\delta_i}^j$, the background value of ${H^j}_i$ vanishes. Since the triad basis transforms under the shear transformation as 
\begin{align}
    e_{i\, C}^{(\alpha)}(t,\, \bm{x}_C) = \bigl[e^{ \frac{C}{2}} \bigr]^{\!j}_{~ i} e_j^{(\alpha)}(t,\, \bm{x})\,, 
\end{align}
we find that ${H^j}_i$ transforms as
\begin{align}
      {H^j}_{i\, C} (t,\, \bm{x}_C) &= 2 \ln \left( \left[e^{ \frac{C}{2}} \right]^{\!l}_{~ i} \left[ e^{H(t, \sbm{x})/2}\right]^{\!j}_{~ l}  \right) \cr
     &=  {H^j}_i (t,\, \bm{x}) + {C^j}_i + \frac{1}{4} {[H(t,\, \bm{x}), C]^j}_i  \cr
     & \qquad  \quad + \frac{1}{48} {\left[ (H(t,\, \bm{x}) - C),[H(t,\, \bm{x}), C]  \right]^j}_i  + \cdots  \, 
\end{align}
with a constant additive shift $+{C^j}_i$, which will turn out to correspond directly to the Weinberg's adiabatic mode~\cite{Weinberg:2003sw}. Here, we have used the Baker-Campbell-Hausdorff formula at the second equality.

We impose the transverse condition of ${H^i}_j$
\begin{align}
   \delta^{ik} \partial_k {H^j}_i (t_*,\, \bm{x}) = 0\,, \label{Cond:transH}
\end{align}
at $t=t_*$ to eliminate the residual gauge degrees of freedom which remain after imposing Eq.~(\ref{Cond:Ni}). Satisfying Eq.~(\ref{Cond:transH}) both before and after the transformation requires $\delta^{ik} \partial_k {C^j}_i(\bm{x})=0$ at the linear perturbation.

Following the usual argument about the Weinberg's adiabatic mode~\cite{Weinberg:2003sw} (see also Refs.~\cite{Tanaka:2017nff, Tanaka:2021dww}), we promote the shear transformation by replacing ${C^i}_j$ with an inhomogeneous parameter $C^i\!_j(\bm{x})$. When we replace ${H^j}_i(t,\, \bm{x})$ with ${H^j}_{i\, C} (t,\, \bm{x}_C)$ and simultaneously $A_{i(\alpha)}(t,\, \bm{x})$ with
\begin{align}
    A_{i(\alpha)\,C}(t,\, \bm{x}_C) = \bigl[e^{ \frac{C}{2}} \bigr]{}^j_{~ i}\, 
    A_{j(\alpha)}(t,\, \bm{x}) = A_{i(\alpha)}(t,\, \bm{x}) + \frac{1}{2} {C^j}_i A_{j(\alpha)}(t,\, \bm{x}) + \cdots  \,,
\end{align}
all the equations of ${\cal E}$ remain invariant at the leading order of the gradient expansion. In other words, when 
\begin{align}
   {H^j}_i(t,\, \bm{x}) = {f^j}_i(t,\, \bm{x})\,, \qquad \delta A_i (t,\, \bm{x}) = f_i(t,\, \bm{x})  \label{Exp:solbefore}
\end{align}
satisfies ${\cal E}$, 
\begin{align}
   & {H^j}_i(t,\, \bm{x}) =  {f^j}_i(t,\, \bm{x}) +  {C^j}_i(\bm{x})  +\cdots\,, \label{Exp:solaftergamma} \\
   &  \delta A_{i(\alpha)} (t,\, \bm{x}) = f_i(t,\, \bm{x}) + \frac{1}{2} {C^j}_i(\bm{x}) \bar{A}_{j(\alpha)}(t,\, \bm{x}) +  \cdots \,,  \label{Exp:solafterA}
\end{align}
should also satisfy ${\cal E}$ at the leading order of the gradient expansion. This indicates that the difference between these two solutions in Eq.~(\ref{Exp:solbefore}) and Eqs.~(\ref{Exp:solaftergamma})/(\ref{Exp:solafterA}), {\it i.e.}, 
\begin{align}
   & H^{j\,{\rm WAM}}_{\,\,i} (t,\, \bm{x}) =   {C^j}_i(\bm{x})  + \cdots \,, \label{Exp:HWAM}\\
   &  \delta A^{\rm WAM}_{i(\alpha)} (t,\, \bm{x}) =  \frac{1}{2} {C^j}_i(\bm{x}) \bar{A}_{j(\alpha)} (t) +  \cdots \,, \label{Exp:AWAM}
\end{align}
solve ${\cal E}$, where the abbreviation denotes the higher order terms in perturbation. Therefore, the validity of the \approxGe condition guarantees that this difference approximately solves the entire equations in the system, ${\cal E}_{\rm all}$, corresponding to the WAM.

In our present setup, the \approxGe  condition does not necessarily hold. In particular, when the gauge fields do not decay at superhorizon scales, the U(1) gauge constraint ${\cal H}_{\rm U(1)}$ does not hold even approximately. Therefore, the ${C^j}_i(\bm{x})$, $\bm{x}$-dependent perturbation discussed above does not satisfy ${\cal H}_{\rm U(1)}$, failing to be an approximate solution at all orders of perturbation. In fact, when we perform the shear transformation (\ref{Exp:sheartrans}) with an inhomogeneous parameter $C^i\!_j(\bm{x})$, there appears a term with $[e^{C/2}]_{~i}^j (\partial_j [e^{-C/2}]^i_{~k})\, \pi_{(\alpha)}^k$ in the gauge constraint ${\cal H}_{\rm U(1)}$, which prevents Eqs.~(\ref{Exp:HWAM}) and (\ref{Exp:AWAM}) from being an approximate solution. Exceptionally at the linear perturbation, ${\cal H}_{\rm U(1)}$ remains unchanged at the leading order of the gradient expansion, even when $\pi_{(\alpha)}^k$ has a non-zero background value, because $\delta^{ki} \partial_k {C^j}_i(\bm{x})=0$ owing to the metric transverse conditions. Therefore, at the linear perturbation, Eqs.~(\ref{Exp:HWAM}) and (\ref{Exp:AWAM}) serve the approximate solution at the leading order of the gradient expansion, corresponding to the WAM.

While the WAM corresponds to the time-independent (approximate) solution of ${H^j}_i$, it is not necessarily the case for the metric perturbation $\delta \gamma_{ij}$. Since ${H^j}_i$ is related to $\delta \gamma_{ij}$ as
\begin{align}
    \delta \gamma_{ij} \equiv \gamma_{ij} - \bar{\gamma}_{ij} \simeq \frac{1}{2} \left( \bar{\gamma}_{il} {H^l}_j  +  {H^l}_i \bar{\gamma}_{lj}  \right) + \cdots \,,  \label{Exp:lineargamma}
\end{align}
the WAM for $\delta \gamma_{ij}$ is given by 
\begin{align}
    \delta {\gamma}^{\rm WAM}_{ij} (t,\, \bm{x}) =  \frac{1}{2} (\bar{\gamma}_{ik}(t) {C^k}_j(\bm{x}) + {C^k}_i (\bm{x}) \bar{\gamma}_{kj}(t)) + \cdots \,. \label{Exp:gammaWAM}
\end{align}
When the background spatial metric $\bar{\gamma}_{ij}$ is time dependent, $\delta \gamma_{ij}$ for the WAM, generated by promoting the shear transformation, becomes time dependent, although $\delta H^j{}_i=C^j{}_i$ is time independent. This is the situation when the background shear is not negligible, e.g., being sourced by (the electric component of) the background gauge fields.

For $D'=1$, where a single gauge field breaks the global rotation symmetry, there exist 4 WAMs generated by the shear transformation, corresponding to the number of the degrees of freedom, {\it i.e.}, 
$$
 3^2 - 1 ({\rm traceless}) - 3 ({\rm transverse}) - 1 ({\rm rotation~sym.}) = 4\,,
$$
for linear perturbations. The number of degrees of freedom of $C^j{}_i$ is 8 because it should be traceless. 
Three of them can be removed by imposing the gauge conditions. Furthermore, here, we should take into account the fact that there remains one rotation symmetry on the plane which is orthogonal to the background gauge field. This transformation does not generate any perturbation. 

For $D' \geq 2$, since the global rotation symmetry of the background spacetime is entirely broken, there exist 
$$
 3^2 - 1 ({\rm traceless}) - 3 ({\rm transverse})  = 5\,,
$$
WAMs for $\gamma_{ij}$ at linear order, where the U(1) gauge conditions can be still fulfilled both before and after the inhomogeneous shear transformation~\cite{Tanaka:2021dww}. A more general argument about the number of the WAMs can be found in Sec.3.5 of the published version of Ref.~\cite{Tanaka:2021dww}. 

\subsection{Explicit confirmation of WAMs and their long-wavelength evolution}\label{SSSec:massgap}
In Sec.~\ref{SSSec:formal}, we have derived the WAMs for $\delta \gamma_{ij}$ in an anisotropic background from a formal argument. From an explicit computation, we can also show that Eqs.~(\ref{Exp:AWAM}) and (\ref{Exp:gammaWAM}) indeed become an approximate solution at the leading order of the gradient expansion, which satisfies all the equations when we ignore the contributions that decay in the late time limit. For a notational brevity, in this subsection, we set $D'=1$, dropping the index of the gauge field.

In the ADM formalism, the $(i,\, j)$ components of the Einstein equation in the $N_i=0$ gauge are given by
\begin{align}
    &\frac{1}{N \sqrt{g}} \partial_t \left( \sqrt{g} P^{ij} \right) + 2 (K^{il} {K^j}_{l} -  K^{ij} K) - \frac{1}{2} g^{ij} ({K^l}_m {K^m}_l -  K^2 ) \cr
    & \qquad  + {^{\rm s}\!R}^{ij} - \frac{1}{2}g^{ij}\, {^{\rm s}\!R} - \frac{1}{N} (D^i D^j - g^{ij} D^2 ) N  = \frac{1}{\Mp^2} T^{ij}\,,  \label{Eq:EijTLgen}
\end{align}
where $D_i$ is the covariant derivative associated with $g_{ij}$, $D^2 \equiv g^{ij} D_i D_j$, ${^{\rm s}\!R}_{ij}$ and ${^{\rm s}\!R}$ are the spatial Ricci tensor and scalar,  $T^{ij}$ is (the spatial component of) the energy-momentum tensor, given by $T_{ij} = g_{ij} {\cal L}_{\rm mat} - 2 \partial {\cal L}_{\rm mat}/\partial g^{ij}$, and $P^{ij}$ is given by
\begin{align}
    P^{ij} \equiv \frac{\partial L}{\partial \dot g_{ij}} =K^{ij} -  K g^{ij} = A^{ij} - \frac{2}{3} K  g^{ij}\,.
\end{align}
The trace part of Eq.~(\ref{Eq:EijTLgen}) reads
\begin{align}
    \frac{1}{N \sqrt{g}} \partial_t (\sqrt{g} K) - \frac{1}{2} K^2  + \frac{3}{4} {A^i}_j {A^j}_i + \frac{1}{4} {^{\rm s}\!R} - \frac{1}{N} D^2 N  =  - \frac{3}{2} \frac{P}{\Mp^2} \,,  \label{Eq:EijT}
\end{align}
and the traceless part reads
\begin{align}
    \frac{1}{N \sqrt{g}} \partial_t \left( \sqrt{g} {A^i}_j \right) + \left[ {{^{\rm s}\!R}^i}_j - \frac{1}{N} D^i D_j N \right]^{\rm TL} = 8 \pi G\,{\Pi^i}_j\,, \label{Eq:ijSap}
\end{align}
where $P$ and ${\Pi^i}_j$ are the isotropic and anisotropic pressures, defined by
\begin{align}
 P \equiv \frac{1}{3} {T^i}_i\,, \qquad    \Pi^{ij} \equiv \left( \delta^i\!_k \delta^j\!_l - \frac{1}{3} g^{ij} g_{kl} \right) T^{kl}\,.
\end{align}
We can further rewrite Eq.~(\ref{Eq:ijSap}) as 
\begin{align}
    \partial^2_\tau \gamma_{ij} + K \partial_\tau \gamma_{ij} - \gamma^{lm} \partial_\tau \gamma_{il} \partial_\tau \gamma_{jm}   + e^{- 2 \psi} \left[{^{\rm s}\!R}_{ij} - \frac{1}{N} D_i D_j N \right]^{\rm TL} = 16 \pi G\, e^{- 2 \psi} {\Pi}_{ij}  \label{Eq:gammaij}
\end{align}
with $d \tau = N dt$.

At the linear perturbation, choosing the $\delta \psi=0$ slicing, we obtain 
\begin{align}
  & {^{\rm s}\!R}_{ij} = \frac{1}{2} \bar{\gamma}^{kl} (\delta \gamma_{lj, ki} + \delta \gamma_{il, kj} - \delta \gamma_{ij, kl})\,, \\
  & {^{\rm s}\!R} = e^{- 2 \bar{\psi}} \bar{\gamma}^{ij} \bar{\gamma}^{kl} \delta \gamma_{ik, jl}\,, 
\end{align}
and 
\begin{align}
     {\Pi}_{ij} = - \frac{1}{N^2} f^2(\phi)  \left[ \dot{A}_i \dot{A}_j - \frac{1}{3} \gamma_{ij} \gamma^{kl} \dot{A}_k \dot{A}_l  \right] \,.  
\end{align}
Here, we have used $\bar{\gamma}^{ij} \delta \gamma_{ij} = 0$, which can be derived by using $\det \gamma = \det \bar{\gamma} = 1$. 

Taking into account only the perturbations of $\gamma_{ij}$ and $A_i$ ({\it i.e.}, ignoring the scalar perturbation), Eq.~(\ref{Eq:gammaij}) reads
\begin{align}
    \ddot{\bar{\gamma}}_{ij} + \bar{K} \dot{\bar{\gamma}}_{ij} - \bar{\gamma}^{lm} \dot{\bar{\gamma}}_{il} \dot{\bar{\gamma}}_{jm} = \frac{2}{\Mp^2}\, e^{- 2 \bar{\psi}} {\bar{\Pi}}_{ij} \label{Eq:gammaij_bg}
\end{align}
for the background and 
\begin{align}
  &  \delta \ddot{\gamma}_{ij} + \bar{K} \delta \dot{\gamma}_{ij} - 2 \bar{\gamma}^{lm} \dot{\bar{\gamma}}_{m\{i} \delta \dot{\gamma}_{j\}l} - \frac{4}{3} \frac{\bar{\rho}_A}{\Mp^2} \delta \gamma_{ij} + 4 {\bar{A}^k}_i {\bar{A}^l}_j \delta \gamma_{kl} - \frac{2}{3 \Mp^2} \bar{\gamma}_{ij} e^{-2 \bar{\psi}} \bar{f}^2 \dot{\bar{A}}_k \dot{\bar{A}}_l \delta \gamma^{kl} \cr
  & = - \frac{2}{\Mp^2} e^{-2 \bar{\psi}} \bar{f}^2 \left( \dot{\bar{A}}_i \delta \dot{A}_j + \delta \dot{A}_i \dot{\bar{A}}_j - \frac{2}{3} \bar{\gamma}_{ij} \bar{\gamma}^{kl} \dot{\bar{A}}_k \delta \dot{A}_l  \right) + {\cal O}(\epsilon^2)\,,\label{Eq:gammaij_linear}
\end{align}
for linear perturbations. 

When we ignore the background shear and the background value of the gauge field, the third to the sixth terms in the first line all vanish, reproducing the usual dispersion relation for GWs with $\omega \to 0$ in the limit $k \to 0$, {\it i.e.}, being gapless. Meanwhile, a violation of the background isotropy due to the gauge field seems to modify the dispersion relation, making GWs gapped. The last three terms in the first line can be interpreted as the mass term. For example, for the solution addressed in Sec.~\ref{Cond:deltaI}, $ {\bar{A}^k}_i {\bar{A}^l}_j$ typically becomes $K^2 \times {\cal O} ((\rho_A/\rho)^2 )$. Then, the fifth term becomes smaller than the fourth term, whose coefficient is of $K^2 \times {\cal O} (\rho_A/\rho)$. The last term does not appear after projecting into the transverse traceless basis at the order of $K^2 \times {\cal O} (\rho_A/\rho)$, where the background spatial metric can be replaced with $\delta_{ij}$. A similar modification of the dispersion relation was reported in Ref.~\cite{Flauger:2017ged}. 

ven by Eqs.~(\ref{Exp:gammaWAM}) and (\ref{Exp:AWAM}) satisfies the U(1) gauge constraint, ${\cal H}_{\rm U(1)}$. Precisely speaking, this solution does not satisfy the momentum constraints ${\cal H}_i$ likewise the WAM for $\psi$. Nevertheless, this error falls off as the inverse of the physical volume, ensuring that Eqs.~(\ref{Exp:AWAM}) and (\ref{Exp:gammaWAM}) indeed provide an approximate solution at the linear perturbation in the large scale limit.

\section{Circular polarization of GWs} \label{Sec:circualrP}
When one wants to consider a model with a parity violation, it may be more convenient to use the circular polarization basis of GWs. Here, let us also provide the formulae in this basis. Operating the circular polarization bases on Eq.~(\ref{Sol:gamma2}), we obtain
\begin{align}
    \gamma^{(\pm_c)}(t_f,\, \bm{k}) &= \gamma^{(\pm_c)}_*(\bm{k}) - \sum_{\alpha=1}^{D'} \sin^2 \Theta_\alpha \,  e^{\pm 2 i \Psi_\alpha}\delta \hat{{\cal I}}_{(\alpha)f}(\bm{k}) \cr
     & \quad - 2 \sqrt{2} \sum_{\alpha=1}^{D'}  \sin \Theta_\alpha e^{\pm i \Psi_\alpha } 
 \bar{\hat{{\cal I}}}_{(\alpha)f} \frac{\delta \pi^j_{(\alpha)}(\bm{k})}{\Bar{\pi}_{(\alpha)}}   \biggl[ \left( 1 - \frac{\sin^2 \Theta_\alpha}{2} \right) e_j^{(\mp_c)}(\hat{\bm{k}})  \cr
     & \qquad \qquad \qquad \qquad \qquad \qquad \qquad \qquad \qquad  \quad  - \frac{\sin^2 \Theta_\alpha}{2}  e^{\pm 2 i \Psi_\alpha }  e_j^{(\pm_c)}(\hat{\bm{k}})  \biggr], \label{Exp:gammapm_linear}
\end{align}
where we have used
\begin{align}
   &  e_i^{(\mp_c)} (\hat{\bm{k}})\bar{\hat{\pi}}^i_{(\alpha)} = \frac{1}{\sqrt{2}} \sin \Theta_\alpha e^{\pm i \Psi_\alpha} \,,  \\
   & e_i^{(\mp_c)}(\hat{\bm{k}}) \delta \hat{\pi}^i_{(\alpha)}(\bm{k}) = \frac{\delta \pi^i_{(\alpha)}(\bm{k})}{\Bar{\pi}_{(\alpha)}}   \biggl[ \left( 1 - \frac{\sin^2 \Theta_\alpha}{2} \right) e_i^{(\mp_c)}(\hat{\bm{k}})    - \frac{\sin^2 \Theta_\alpha}{2}  e^{\pm 2 i \Psi_\alpha }  e_i^{(\pm_c)}(\hat{\bm{k}})  \biggr]. 
\end{align}
When we can ignore the metric perturbations at $t=t_*$, inserting Eq.~(\ref{Exp:deltaI}) into Eq.~(\ref{Exp:gammapm_linear}), we obtain
\begin{align}
  &  \gamma^{(\pm_c)}(t_f,\, \bm{k}) - \gamma^{(\pm_c)}_*(\bm{k}) \cr
  & = - \sum_{\alpha=1}^{D'} \sin^2 \Theta_\alpha e^{\pm 2 i \Psi_\alpha} \left( \frac{\partial \bar{\hat{{\cal I}}}_{(\alpha)f}}{\partial \bar{\Phi}^I_*}  \right)_{\!\!\pi_{(\alpha)}} \hspace{-5pt} \delta \Phi^I_*(\bm{k}) \cr
  & \qquad - 2 \sqrt{2} \sum_{\alpha=1}^{D'}  \sin \Theta_\alpha e^{\pm i \Psi_\alpha } 
 \bar{\hat{{\cal I}}}_{(\alpha)f} \frac{\delta \pi^j_{(\alpha)}(\bm{k})}{\Bar{\pi}_{(\alpha)}}   \biggl[ \left( 1 + \frac{{\cal D} \bar{\hat{{\cal I}}}_{(\alpha)f}}{\bar{\hat{{\cal I}}}_{(\alpha)f}} \frac{\sin^2 \Theta_\alpha}{2} \right) e_j^{(\mp_c)}(\hat{\bm{k}})  \cr
     & \qquad \qquad \qquad \qquad \qquad \qquad \qquad \qquad \qquad \qquad   +\frac{{\cal D} \bar{\hat{{\cal I}}}_{(\alpha)f}}{\bar{\hat{{\cal I}}}_{(\alpha)f}} \frac{\sin^2 \Theta_\alpha}{2}  e^{\pm 2 i \Psi_\alpha }  e_j^{(\pm_c)}(\hat{\bm{k}})  \biggr]\!. \label{Exp:gammapm_linear2}
\end{align}
As was commented in footnote \ref{footnote:replacement}, the expression remains the same even if we expand $\gamma^{(\pm_c)}(t_f,\, \bm{k})$ in terms of $\delta \dot{A}_{i(\alpha)*}$.  

Similarly to Sec.~\ref{Sec:PowerS}, assuming that the contributions of the gauge fields to the background evolution had been negligibly small until $t=t_*$ and using the power spectrums (\ref{Exp:PA}) in the circular polarization bases, here let us also compute the power spectrums of GWs in the circular polarization basis. In the presence of the Chern-Simons terms in Eq.~(\ref{Exp:Lmattergeneral}), the amplitudes of $P_*^{(\alpha) \pm_c}(k)$ can be different.  The cross-correlation between the two polarization modes of the gauge fields at $t=t_*$ is set to 0, since it vanishes at the linear perturbation. In this case, the initial time $t_*$ should be set after the spatial gradient terms which appear from the Chern-Simons terms become negligible.

Using Eq.~(\ref{Exp:gammapm_linear2}) and (\ref{Ep:ThetaPsiparity}), we obtain
\begin{align}
    &\langle \gamma^{(\pm_c)} (t_f,\, \bm{k}) \gamma^{(\pm_c)} (t_f,\, \bm{p}) \rangle \cr
    & = \delta (\bm{k} + \bm{p}) \frac{2}{k^3} \left( \frac{H_{*}}{\Mp} \right)^2\! \left( 1 +  \sum_{\alpha=1}^{D'} g_{t (\alpha)}^{\pm_c} (\bm{k}) \sin^2 \Theta_\alpha (\hat{\bm{k}}) + \Delta_\phi   \right)  \,, \label{Def:gt_cp}
\end{align}
with 
\begin{align}
    & g_{t (\alpha)}^{\pm_c}(\bm{k}) = g_{t(\alpha)}  \left\{ \left(1 + \frac{{\cal D}\bar{\hat{{\cal I}}}_{(\alpha)f}}{\bar{\hat{{\cal I}}}_{(\alpha)f}} \frac{\sin^2 \Theta_\alpha(\hat{\bm{k}})}{2} \right)^2 {\cal P}^{\pm_c}_{(\alpha)*}(k) + \frac{1}{4} \left(\frac{{\cal D}\bar{\hat{{\cal I}}}_{(\alpha)f}}{\bar{\hat{{\cal I}}}_{(\alpha)f}} \right)^2 \sin^4 \Theta_\alpha(\hat{\bm{k}}) {\cal P}^{\mp_c}_{(\alpha)*} (k)  \right\}\,, \label{Exp:gt_cp} \\
    & \Delta_\phi  =  \frac{1}{2}  \sum_{\alpha=1}^{D'}  g_{t(\phi;\alpha\alpha)}  \sin^4 \Theta_\alpha  + \sum_{\alpha=1}^{D'}\sum_{\beta=1}^{\alpha -1} g_{t(\phi;\alpha\beta)}  \sin^2 \Theta_\alpha \sin^2 \Theta_\beta \cos 2(\Psi_\alpha(\hat{\bm{k}}) - \Psi_\beta (\hat{\bm{k}})))  \,,\label{Exp:Deltaphipm}
\end{align}
where $g_{t(\alpha)}$ and $g_{t(\phi;\alpha\alpha)}$ are defined by Eqs.~(\ref{Exp:gt}) and (\ref{Exp:gtphi}).Here, ${\cal P}_{(\alpha)*}^{\pm_c}$ measures the amplitude of the gauge fields at $t=t_*$ by comparing it with those for $g_{(\alpha)} = 0$, given in Eq.~(\ref{Eq:ampA}) as 
\begin{align}
    {\cal P}_{(\alpha)*}^{\pm_c}(k) \equiv \frac{P_{(\alpha)*}^{\pm_c}(k)/\dot{\bar{A}}_{(\alpha)*}^2}{\frac{1}{12 k^3} \frac{\Bar{\rho}_{*}}{\Bar{\rho}_{A (\alpha)*}} \biggl( 1 + \Bigl( \frac{\dot{\bar{f}}_{(\alpha)*}}{H_* \bar{f}_{(\alpha)}} \Bigr)^2 \biggr) \left( \frac{H_{*}}{\Mp} \right)^2}\,. 
\end{align}
For $g_{(\alpha)} = 0$, we find ${\cal P}_{(\alpha)*}^{+_c} = {\cal P}_{(\alpha)*}^{-_c}$. Equation (\ref{Def:gt_cp}) with Eqs.~(\ref{Exp:gt_cp}) and (\ref{Exp:Deltaphipm}) shows that how the circular polarization of the gauge fields at $t=t_*$ is transferred to GWs at $t=t_f$. The scalar field(s) contribute to the two circular polarization modes of GWs in the same way. As discussed in Sec.~\ref{SSSec:parity}, in the absence of the circular polarization at $t=t_*$, the auto power spectrums of the two circular polarization modes $t=t_f$ coincide as discussed at the end of Sec.~\ref{SSec:generalformulaPS}.

Similarly, we obtain the cross correlation between the two polarization modes as
\begin{align}
     \langle \gamma^{(\pm_c)} (t_f,\, \bm{k}) \gamma^{(\mp_c)} (t_f,\, \bm{p}) \rangle = 
     \delta (\bm{k} + \bm{p}) \frac{2}{k^3} \left( \frac{H_{*}}{\Mp} \right)^2 \left(  \,\sum_{\alpha=1}^{D'} g_{t (\alpha)}^{{\rm cr}\pm_c} (\bm{k}) \sin^2 \Theta_\alpha (\hat{\bm{k}})+ \Delta^{{\rm cr}\pm_c}_\phi(\bm{k})   \right)
\end{align}
with
\begin{align}
    &g_{t (\alpha)}^{{\rm cr}\pm_c} (\bm{k}) =  g_{t(\alpha)} \sin^2 \Theta_\alpha(\hat{\bm{k}}) e^{\pm 4i \Psi_\alpha(\hat{\sbm{k}})}  \cr
    & \qquad \qquad \qquad \times \frac{1}{2} \frac{{\cal D}\bar{\hat{{\cal I}}}_{(\alpha)f}}{ \bar{\hat{{\cal I}}}_{(\alpha)f}} \left( 1 + \frac{1}{2} \frac{{\cal D}\bar{\hat{{\cal I}}}_{(\alpha)f}}{\bar{\hat{{\cal I}}}_{(\alpha)f}} \sin^2 \Theta_\alpha(\hat{\bm{k}})  \right) \left[ {\cal P}_{(\alpha)*}^{+_c}(k) + {\cal P}_{(\alpha)*}^{-_c}(k) \right], \\
    & \Delta_{\phi}^{\rm cr, l} (\bm{k})  = \frac{1}{2} \sum_{\alpha=1}^{D'} \sum_{\beta=1}^{D'}  g_{t(\phi;\alpha\beta)} \sin^2 \Theta_\alpha(\hat{\bm{k}})  \sin^2 \Theta_\beta(\hat{\bm{k}}) e^{\pm 2i (\Psi_\alpha(\hat{\sbm{k}})+ \Psi_\beta(\hat{\sbm{k}}))}.
\end{align}
The complex conjugate of $\langle \gamma^{(\pm_c)} (t_f,\, \bm{k}) \gamma^{(\mp_c)} (t_f,\, \bm{p}) \rangle$ becomes $\langle \gamma^{(\mp_c)} (t_f,\, \bm{k}) \gamma^{(\pm_c)} (t_f,\, \bm{p}) \rangle$ as a consequence of $\gamma^{(\pm_c)\dagger}(t_f,\, \bm{k}) =\gamma^{(\pm_c)}(t_f,\, -\bm{k}) $.

\acknowledgments
T.~T. is supported by Grant-in-Aid for Scientific Research under Contract Nos. JP23H00110,  JP20K03928, JP24H00963, and JP24H01809. Y.~U. is supported by Grant-in-Aid for Scientific Research under Contract Nos.~JP19H01894, JP21KK0050, and JP23K25873, and JST FOREST Program under Contract No.~JPMJFR222Y. This research was in part supported by the Munich Institute for Astro-, Particle and BioPhysics (MIAPbP), which is funded by the Deutsche Forschungsgemeinschaft (DFG, German Research Foundation) under Germany´s Excellence Strategy – EXC-2094 – 390783311.

\bibliography{refst}

\end{document}